\documentclass[acmsmall, nonacm, screen]{preprint}

\usepackage{float}        
\usepackage{wrapfig}     
\usepackage{tabularx}    
\usepackage{multirow}
\usepackage{pgfplots}
\pgfplotsset{compat=1.18}
\usepackage{enumitem}
\usepackage{pifont}

\usepackage{tikz}
\usetikzlibrary{positioning, arrows.meta, fit, backgrounds, calc}
\usepackage{graphicx}
\providecommand{\Description}[1]{}   
\usepackage{booktabs}
\definecolor{ethcolor}{HTML}{2F3A4A}
\definecolor{bsccolor}{HTML}{B85C38}
\definecolor{basecolor}{HTML}{2E6FCE}
\colorlet{combocolor}{blue!60}

\usepackage{graphicx}
\usepackage{subcaption}
\usepackage{makecell}

\usepackage{xcolor}

\theoremstyle{definition}

\usepackage{tcolorbox}
\usepackage{fontawesome5}   

\tcbuselibrary{skins, breakable}

\newtcolorbox{insightbox}[1][Insight]{
  breakable,
  enhanced,
  colback    = blue!5,
  colframe   = blue!40!gray,
  leftrule   = 3.5pt,
  toprule    = 0pt,
  bottomrule = 0pt,
  rightrule  = 0pt,
  arc        = 2pt,
  left       = 26pt,
  right      = 8pt,
  top        = 4pt,
  bottom     = 4pt,
  overlay    = {
    \fill[blue!15] (frame.south west) rectangle ([xshift=22pt]frame.north west);
    \node[
      rotate        = 90,
      anchor        = center,
      font          = \fontsize{7.5}{9}\selectfont\bfseries,
      text          = blue!60!black,
    ] at ([xshift=11pt]frame.west |- frame.center)
      {\faLightbulb[regular]\enspace #1};
  },
}

\newcommand{\mypie}[2]{%
  \tikz[baseline=-0.5ex]{%
    \fill[gray!25] (0,0) circle (0.13cm);%
    \fill[#1] (0,0) -- (90:0.13cm) arc (90:90-#2*3.6:0.13cm) -- cycle;%
  }%
}

\hypersetup{
    colorlinks=true,
    linkcolor=blue,
    filecolor=magenta,      
    urlcolor=teal,
    }

\setcopyright{none}
\renewcommand\footnotetextcopyrightpermission[1]{}
\authorsaddresses{}  

\begin{document}
\title[An Empirical Study of ERC-8004]
{Can Trustless Agents Be Trusted? An Empirical Study of the ERC-8004 Decentralized AI Agent Ecosystem}


\author{Xihan Xiong}
\affiliation{%
  \institution{Imperial College London}
  \country{UK}
  }
\author{Zelin Li}
\affiliation{%
  \institution{Ohio State University}
  \country{USA}
  }

\author{Wei Wei}
\affiliation{%
  \institution{University of Bristol}
  \country{UK}
  }
  \author{Qin Wang}
\affiliation{%
  \institution{CSIRO}
  \country{Australia}
  }

  \author{William Knottenbelt}
\affiliation{%
  \institution{Imperial College London}
  \country{UK}
  }

\author{Zhipeng Wang}
\affiliation{%
  \institution{The University of Manchester}
  \country{UK}
  }
  
\renewcommand{\shortauthors}{Xiong et al.}

\begin{abstract}
As autonomous AI agents increasingly transact across organizational boundaries, a fundamental trust challenge emerges: how can an agent assess whether an unknown counterpart is trustworthy?
The ERC-8004 protocol addresses this challenge with the first permissionless trust layer for AI agent economies, built around three on-chain registries for Identity, Reputation, and Validation. 
Despite its rapid adoption, the protocol has not been studied empirically, leaving it unclear whether the information it records provides a trustworthy basis for decision-making.
To address this gap, we present the first empirical study of ERC-8004 across three chains: Ethereum, BNB Smart Chain (BSC), and Base, covering the period from protocol deployment through May 13, 2026. We crawl on-chain Identity and Reputation events, off-chain files, and x402 payment transactions.

On the identity side, we find that most registrations are placeholders rather than active agents, with only a small fraction ($3\%$, $4\%$, and $15\%$ across Ethereum, BSC, and Base) exposing a valid ERC-8004 registration file with at least one live service endpoint.
On the reputation side, we show that the Registry, as currently deployed, cannot function as a trust signal: values are not commensurable, feedback records are rarely grounded in verifiable interactions, and reputation can be manipulated at minimal cost. Consistent with these design weaknesses, we find that a substantial fraction of reviewers ($73.5\%$, $59.2\%$, and $90.6\%$ across Ethereum, BSC, and Base) exhibit coordinated Sybil behavior.
After removing Sybil-flagged feedback, $15.8\%$, $77.9\%$, and $86.8\%$ of rated agents, respectively, are left with no valid feedback.
We then turn these findings into concrete recommendations for future revisions of ERC-8004. Our study yields actionable protocol-design implications and establishes an empirical baseline for research on AI agent markets.
\end{abstract}


\maketitle

\section{Introduction}
\label{sec:intro}

Autonomous AI agents are evolving from passive assistants into active economic participants capable of executing transactions and operating continuously without human supervision~\cite{park2023generative,kim2026sok,sapkota2025ai, lui2024sok}. This transition creates a coordination challenge: agents from different organizations, built on heterogeneous frameworks, and operating across untrusted networks must nonetheless discover one another, assess each other's reliability, and transact without pre-existing relationships~\cite{guo2026agent}.

The ERC-8004 (``Trustless Agents'') protocol~\cite{erc8004} fills this gap. Introduced on the Ethereum mainnet on January 29, 2026, it establishes an on-chain trust layer between agent communication protocols (e.g., A2A~\cite{google2025a2a} and MCP~\cite{anthropic2024mcp}) and payment rails (e.g., x402~\cite{coinbase2025x402}). The protocol defines three registries, each deployed once per chain. The \emph{Identity} Registry assigns agents portable ERC-721-based identities. The \emph{Reputation} Registry records publicly accessible feedback that can be consumed by users, agents, or higher-level reputation systems. The \emph{Validation} Registry stores independent attestations of agent performance. Together, these registries enable agents to evaluate previously unknown counterparts before committing value, without relying on a centralized intermediary.

Adoption has been real and fast. In its first few months, ERC-8004 drew more than $170$k registered agents (Figure~\ref{fig:cum_agents}) across Ethereum, BNB Smart Chain (BSC), and Base, with a total reputation market of over $150$k feedback records (Table~\ref{tab:rep_participation}). 
Despite this rapid growth and its ambition to serve as the first open trust layer for agent economies, the protocol has not yet been studied empirically. We do not know who is using it, whether its registered identities correspond to functional agents, or whether the reputation it exposes can be trusted. Yet these are precisely the questions autonomous agents must answer before deciding whether to interact with an unknown counterpart.

To address this gap, we present the first cross-chain empirical study of ERC-8004 as deployed in practice. We collect every event from the Identity and Reputation Registries across all three chains that host the protocol, from each chain's deployment through 13 May 2026, with the referenced off-chain registration files, gas costs, and x402 payment transactions. Note that the Validation Registry had no confirmed mainnet deployment during our observation period and is therefore outside the scope of this study. Using this dataset, we ask three fundamental questions: Who registers agents? Are the registered identities meaningful? Can the Reputation Registry serve as a trustworthy trust signal? 
We summarize our main contributions as follows:

\begin{itemize}[topsep=2pt, leftmargin=*, itemsep=3pt]
  \item \textbf{Multi-chain dataset (\S\ref{sec:data}).} We assemble the first comprehensive dataset of ERC-8004 activity: every Identity and Reputation event on Ethereum, BSC, and Base, enriched with off-chain registration and feedback files, per-transaction gas costs, and x402 settlements.
  Researchers may request access to the dataset to support reproducibility and follow-on research.

  \item \textbf{Identity and adoption analysis (\S\ref{sec:identity_analysis}).} We analyze who registers and what they register. Batch-minted placeholders and templated deployments dominate registration, and ownership is highly concentrated on Ethereum (Gini $0.733$) and Base (Gini $0.708$). Platform-mediated registration and identity distribution (e.g., Nookplot) help explain this concentration.
  We also find that most identities never become active agents, with missing URIs for $53\%$ on Ethereum, $9\%$ on BSC, and $37\%$ on Base. 
 Across chains, only $3\%$ to $15\%$ expose a valid ERC-8004 registration file with at least one declared service endpoint. This reveals a substantial gap between on-chain registration counts and genuine agent deployment, suggesting that raw registration volume is a poor proxy for ecosystem maturity. Agent functional profiles also differ across chains: BSC is dominated by trading-related applications, while Base hosts the broadest range of advertised functions.

  \item \textbf{Reputation analysis, including security (\S\ref{sec:reputation_analysis}, \S\ref{sec:rep_security}).} We study the Reputation Registry as a trust signal. As deployed, it meets none of the four necessary conditions for a trustworthy score: values are not commensurable, the aggregated score can be moved by a single input, feedback records are not tied to verifiable interactions, and a reputation can be fabricated or destroyed at minimal cost (median cost: \$$0.055$/\$$0.0042$/\$$0.0027$ on Ethereum/BSC/Base). 
  Measuring manipulation in the wild, we identify Sybil-flagged reviewers accounting for $73.5\%$, $59.2\%$, and $90.6\%$ of reviewers on Ethereum, BSC, and Base, respectively, affecting the displayed reputations of $26.4\%$, $81.4\%$, and $96.2\%$ of rated agents. Removing their feedback leaves $15.8\%$, $77.9\%$, and $86.8\%$ of rated agents on Ethereum, BSC, and Base without any valid feedback remaining to support their reputation.

  \item \textbf{Recommendations and disclosure (\S\ref{sec:recommendations})}. We turn each finding into an implementable design change for the next revision of ERC-8004. We have also responsibly disclosed our findings to the ERC-8004 proposer (Ethereum Team) and incorporated their feedback into our paper.
\end{itemize}

\section{Technical Warm-ups}
\label{sec:background}

\noindent\textbf{From AI assistants to economic agents}.
LLMs increasingly act as autonomous \emph{agents} that plan, use tools, and delegate tasks~\cite{park2023generative,kim2026sok}.
With wallets and programmatic payments, they can trade services across organizational boundaries~\cite{tomasev2025virtual}.
This creates a trust problem among previously unknown agents.
Existing protocols address communication, but not trust: Anthropic's Model Context Protocol~(MCP)~\cite{anthropic2024mcp} standardizes agent-to-tool interaction, while Google's Agent-to-Agent (A2A) protocol~\cite{google2025a2a} supports discovery, messaging, and task lifecycles via \emph{AgentCards}.
Both define \emph{how} parties communicate, but leave \emph{whether} they can be trusted to the application layer~\cite{erc8004}.

\vspace{0.1cm}
\noindent\textbf{The trust gap and the case for an on-chain layer}.
Existing remedies address only parts of this gap. Centralized reputation platforms rely on trusted operators; W3C Decentralized Identifiers (DIDs)~\cite{sporny2022did} establish identity, not behavioral reputation; and web PKI authenticates domains, not conduct.
A permissionless blockchain offers censorship-resistant identifiers and an append-only feedback ledger directly accessible to smart contracts.
Combined with machine-native payment rails such as x402~\cite{coinbase2025x402,li2026five,ling2026free}, which settles payments within the HTTP request/response cycle, it enables a shared identity, reputation, and payment stack for autonomous commerce.

\vspace{0.1cm}
\noindent\textbf{ERC-8004 in the agentic-internet stack}.
The ERC-8004 protocol~\cite{erc8004} fills this gap. It inserts a \emph{trust layer} between the communication protocols (A2A/MCP) and value settlement (x402): communication makes agents legible, the trust layer makes them accountable, and settlement makes them payable (Figure~\ref{fig:stack}). Its trust is deliberately pluggable and tiered, scaling with the value at risk, from lightweight client reputation for low-stakes tasks to crypto-economic or cryptographic validation (stake-secured re-execution, zkML proofs \cite{chen2024zkml}, TEE attestation ~\cite{costan2016intel}) for high-stakes ones~\cite{erc8004}.

\section{System Model: The ERC-8004 Protocol}
\label{sec:protocol}

The ERC-8004 system~\cite{erc8004} contains three on-chain singleton registries, deployed once per EVM-compatible chain. The {Identity} Registry gives every agent a portable handle. The {Reputation} Registry collects client feedback about agents. The {Validation} Registry records independent checks of an agent's work. The Identity Registry is the anchor of the design; the other two keep a reference to its address and ask it, before accepting any write, whether the caller is really allowed to act for the agent in question. Technically, the standard builds on ERC-721 for the identity token~\cite{entriken2018erc721}, EIP-155 for cross-chain identifiers~\cite{buterin2016eip155}, and EIP-712/ERC-1271 for signature verification~\cite{bloemen2017eip712, giordano2018eip1271}.

\begin{figure}[!]
\centering
    \includegraphics[width=\linewidth]{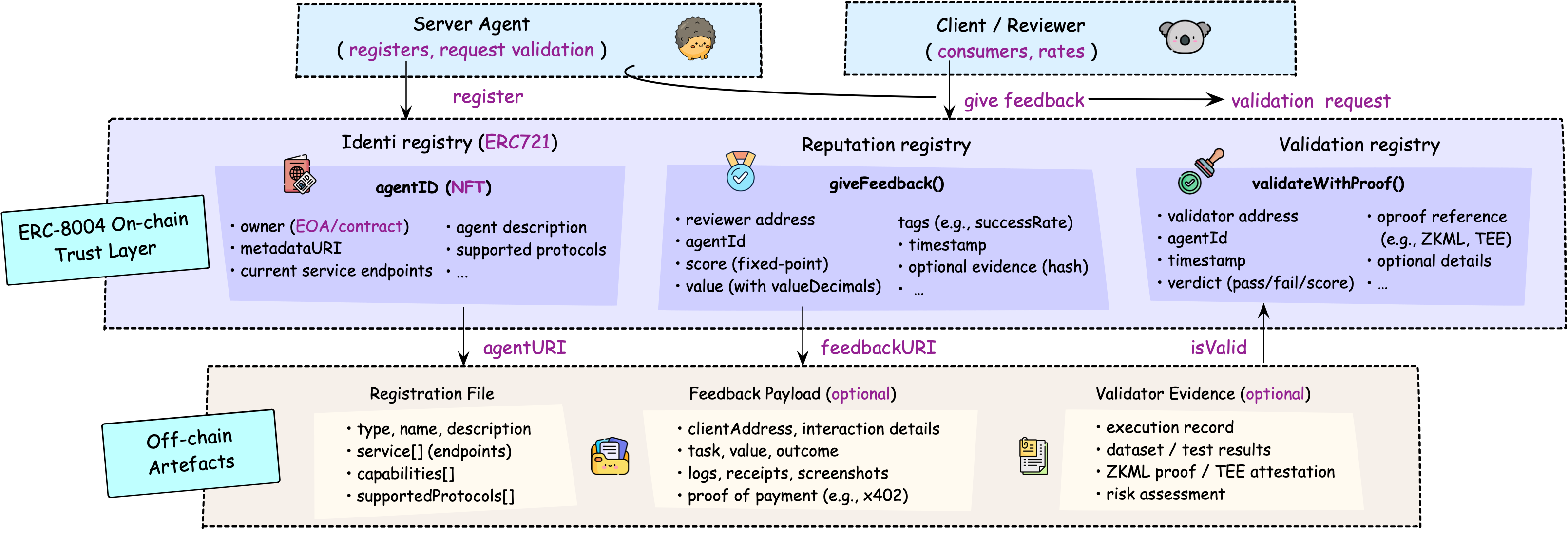}
    \caption{ERC-8004 protocol architecture~\cite{erc8004}. Three on-chain singleton registries, Identity (ERC-721), Reputation, and Validation, anchor only \emph{pointers and commitments} (URIs and hashes), while content-heavy artifacts (registration files, feedback payloads, validator evidence) live off-chain (e.g., on IPFS/HTTPS).}
    \label{fig:architecture}
\end{figure}

\smallskip
The \textbf{\textit{Identity Registry}} extends ERC-721~\cite{entriken2018erc721} with \texttt{URIStorage}, so every agent is a transferable NFT that existing wallets and indexers can browse. An agent ($a$) is identified by its chain namespace, chain ID, Registry address, and an~\texttt{agentId} assigned incrementally at registration. Its ERC-721 \texttt{tokenURI}, which we call the \texttt{agentURI} ($u_a$), resolves to the off-chain registration file ($R_a$):
\begin{equation}
a \;=\; \langle\, \mathtt{eip155} : \mathit{chainId} : \mathit{registry} ,\ \mathit{agentId} \,\rangle
\label{eq:registration}
\end{equation}

Equation~\ref{eq:registration} defines an agent's canonical identifier across chains and registries. Here, \texttt{eip155}~\cite{buterin2016eip155} denotes the namespace for EVM-compatible chains, while \textit{chainId} is the EIP-155 chain identifier.

\vspace{0.1em}
\noindent\textbf{Registration and activation.} Since $u_a$ may be empty at mint and set later,  we distinguish minting from
\emph{activation} (the first assignment of a non-empty $u_a$). This gap drives the identity-activity analysis of \S\ref{sec:Identity-Activity Gap}; replaying \texttt{URIUpdated} events recovers each agent's current URI.

\vspace{0.1em}
\noindent\textbf{Payment wallet.} A reserved metadata key designates the agent's payout address $w_a$. It is initialized to the owner's address, changed only by a signature proving control of the new wallet (EIP-712 for Externally Owned Accounts (EOAs), ERC-1271 for contract wallets~\cite{bloemen2017eip712, giordano2018eip1271}), and cleared on transfer. We use agent $a$'s payout address $w_a$ to attribute x402 settlement transactions to agent $a$ (\S\ref{sec:data}).

\vspace{0.1em}
\noindent\textbf{Off-chain registration file.} 
The registration file $R_a$ at $u_a$ is the agent's public ``passport'', a JSON document declaring its identity and capabilities. We call $R_a$ \emph{compliant} iff $\mathrm{type}(R_a)=\tau^\star$, the canonical ERC-8004 identifier, a criterion we use in \S\ref{sec: service_and_quality}. Alongside descriptive metadata,
$R_a$ carries a \texttt{services} array of typed endpoints (e.g., A2A, MCP, HTTPS, OASF~\cite{agntcy2025oasf}, ENS, DID); a \texttt{registrations} array declaring the agent's identities on other chains, which provides self-declared, unverified cross-chain links; and optional \texttt{supportedTrust} and \texttt{x402Support} flags.
An empty \texttt{supportedTrust} marks an agent that uses ERC-8004 for discovery only, not trust.

\smallskip
The \textbf{\textit{Reputation Registry}} records client feedback as compact on-chain signals,
exposing reputation to other contracts while delegating richer evidence to
off-chain files. 

\vspace{0.1em}
\noindent\textbf{Feedback.} A feedback item $f \;=\; (a,\, c,\, v,\, d,\, t_1,\, t_2)$ is issued by a client address ($c$) for an agent ($a$) and carries a signed fixed-point score ($v$), given by a value with \texttt{valueDecimals} ($d$), plus two free-form tags ($t_1,t_2$) that name what the score measures (e.g., \texttt{uptime}, \texttt{successRate}, \texttt{liveness}).

\vspace{0.1em}
\noindent\textbf{Giving, revoking, and responding.} When giving feedback (\texttt{giveFeedback()}), the Registry forbids self-promotion: the agent's controllers or operators cannot rate $a$ (i.e., $c \notin \{\mathrm{owner}(a)\} \cup \mathrm{operators}(a)$). Replaying \texttt{NewFeedback} events recovers $\mathcal{F}_a$, the feedback set of $a$. A client can revoke its own feedback record (\texttt{revokeFeedback()}), which flags but never deletes it, so revoked signals remain auditable. Any party may append a response to existing feedback (\texttt{appendResponse()}).

\vspace{0.1em}
\noindent\textbf{Off-chain feedback file.}  Only $\langle v,d,t_1,t_2\rangle$ and the revocation flag are stored; the endpoint, the feedback file URI ($\phi_f$), and its hash are emitted but not persisted. The file $F_f$ at $\phi_f$ extends the on-chain record with mandatory provenance (agent, client, timestamp, value) and optional context: the MCP tool, A2A task/skill, and a \texttt{proofOfPayment} object (e.g., x402 txHash or nonce).

\smallskip
The \textbf{\textit{Validation Registry}} records independent checks of an agent's work as tuples $(\mathit{validator}, a, b)$ with verdict $b \in \{0,1\}$. It is pluggable~\cite{erc8004}: a validator may use stake-secured re-execution, a zkML proof \cite{chen2024zkml}, or a TEE attestation~\cite{costan2016intel}, with assurance matched to the value at risk~\cite{erc8004}, forming the high-assurance tier of the protocol's trust model. As of this writing, we observed no mainnet deployment of this Registry on the chains we study, so we focus on the Identity and Reputation registries; we describe it here only to situate reputation within the intended trust hierarchy.

\vspace{0.1cm}
Overall, the three registries support an agent’s lifecycle from discovery to feedback (Figure~\ref{fig:architecture}). An agent first registers its identity and publishes its capabilities and service endpoints. A prospective client or marketplace can then consult prior reputation feedback and, where available, independent validation evidence to decide whether to select the agent for a task. After execution, the client can submit feedback to the Reputation Registry, while validators can record independent assessments through the Validation Registry. These records may inform subsequent clients’ selection decisions.

\section{Data}
\label{sec:data}

We collect data from three EVM-compatible chains: Ethereum~(ETH), BNB Smart Chain~(BSC), and Base, which had the highest registration and feedback volumes during our observation period.
 
\begin{table}[htb]
  \centering
  \caption{Data collection time windows.
    All end blocks correspond to the same calendar moment (May~13,~2026, 21:49:47~UTC).
    Each chain's start block precedes the earliest contract deployment to avoid missing events.}
  \label{tab:chain_params}
  \resizebox{\linewidth}{!}{
  \begin{tabular}{ccccp{1.7cm}p{1.7cm}r}
    \toprule
    {Chain} & {Chain ID} & {Identity Registry} & {Reputation Registry} & {Start Block} & {End Block} & \multicolumn{1}{c}{Approx.\ Window} \\
    \cmidrule(rl){2-2}\cmidrule(rl){3-3}\cmidrule(rl){4-4}\cmidrule(rl){5-6}\cmidrule(rl){7-7}
    
    ETH & 1  & \href{https://etherscan.io/address/0x8004A169FB4a3325136EB29fA0ceB6D2e539a432}{0x8004...a432}   & \href{https://etherscan.io/address/0x8004BAa17C55a88189AE136b182e5fdA19dE9b63}{0x8004...9b63}  & $24{,}339{,}000$ & $25{,}089{,}000$ & Jan~29, 2026 -- May~13, 2026 \\
    BSC      & 56   & \href{https://bscscan.com/address/0x8004A169FB4a3325136EB29fA0ceB6D2e539a432}{0x8004...a432}   & \href{https://bscscan.com/address/0x8004BAa17C55a88189AE136b182e5fdA19dE9b63}{0x8004...9b63}   & $79{,}027{,}200$ & $98{,}121{,}735$ & Feb~3, 2026 -- May~13, 2026  \\
    Base     & 8453  & \href{https://basescan.org/address/0x8004A169FB4a3325136EB29fA0ceB6D2e539a432}{0x8004...a432}   & \href{https://basescan.org/address/0x8004BAa17C55a88189AE136b182e5fdA19dE9b63}{0x8004...9b63}  & $41{,}663{,}700$ & $45{,}959{,}820$ & Feb~3, 2026 -- May~13, 2026   \\
    \bottomrule
  \end{tabular}
}

\smallskip
\footnotesize
$^\dagger$ We do not study the Validation Registry because its mainnet deployment is pending as of this writing.
\end{table}

\noindent\textbf{Identity and reputation data}.
For each chain, we crawl on-chain events emitted by both registries and fetch the off-chain content referenced by them. Table~\ref{tab:datasets} in Appendix~\ref{app:data_more} gives a full inventory.

\vspace{0.2em}
\noindent\textit{{Identity Registry.}}
We collect the following event types from the Identity Registry.
The \texttt{Registered} event is emitted at agent creation and serves as the canonical record of each agent's \texttt{agentId}, registering wallet, and initial URI.
Alongside each \texttt{Registered} event, the contract also emits an ERC-721 \texttt{Transfer} from the zero address, which we record separately as the mint signal.
Post-mint ownership changes are captured as secondary \texttt{Transfer} events. 
Calls to \texttt{setAgentURI()} emit \texttt{URIUpdated} events; replaying these in block order over the initial \texttt{Registered}
URI yields each agent's current URI at \texttt{END\_BLOCK}.
Finally, \texttt{MetadataSet} events record arbitrary key--value metadata attached to an agent, most notably the \texttt{agentWallet} key that designates a separate payment address.
For each agent, we also fetch the off-chain registration file at its current URI. 

\vspace{0.2em}
\noindent\textit{{Reputation Registry.}}
We collect three event types from the Reputation Registry.
\texttt{NewFeedback} events constitute the primary reputation data: each record carries a signed fixed-point score, decimal precision, semantic tags (\texttt{tag1}, \texttt{tag2}), the rated service endpoint, and a \texttt{feedbackURI} pointing to an off-chain feedback document. 
\texttt{FeedbackRevoked} events record the withdrawal of a previously submitted feedback; each revocation is linked to its original \texttt{NewFeedback} record via the composite key (\texttt{agentId}, \texttt{clientAddress}, \texttt{feedbackIndex}), and the resulting revocation status is propagated back to the feedback
record.
\texttt{ResponseAppended} events record replies to existing feedback records, each referencing a \texttt{responseURI}.
For each feedback and response, we also fetch and parse the referenced off-chain file to extract MCP tools, A2A tasks, and x402 payment proofs.

\smallskip
\noindent\textbf{Gas data}.
We measure gas for five event types:
\texttt{Registered}, \texttt{URIUpdated}, secondary \texttt{Transfer},
\texttt{NewFeedback}, and \texttt{FeedbackRevoked}.
To enable cross-chain comparison, native-token costs (i.e., \texttt{gasCostInETH} for Ethereum and Base;
\texttt{gasCostInBNB} for BSC) are converted to USD-equivalent cost (\texttt{gasCostInUSD}) using hourly ETH/USDT and BNB/USDT closing-price data from Binance, matched to each transaction by nearest-candle lookup within a $\pm2$-hour tolerance window.

\smallskip
\noindent\textbf{x402 settlements}.
We collect x402 payments related to ERC-8004 agents. Because x402 payment routing is complex, we defer details of data collection and attribution challenges to Appendix~\ref{app:x402}.

\section{Agent Identity and Adoption}
\label{sec:identity_analysis}

This section analyzes ERC-8004 agent identity and adoption trends across Ethereum, BSC, and Base to characterize the structure and maturity of this emerging system.

\subsection{Agent Registration Dynamics}
\label{sec:registration}

\begin{figure}[htb]
\centering
\begin{minipage}[t]{0.48\textwidth}
    \centering
    \includegraphics[width=\linewidth]{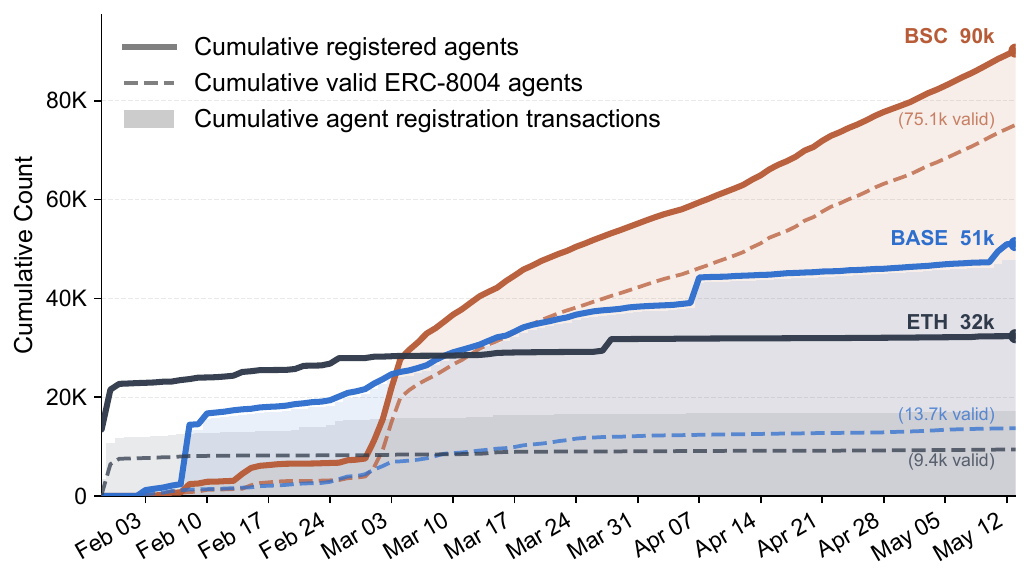}
    \caption{Cumulative agent registrations (solid), valid ERC-8004 registrations (dashed), and unique registration transactions (shaded) across chains.}
    \label{fig:cum_agents}
\end{minipage}
\hfill
\begin{minipage}[t]{0.48\textwidth}
    \centering
    \includegraphics[width=\linewidth]{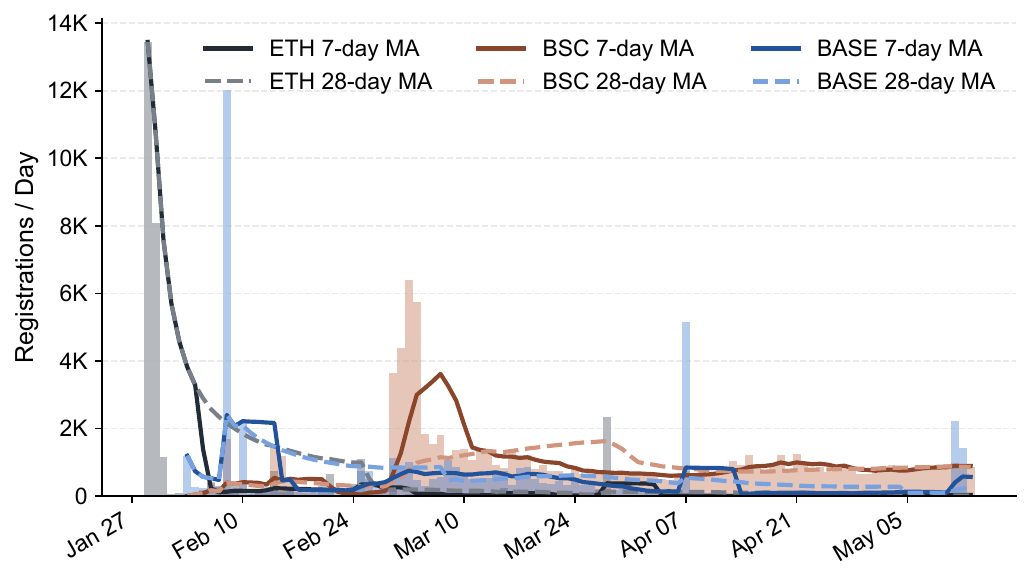}
    \caption{Daily agent registrations (bars) from protocol deployment date, with $7$-day moving averages (solid line) and $28$-day moving averages (dashed line).}
    \label{fig:daily_activity_comparison}
\end{minipage}

\end{figure}

\noindent\textbf{Cumulative registration trend}. Figure~\ref{fig:cum_agents} shows the cumulative agent registrations across three chains from protocol deployment to May 13, 2026.
Ethereum, as the genesis chain of ERC-8004, was the first to launch. BSC and Base launched approximately four days later, and both chains grew rapidly, ultimately surpassing Ethereum in total registration volume. BSC registered the largest absolute number of agents ($90$k), followed by Base ($51$k) and Ethereum ($32$k).
On all three chains, the cumulative agent count (lines) exceeds the cumulative transaction count (shaded areas), indicating that a non-trivial fraction of agents were registered via batch transactions (examined in \S\ref{sec:batch}).

The dashed lines in Figure~\ref{fig:cum_agents} distinguish valid ERC-8004 registrations from the total registered population. Only $29.3\%$ of Ethereum agents, $83.3\%$ of BSC agents, and $26.9\%$ of Base agents have a valid ERC-8004 registration file. The much larger gap on Ethereum and Base suggests that many registrations are placeholders or otherwise non-compliant. We investigate this gap in \S\ref{sec: service_and_quality}.

\noindent\textbf{Daily registration trend}.
Figure~\ref{fig:daily_activity_comparison} compares daily registration activity across the three chains, with $7$-day and $28$-day moving averages. 
The three ecosystems exhibit distinct growth trajectories. Ethereum experienced a pronounced launch-day surge of over $13$k registrations, followed by a rapid decline to relatively low activity levels, producing the apparent plateau. Higher transaction costs and differences in x402 support on Ethereum may have contributed to this slowdown, although our data do not establish causality.
In contrast, BSC showed a delayed but sustained growth phase, peaking at over $6$k daily registrations before stabilizing at around $1$k per day. Base combined both patterns: an initial launch spike was followed by a second major surge in April.

\subsection{Batch Registration and Ownership Concentration}
\label{sec:batch}

\begin{table}[htb]
\centering
\caption{Comparison of batch registration activities across Ethereum, BSC, and Base.}
\label{tab:batch_activity_compare}
\setlength{\tabcolsep}{6pt}
\resizebox{0.95\textwidth}{!}{%
\begin{tabular}{c r @{\hspace{4pt}} l  r @{\hspace{4pt}} l  r @{\hspace{4pt}} l  c  r@{\hspace{4pt}}c  c c}
\toprule
Chain & \multicolumn{2}{c}{Total Agents} & \multicolumn{2}{c}{Batch Txs} & \multicolumn{2}{c}{Batch Agents} & Batch Tx \% & \multicolumn{2}{c}{Batch Agent \%} & Max Size & Median Size\\
\cmidrule(rl){2-3}\cmidrule(rl){4-5}\cmidrule(rl){6-7}\cmidrule(rl){8-8}\cmidrule(rl){9-10}\cmidrule(rl){11-11}\cmidrule(rl){12-12}

ETH & 32{,}343 & \tikz[baseline=-0.6ex]\fill[ethcolor] (0,0) rectangle (0.22cm,0.22cm); & 446 & \tikz[baseline=-0.6ex]\fill[ethcolor] (0,0) rectangle (0.42cm,0.22cm); & 15{,}607 & \tikz[baseline=-0.6ex]\fill[ethcolor] (0,0) rectangle (0.96cm,0.22cm); & 2.6 & 48.3 & \tikz[baseline=-0.6ex]{\fill[gray!20] (0,0) circle (0.18cm);\fill[ethcolor] (0,0) -- (0:0.18cm) arc (0:173.9:0.18cm) -- cycle;} & 182 & 10 \\
BSC & 90{,}145 & \tikz[baseline=-0.6ex]\fill[bsccolor] (0,0) rectangle (0.62cm,0.22cm); & 11 & \tikz[baseline=-0.6ex]\fill[bsccolor] (0,0) rectangle (0.01cm,0.22cm); & 22 & \tikz[baseline=-0.6ex]\fill[bsccolor] (0,0) rectangle (0.0cm,0.22cm); & 0.0 & 0.0 & \tikz[baseline=-0.6ex]\draw[gray!30] (0,0) circle (0.18cm); & 2 & 2 \\
Base & 50{,}985 & \tikz[baseline=-0.6ex]\fill[basecolor] (0,0) rectangle (0.35cm,0.22cm); & 829 & \tikz[baseline=-0.6ex]\fill[basecolor] (0,0) rectangle (0.77cm,0.22cm); & 3{,}966 & \tikz[baseline=-0.6ex]\fill[basecolor] (0,0) rectangle (0.24cm,0.22cm); & 1.7 & 7.8 & \tikz[baseline=-0.6ex]{\fill[gray!20] (0,0) circle (0.18cm);\fill[basecolor] (0,0) -- (0:0.18cm) arc (0:28.1:0.18cm) -- cycle;} & 40 & 3 \\
\bottomrule
\end{tabular}
}
\end{table}

\noindent\textbf{Batch registration behavior}. As noted in \S\ref{sec:registration}, the gap between cumulative agents and cumulative transactions visible in Figure~\ref{fig:cum_agents} indicates that a subset of agents were registered through batch transactions. Table~\ref{tab:batch_activity_compare} quantifies this behavior across the three chains.
On Ethereum, batch registration is both prevalent and substantial. Although batch transactions account for only $2.6\%$ of all registration transactions ($446$ out of $17{,}182$), they are responsible for $48.3\%$ of all registered agents ($15,607$ out of $32,343$). BSC presents a stark contrast. Only $11$ batch transactions were observed, registering a total of $22$ agents.  On Base, batch registration occupies an intermediate position. A total of $829$ batch transactions registered $3{,}966$ agents ($7.8\%$ of the Base total registered agents).

\begin{figure}[htb]
    \centering
    \begin{subfigure}{0.32\linewidth}
        \centering
        \includegraphics[width=\linewidth]{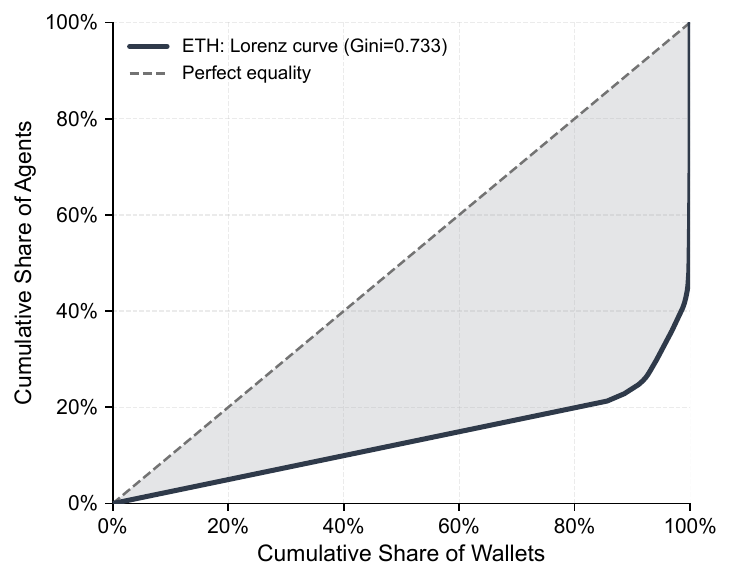}
    \end{subfigure}
    \hfill
    \begin{subfigure}{0.32\linewidth}
        \centering
        \includegraphics[width=\linewidth]{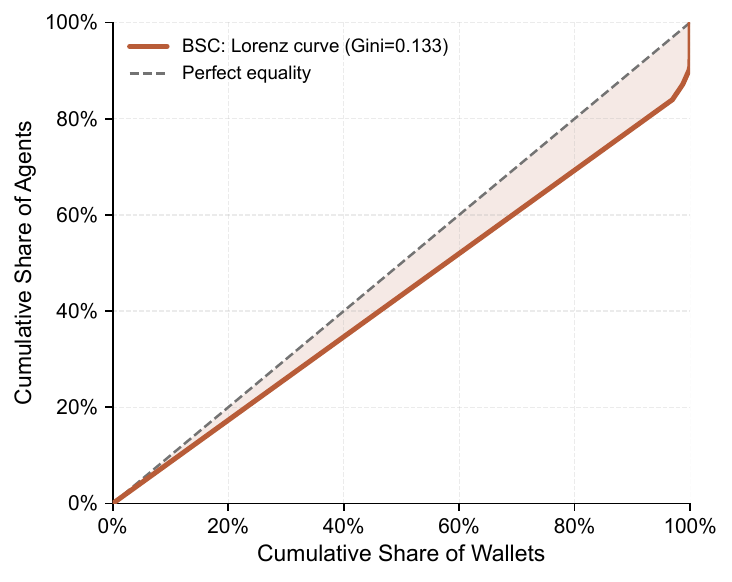}
    \end{subfigure}
    \hfill
    \begin{subfigure}{0.32\linewidth}
        \centering
        \includegraphics[width=\linewidth]{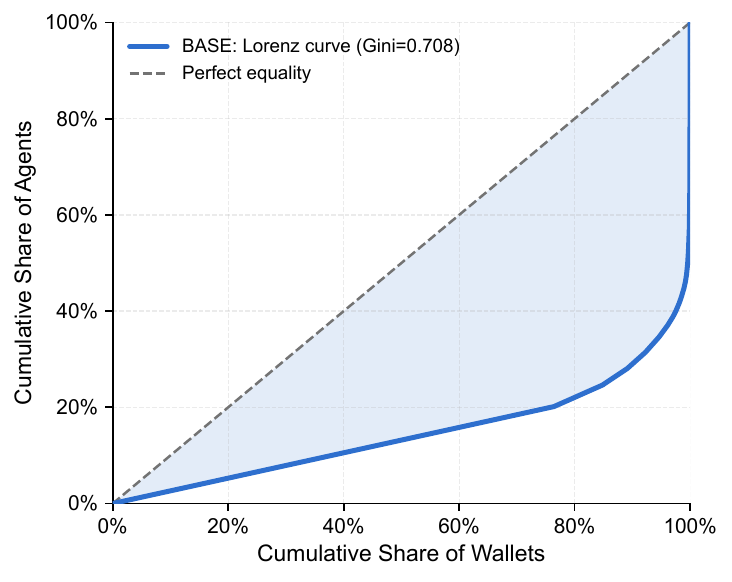}
    \end{subfigure}
   \caption{Ownership concentration of agents across Ethereum, BSC, and Base (Lorenz curves). 
   }
    \label{fig:lorenz_comparison}
\end{figure}

\noindent\textbf{Ownership concentration}. 
The prevalence of batch registration raises the question of whether agent ownership is concentrated among a few wallets. Figure~\ref{fig:lorenz_comparison} reveals cross-chain differences.

Ethereum and Base exhibit severe ownership concentration, with Gini coefficients of $0.733$ and $0.708$ respectively. 
On Ethereum, the top $1\%$ of wallets own $58.5\%$ of registered agents ($54.9\%$ on Base), while the top $10\%$ control more than $70\%$ on both chains, indicating that a small number of deployers dominate registrations.
In contrast, BSC exhibits a much more even ownership structure (Gini coefficient: $0.133$). The top $1\%$ and $10\%$ of wallets own just $12.1\%$ and $22.1\%$ of registered agents respectively, consistent with their near absence of batch registration ($11$ batch transactions registering only $22$ agents). While repeated single-agent registrations can still create concentration, the lack of large-scale batch activity is associated with a more even ownership distribution.

\begin{table*}[htb]
\centering\small
\setlength{\tabcolsep}{3pt}
\caption{Representative registration recipients and \#~agent NFTs received in registration transactions.}
\label{tab:who-registers}
\footnotesize
\begin{tabular}{cclcr}
\toprule
Chain & \textbf{Register Recipient} & Recipient address & \textbf{Identities} & \multicolumn{1}{c}{\textbf{Registration role}}  \\

\cmidrule(lr){2-2}\cmidrule(lr){3-3}\cmidrule(lr){4-4}\cmidrule(lr){5-5}

BASE & \href{https://github.com/nookprotocol/nookplot}{Nookplot} & \href{https://basescan.org/address/0x9f1045d983a6ac1faea82fe9314b47de73515d1a}{\texttt{0x9f10\ldots{}5d1a}} & 6,975 & Registration and distribution\\
BASE & \href{https://www.uniagent.world/}{UniAgent} & \href{https://basescan.org/address/0x83a0c82bb1cdb8ffa6e775f84fda30a2993087a0}{\texttt{0x83a0\ldots{}87a0}} & 1,243 & Asset-linked identities\\
BASE & \href{https://basescan.org/address/0x67722c823010ceb4bed5325fe109196c0f67d053\#code}{Olas} & \href{https://basescan.org/address/0x67722c823010ceb4bed5325fe109196c0f67d053}{\texttt{0x6772\ldots{}d053}} & 605 & Service identity mapping\\
BASE & \href{https://docs.execution.market/contracts/addresses}{Execution Market} & \href{https://basescan.org/address/0x103040545ac5031a11e8c03dd11324c7333a13c7}{\texttt{0x1030\ldots{}13c7}} & 194 & Sponsored NFT delivery\\
BASE & \href{https://dexter.cash/}{Dexter} & \href{https://basescan.org/address/0x40272e2eac848ea70db07fd657d799bd309329c4}{\texttt{0x4027\ldots{}29c4}} & 144 & Payment/MCP association\\
BASE & \href{https://github.com/seedstr/seed-agent}{Seedstr} & \href{https://basescan.org/address/0x9030cd7bf686cb15b1813bd507df9d27d32f9ef0}{\texttt{0x9030\ldots{}9ef0}} & 87 & Registration-platform link\\

ETH & Unlabeled & \href{https://etherscan.io/address/0x806eca4d9e4cebea43df3d0fbb4867aa59422c7a}{\texttt{0x806e\ldots{}2c7a}} & 2,500 & Uses a shared batch minter\\
BSC & Unlabeled & \href{https://bscscan.com/address/0x5b77026f2d449a8139aeec8994d8e0deb5a05dc7}{\texttt{0x5b77\ldots{}5dc7}} & 3,200 & Single-agent registrations\\

BSC &
\href{https://docs.pieverse.io/getting-started/core-concepts}{Purr-Fect Claw} &
\href{https://bscscan.com/address/0xebf9e34b26787a04072551c5301510ad6cf00c11}
{\texttt{0xebf9\ldots{}0c11}} &
\multirow{2}{*}{478} &
Cloud-instance identity registration\\

BSC &
\href{https://www.pieverse.io/}{Pieverse} &
\href{https://bscscan.com/address/0x379a3a24860080a1dfeb76a2ae9af24198307a85}
{\texttt{0x379a\ldots{}7a85}} &
&
Platform-managed agent registration\\

\cmidrule(lr){2-3}\cmidrule(lr){4-5}

\multicolumn{3}{@{}r}{\textbf{All} listed cases (10 addresses)} & 15,426 & 31.96\% (\mypie{bsccolor}{31.96}) of the top-$50$ sample\\

\bottomrule
\end{tabular}
\end{table*}

\noindent\textbf{Register recipients}.
To understand who registered agents, we analyze the top $50$ registration recipients on each chain, covering $48{,}260$ agents. A registration recipient is the address holding the agent NFT after the mint transaction. We track subsequent transfers and assign project labels using official address lists or verified contracts. We mark cases without sufficient evidence as \textit{Unlabeled}.

Table~\ref{tab:who-registers} reveals two main patterns. Some accounts distribute identities:
\href{https://docs.execution.market/contracts/addresses}{Execution Market}'s registration account transfers $125$ of $194$ NFTs to $40$ addresses, while the
\href{https://github.com/nookprotocol/nookplot}{Nookplot}-associated account transfers $6{,}909$ of $6{,}975$ NFTs to distinct addresses. 
Others retain identities in shared contracts:
\href{https://basescan.org/address/0x67722c823010ceb4bed5325fe109196c0f67d053\#code}{Olas} holds $605$ NFTs linked to distinct on-chain service IDs, and
\href{https://www.uniagent.world/}{UniAgent} holds $1{,}243$ NFTs to link agent identities to assets.
Metadata and domains also associate recipients with
\href{https://docs.pieverse.io/getting-started/core-concepts}{Purr-fect/Pieverse},
\href{https://dexter.cash/}{Dexter}, and
\href{https://github.com/seedstr/seed-agent}{Seedstr}, although these associations do not establish address control.

The seven platform-associated cases span eight addresses and $9{,}726$ agents ($20.15\%$ of the sample; $5.61\%$ of all $173{,}473$ registrations).
Including the largest unlabeled recipient on each of Ethereum and BSC, Table~\ref{tab:who-registers} covers $15{,}426$ agents ($31.96\%$ of the sample; $8.89\%$ overall).
These cases help explain registration concentration: a platform account or contract can register and distribute identities, or link existing services and assets to ERC-8004, without operating every agent itself.

\subsection{The Identity-Activity Gap}
\label{sec:Identity-Activity Gap}

\begin{figure}[htb]
\centering

\begin{minipage}[t]{0.48\textwidth}
    \centering
    \includegraphics[width=\linewidth]{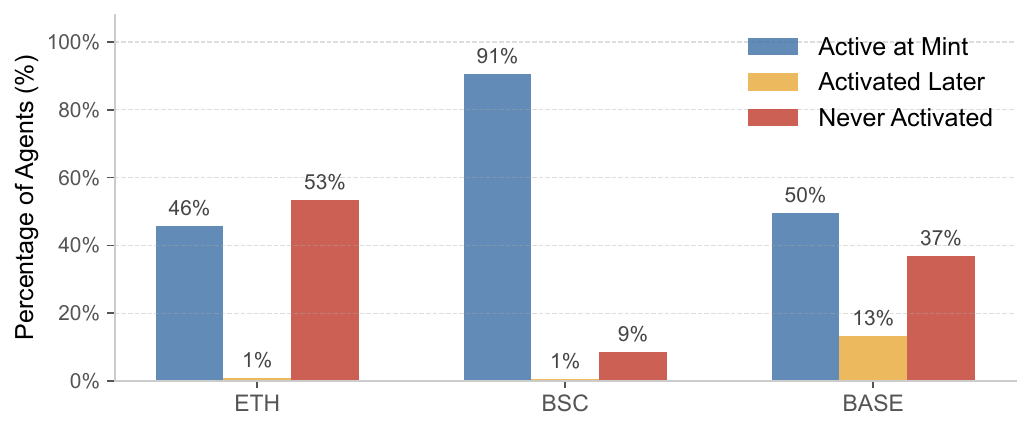}
    \caption{Distribution of agent URI activation status at the end of the observation period, by chain.}
    \label{fig:activation_grouped}
\end{minipage}
\hfill
\begin{minipage}[t]{0.48\textwidth}
    \centering
    \includegraphics[width=\linewidth]{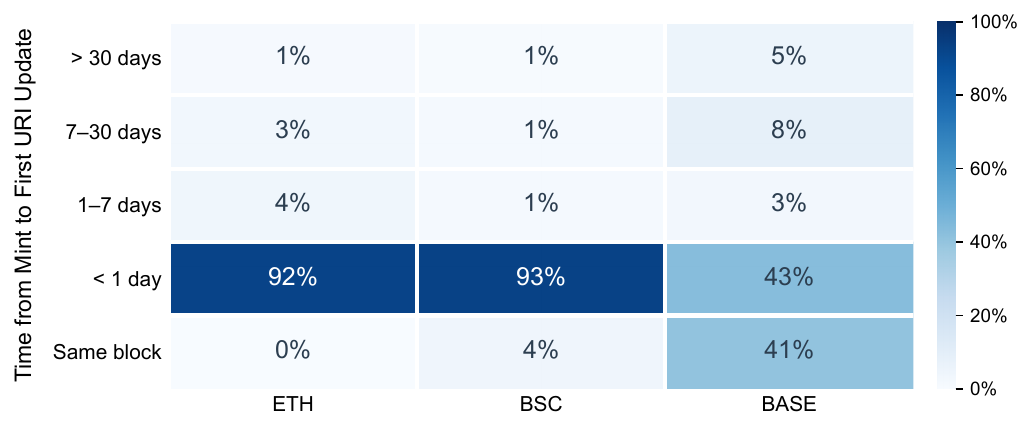}
    \caption{Time-to-activation distribution among agents with post-mint URI updates across the three chains.}
    \label{fig:activation_lag_heatmap}
\end{minipage}

\end{figure}

To characterize the identity-activity gap, we classify each registered agent into one of three categories based on its URI history: \textit{Active at Mint} (URI was set at registration time), \textit{Activated Later} (URI was initially empty but updated later via \texttt{setAgentURI}), and \textit{Never Activated} (URI remains empty as of the end of the observation period). Figure~\ref{fig:activation_grouped} shows their distribution across chains.

Ethereum and Base exhibit a pronounced identity--activity gap. On Ethereum, $53\%$ of agents were never activated, compared with $37\%$ on Base, whereas BSC shows the opposite pattern: $91\%$ of agents were active at mint and only $9\%$ were never activated. Although post-mint activation is possible, it accounts for only $1\%$--$13\%$ of registrations across the three chains.

We therefore examine the activation lag for these \textit{Activated Later} agents. Figure~\ref{fig:activation_lag_heatmap} shows that activation occurs almost entirely within the first day: $92\%$ of Ethereum agents and $93\%$ of BSC agents update their URI within one day, while Base exhibits an even tighter coupling, with $41\%$ updated in the same block and another $43\%$ within one day. Delayed activation is rare, suggesting that agents not activated shortly after registration are unlikely to become active later.

\subsection{Agent Quality and Services}
\label{sec: service_and_quality}

\begin{figure}[htb]
\centering
\begin{minipage}[t]{0.48\textwidth}
    \centering
    \includegraphics[width=\linewidth]{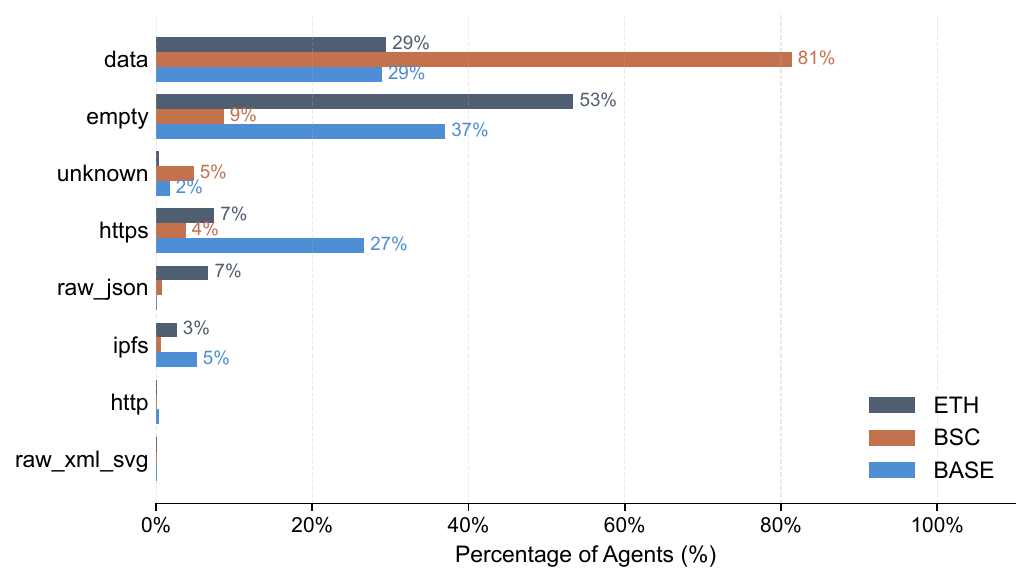}
    \caption{Distribution of agent URI scheme by chain.}
    \label{fig:uri_scheme}
\end{minipage}
\hfill
\begin{minipage}[t]{0.48\textwidth}
    \centering
    \includegraphics[width=\linewidth]{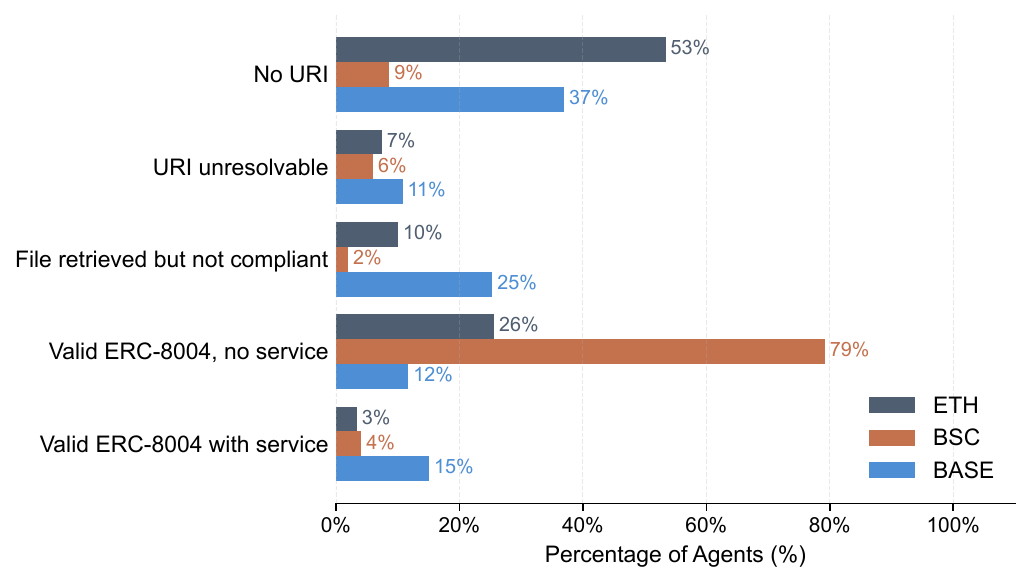}
    \caption{Distribution of agent quality by chain.}
    \label{fig:quality_taxonomy}
\end{minipage}
\end{figure}

\noindent\textbf{Agent URI scheme}. To understand how agents expose their registration files, we examine the URI scheme of each agent's current URI. We distinguish eight categories, including web-hosted (\texttt{https}, \texttt{http}), decentralized (\texttt{ipfs}), inline (\texttt{data}), raw content, unknown, and empty URIs. Figure~\ref{fig:uri_scheme} shows their distribution across chains.
URI usage differs substantially across chains. BSC is dominated by inline \texttt{data} URIs, which account for $81\%$ of all agents. In contrast, Ethereum and Base exhibit more diverse usage patterns: \texttt{data} URIs account for about $29\%$ on both chains, while Base makes much greater use of web-hosted files (\texttt{https}, $27\%$) and is the only chain with meaningful IPFS adoption ($5\%$). Empty URIs remain common, affecting $53\%$ of Ethereum, $37\%$ of Base, and $9\%$ of BSC agents.

\vspace{0.1em}
\noindent\textbf{Agent quality}.
We classify each registered agent into five categories based on the reachability and compliance of its registration file at the end block of the observation period: \emph{No URI} (no URI set), \emph{URI unresolvable} (URI present but not retrievable), \emph{file retrieved but not compliant} (the \texttt{type} field is not ERC-8004 compliant),\footnote{In this study, a registration file is considered ERC-8004 compliant only if the \texttt{type} field exactly matches the canonical identifier \texttt{https://eips.ethereum.org/EIPS/eip-8004\#registration-v1} specified by ERC-8004~\cite{erc8004}.} 
\emph{Valid ERC-8004, no service} (valid registration file with no declared service), and \emph{Valid ERC-8004 with service} (valid registration file with at least one declared service).

Agent quality varies substantially across chains (Figure~\ref{fig:quality_taxonomy}), but a common pattern emerges: most registered identities do not correspond to functional deployments. Ethereum and Base are primarily limited by missing metadata, with $53\%$ and $37\%$ of agents lacking a URI. In contrast, BSC exhibits much higher ERC-8004 compliance, with $79\%$ of agents providing a valid registration file, although only $4\%$ declare a service endpoint. Base achieves the highest share of service-enabled agents ($15\%$), but also the largest proportion of non-compliant files ($25\%$). Overall, functional agents remain uncommon, accounting for just $3\%$, $4\%$, and $15\%$ of agents on Ethereum, BSC, and Base respectively.
We also provide a finer-grained assessment of agent quality in Appendix~\ref{app:quality}.

\begin{insightbox}
    Most registered agents are not functional deployments. Only $3\%$, $4\%$, and $15\%$ of agents on Ethereum, BSC, and Base expose both a valid ERC-8004 registration file and a declared service. Ethereum and Base are primarily constrained by missing URIs, whereas BSC achieves high registration-file compliance but rarely advertises agent capabilities.
\end{insightbox}

\begin{figure}[htb]
    \centering
    \begin{subfigure}{0.325\linewidth}
        \centering
        \includegraphics[width=\linewidth]{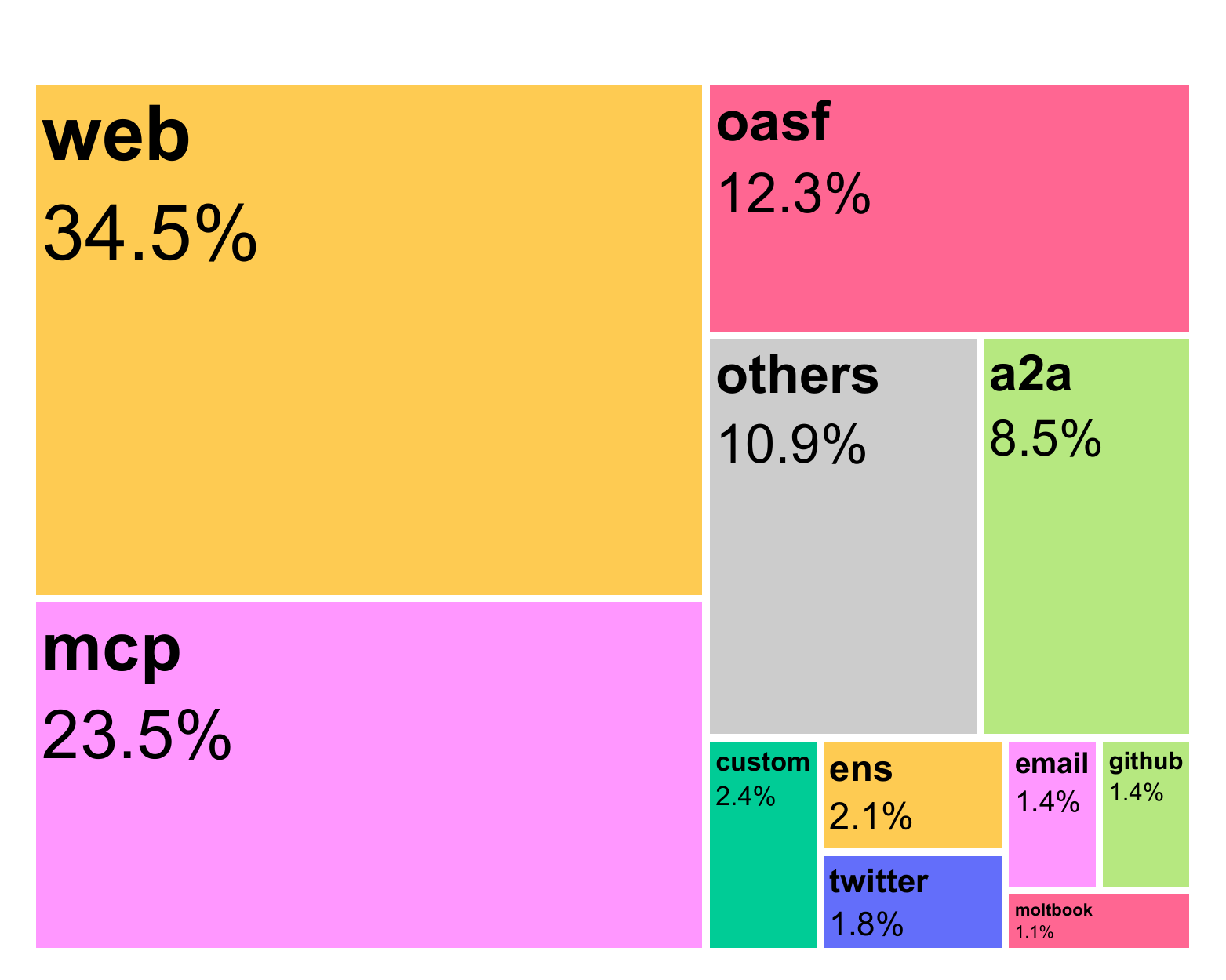}
        \caption{ETH}
        \label{fig:fig_service_treemap_ETH}
    \end{subfigure}
    \hfill
    \begin{subfigure}{0.325\linewidth}
        \centering
        \includegraphics[width=\linewidth]{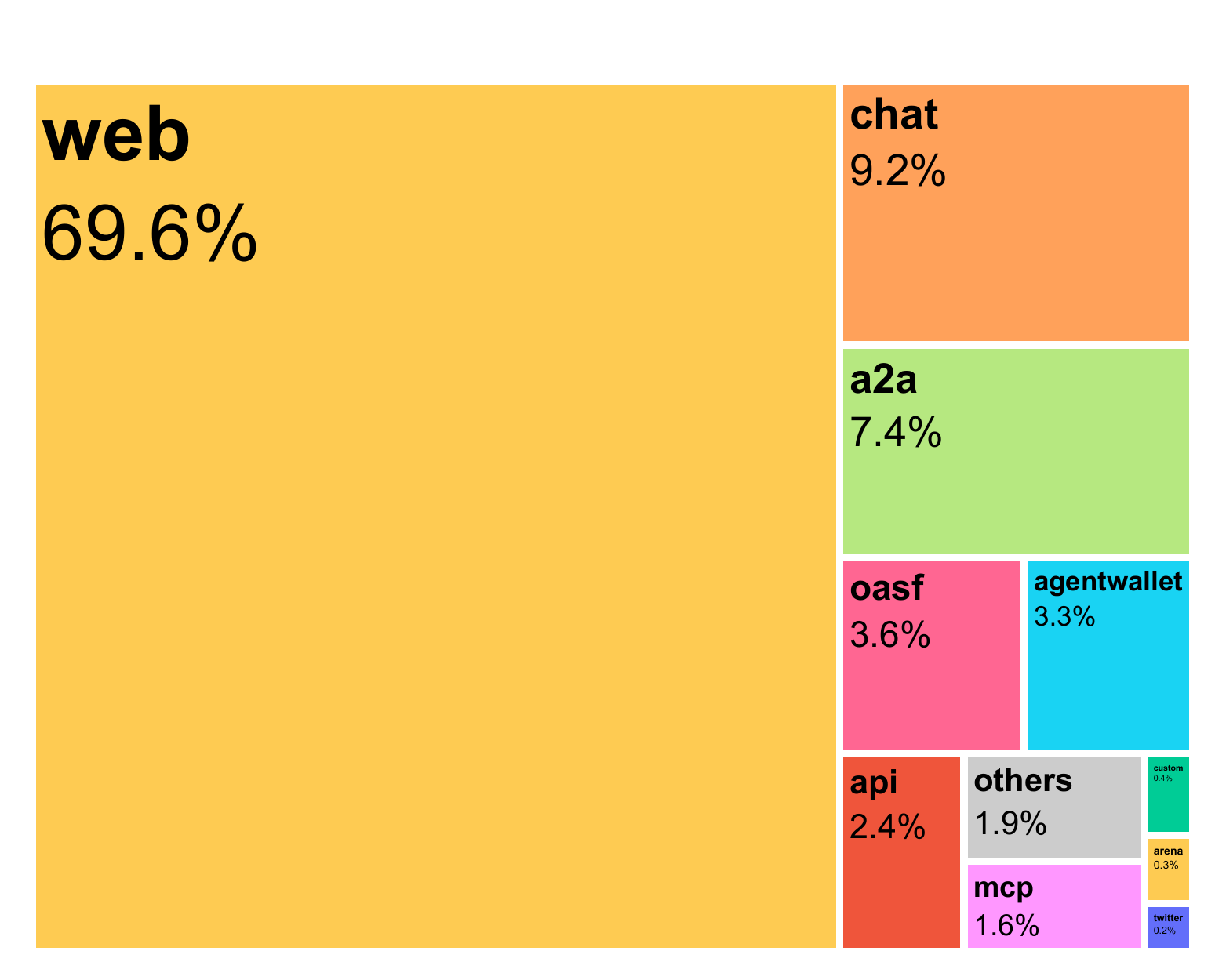}
        \caption{BSC}
        \label{fig:fig_service_treemap_BSC}
    \end{subfigure}
    \hfill
    \begin{subfigure}{0.325\linewidth}
        \centering
        \includegraphics[width=\linewidth]{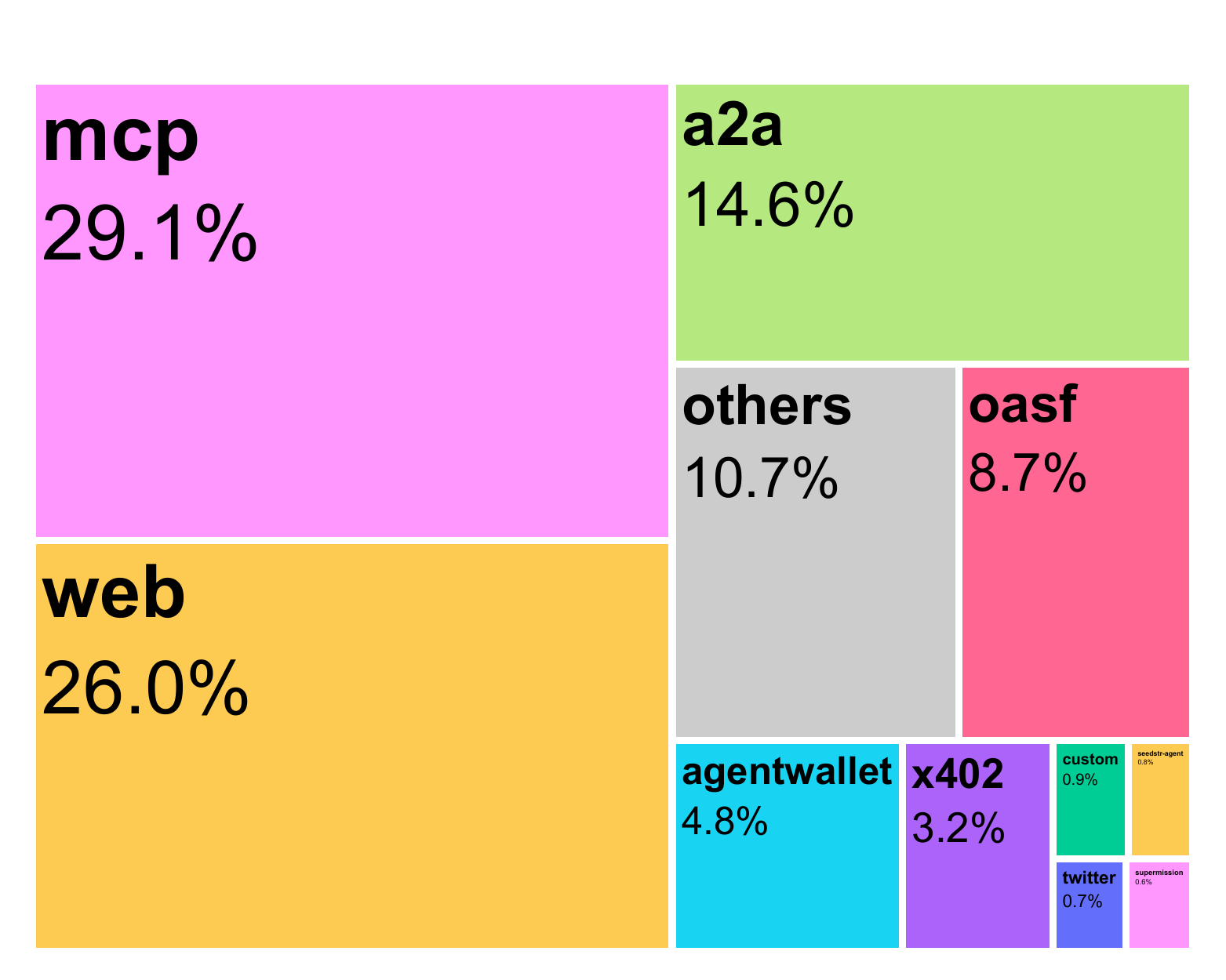}
        \caption{Base}
        \label{fig:fig_service_treemap_BASE}
    \end{subfigure}
   \caption{ Distribution of declared service types among valid ERC-8004 agents with active services.}
    \label{fig:service_comparison}
\end{figure}

\noindent\textbf{Service type}. For valid ERC-8004 agents with active services, we further analyze their declared service type.
Figure~\ref{fig:service_comparison} presents the distribution of declared service types among these agents, showing the top 10 service types by count with an aggregated \textit{others} category. Note that the counts reflect individual service declarations because a single agent may declare multiple service types.

Figure~\ref{fig:service_comparison} reveals substantial cross-chain differences in service composition. BSC is clearly web-centric, with Web services accounting for nearly $70\%$ of all declarations. In contrast, Ethereum and Base exhibit more balanced ecosystems. Ethereum combines strong adoption of both Web and MCP services, whereas Base shows the strongest presence of agent-native protocols, with MCP and A2A accounting for nearly half of all declarations. These patterns suggest that BSC focuses on conventional services, while Base places greater emphasis on agent interoperability. Ethereum occupies a middle ground, blending traditional Web services with emerging agent-native protocols.

\subsection{Agent Functional Profile}
\label{sec:identity_function}
To understand what agents do downstream, we examine the tasks they are intended to perform for users.
In practice, actual task execution is largely unobservable: it occurs off-chain, while on-chain references (e.g., job IDs) reveal little about task content. We therefore use what agents \emph{claim} to do as a proxy for their intended tasks, drawing on the text descriptions in their registration files.

\vspace{0.1cm}
\noindent\textbf{Data.} $98{,}702$ ($56.9\%$) of registered agents on the three chains have usable descriptions; the rest have unretrievable files or descriptions that are empty, gibberish, placeholders, or too short.
Descriptions are highly repetitive: only $14{,}530$ distinct texts remain, and the top $50$ cover $77.3\%$ of agents.

\begin{figure}[htb]
    \centering
    \includegraphics[width=0.8\linewidth]{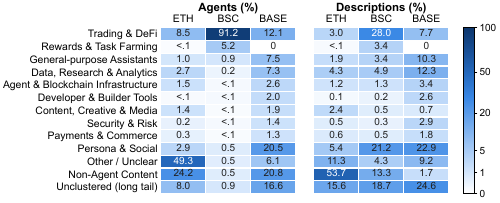}
    \caption{Category shares by chain. Left: agents\% with a usable description. Right: description groups\%.}
    \label{fig:agent_function_taxonomy}
\end{figure}

\vspace{0.1cm}
\noindent\textbf{Method.}
We clean the text and merge exact and near-identical descriptions (cosine similarity $\ge 0.95$).
We embed each group using the multilingual model \texttt{bge-m3}, reduce dimensionality with UMAP, and cluster with HDBSCAN.
We label each cluster and the dominant templates by the agent's stated function, yielding $12$ categories (Figure~\ref{fig:agent_function_taxonomy}).
We also extend labels to the long tail by propagating them using a trained classifier. Appendix~\ref{app:agent_function_1} provides further methodological details.

\noindent\textbf{Results.}
First, a few templates dominate agent counts. Among agents with usable descriptions, \emph{Trading \& DeFi} accounts for $67.1\%$, largely driven by two templates covering $85.2\%$ of this population on BSC.
Second, usable descriptions do not always convey a function: $17.5\%$ are generic, non-agent, or unclustered text, and another $5.1\%$ describe personas without concrete services.
Third, the chains show different functional profiles. $91.2\%$ of agents with usable descriptions on BSC advertise trading functions. On ETH, $49.3\%$ provide generic or vague text and $24.2\%$ provide non-agent text; only $15.7\%$ state a concrete function. Base is the most diverse and hosts most agents advertising specific services such as data analytics, developer tools, security, and agent infrastructure.

We also analyze agent functional profiles in Appendix~\ref{app:agent_function_2} using service-level descriptions.

\section{The Reputation Market}
\label{sec:reputation_analysis}
This section characterizes the ERC-8004 reputation market as it operates in practice.

\subsection{Market Overview}
\label{sec:reputation_overview}

\noindent\textbf{Market scale and participation.} Table~\ref{tab:rep_participation} summarizes reputation market statistics. Base dominates in overall scale, accounting for $122{,}798$ feedback records submitted by $3{,}073$ reviewers across $28{,}592$ rated agents. BSC and Ethereum are considerably smaller, with $29{,}444$ and $3{,}058$ records respectively.

\begin{table}[htb]
\centering
\caption{Reputation market participation summary.}
\label{tab:rep_participation}
\setlength{\tabcolsep}{6pt}
\resizebox{0.95\textwidth}{!}{%
\begin{tabular}{c r @{\hspace{4pt}} l  r @{\hspace{4pt}} l  r @{\hspace{4pt}} l  ccc}
\toprule
Chain & \multicolumn{2}{c}{Total Feedback} & \multicolumn{2}{c}{Rated Agents} & \multicolumn{2}{c}{Uniq. Reviewers} & Avg. FB/Agent & Avg. FB/Rev. & Avg. Agents/Rev. \\
\cmidrule(rl){2-3}\cmidrule(rl){4-5}\cmidrule(rl){6-7}
\cmidrule(rl){8-8}\cmidrule(rl){9-9}\cmidrule(rl){10-10}

ETH & 3{,}058 & \tikz[baseline=-0.6ex]\fill[ethcolor] (0,0) rectangle (0.02cm,0.22cm); & 1{,}559 & \tikz[baseline=-0.6ex]\fill[ethcolor] (0,0) rectangle (0.05cm,0.22cm); & 618 & \tikz[baseline=-0.6ex]\fill[ethcolor] (0,0) rectangle (0.2cm,0.22cm); & 2.0 & 4.9 & 2.5 \\
BSC & 29{,}444 & \tikz[baseline=-0.6ex]\fill[bsccolor] (0,0) rectangle (0.23cm,0.22cm); & 4{,}310 & \tikz[baseline=-0.6ex]\fill[bsccolor] (0,0) rectangle (0.15cm,0.22cm); & 76 & \tikz[baseline=-0.6ex]\fill[bsccolor] (0,0) rectangle (0.02cm,0.22cm); & 6.8 & \textcolor{bsccolor}{387.4} & \textcolor{bsccolor}{56.7} \\
Base & 122{,}798 & \tikz[baseline=-0.6ex]\fill[basecolor] (0,0) rectangle (0.95cm,0.22cm); & 28{,}592 & \tikz[baseline=-0.6ex]\fill[basecolor] (0,0) rectangle (1.0cm,0.22cm); & 3{,}073 & \tikz[baseline=-0.6ex]\fill[basecolor] (0,0) rectangle (0.98cm,0.22cm); & 4.3 & 40.0 & 9.3 \\
\bottomrule
\end{tabular}
}
\end{table}

Reviewer participation patterns differ markedly across chains. BSC exhibits an unusually high degree of \textit{concentration}: its $76$ reviewers submit an average of $387.4$ feedback records each and evaluate $56.7$ agents on average, compared with $40$ records and $9.3$ agents on Base, and just $4.9$ records and $2.5$ agents on Ethereum. This suggests that reputation formation on BSC is shaped by a small set of reviewers, raising questions about the quality of the resulting reputation signals.

\begin{figure}[htb]
    \centering
    \begin{subfigure}{0.325\linewidth}
        \centering
        \includegraphics[width=\linewidth]{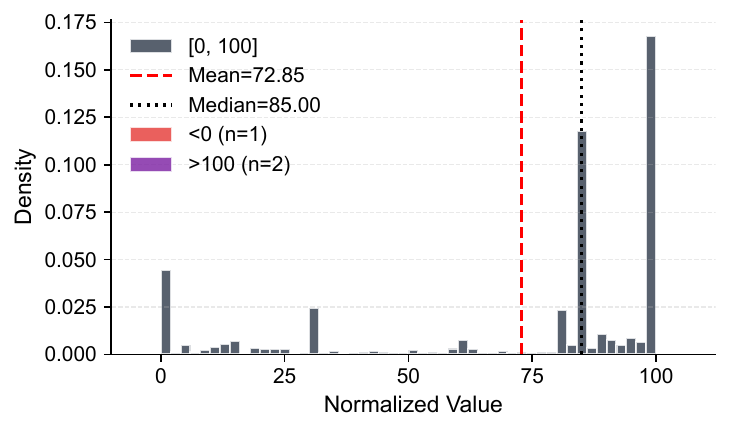}
        \caption{ETH}
        \label{fig:value_distribution_ETH}
    \end{subfigure}
    \hfill
    \begin{subfigure}{0.325\linewidth}
        \centering
        \includegraphics[width=\linewidth]{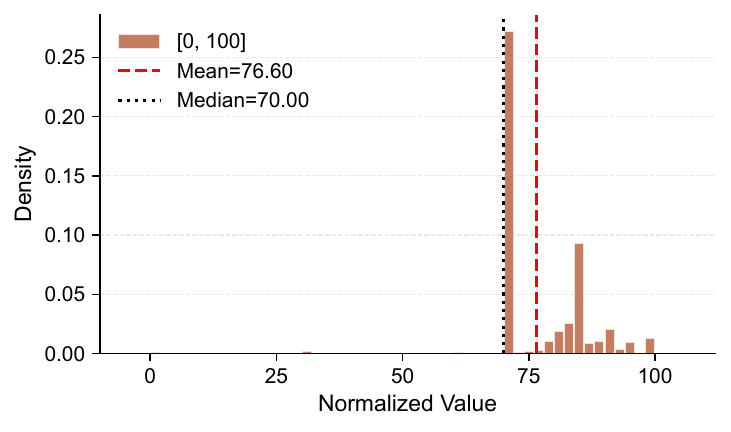}
        \caption{BSC}
        \label{fig:value_distribution_BSC}
    \end{subfigure}
    \hfill
    \begin{subfigure}{0.325\linewidth}
        \centering
        \includegraphics[width=\linewidth]{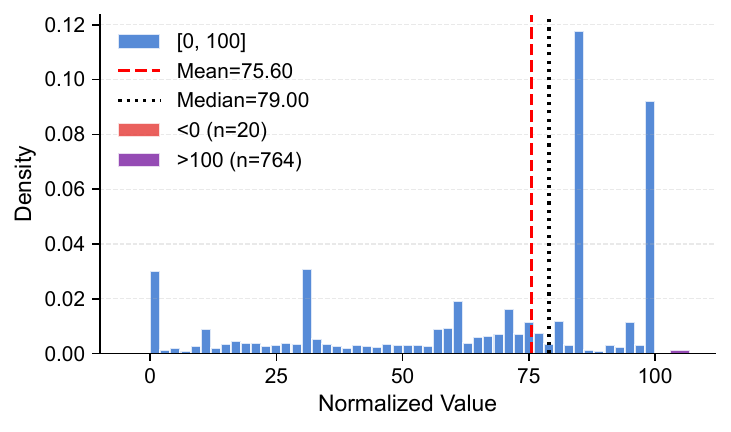}
        \caption{Base}
        \label{fig:value_distribution_BASE}
    \end{subfigure}
   \caption{Distribution of normalized feedback values across three chains. The main histogram covers the [$0$,$100$] range; red and purple bars at the left and right margins indicate the proportion of feedback with values below zero and above $100$ respectively. Vertical lines mark the mean (dashed red) and median (dotted black).}
    \label{fig:value_distribution}
\end{figure}

\noindent\textbf{Feedback value.} Figure~\ref{fig:value_distribution} shows the distribution of normalized feedback scores across the three chains, where each score is computed as $\texttt{value}/10^{\texttt{decimals}}$. Because the protocol does not restrict feedback values to the $[0,100]$ range, scores outside this interval occur in the data (\S\ref{sec:rep:semantics}). 

Across all three chains, feedback scores are concentrated toward the upper end of the scale, but the distribution differs by chain. Ethereum exhibits a pronounced spike at $100$ with a smaller cluster of very low ratings. BSC shows the most concentrated distribution, with most feedback falling between $70$ and $80$ and very few out-of-range values. Base displays the greatest dispersion, combining substantial mass at both $100$ and near-zero scores. 
Scores outside $[0, 100]$ are rare on ETH and absent on BSC, but non-negligible on Base ($n=1$ below zero, $n=2$ above $100$), suggesting that a small fraction of reviewers use the value field to record arbitrary numerical quantities.

\begin{figure}[htb]
    \centering
    \begin{subfigure}{0.325\linewidth}
        \centering
        \includegraphics[width=\linewidth]{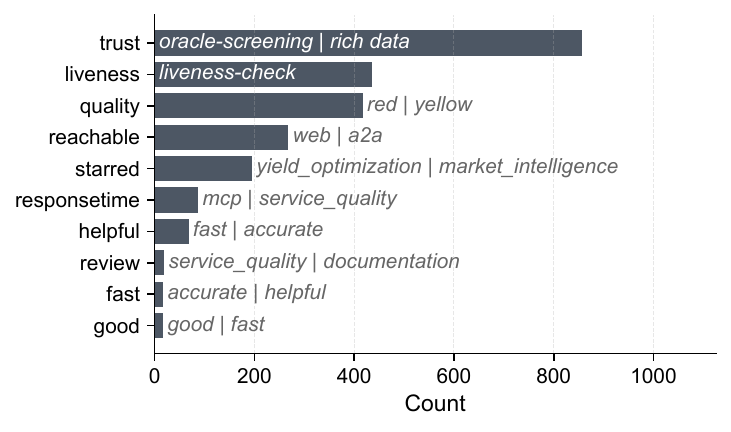}
        \caption{ETH}
        \label{fig:fig_tag1_tag2_ETH}
    \end{subfigure}
    \hfill
    \begin{subfigure}{0.325\linewidth}
        \centering
        \includegraphics[width=\linewidth]{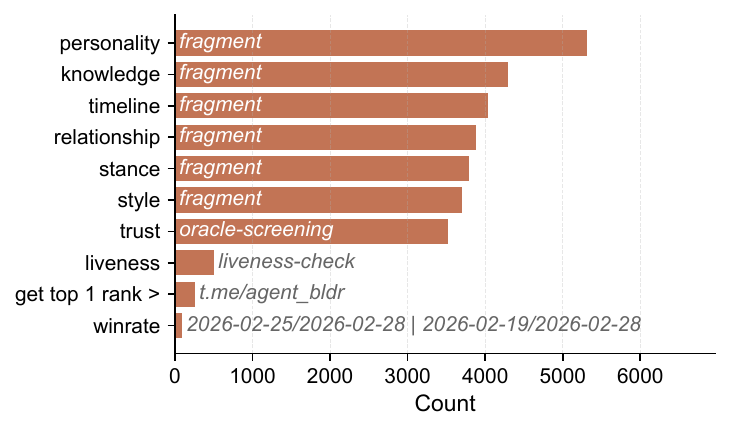}
        \caption{BSC}
        \label{fig:fig_tag1_tag2_BSC}
    \end{subfigure}
    \hfill
    \begin{subfigure}{0.325\linewidth}
        \centering
        \includegraphics[width=\linewidth]{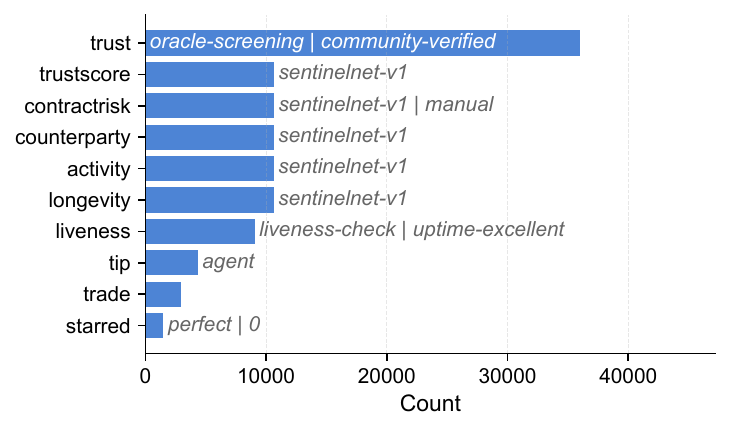}
        \caption{Base}
        \label{fig:fig_tag1_tag2_BASE}
    \end{subfigure}
   \caption{Top 10 feedback evaluation dimensions (\texttt{tag1}) with their two most frequent sub-dimensions (\texttt{tag2}).}
    \label{fig:tag_vocabulary}
\end{figure}

\noindent\textbf{Tag semantics.} Figure~\ref{fig:tag_vocabulary} reports the ten most frequent \texttt{tag1} values on each chain, with the two most common \texttt{tag2} sub-dimensions annotated inside each bar.
Ethereum exhibits the most diverse vocabulary, spanning trust, liveness, quality, reachability, and response-time assessments. 
BSC displays a highly repetitive pattern. Multiple top-level dimensions share the same \texttt{fragment} sub-dimension, and several tags contain placeholder-like or machine-generated values. This is consistent with templated or automated feedback submission rather than independent reviewer judgment.
Base occupies a different position. While \texttt{trust} remains the dominant dimension, much of the vocabulary is centered on risk assessment, including \texttt{trustscore}, \texttt{contractrisk}, and \texttt{counterparty}. 

\smallskip
\noindent\textbf{Cross-Chain Reputation.} We compare reputation scores of agents registered on multiple chains and find no significant cross-chain correlation, suggesting fragmented reputation silos (Appendix~\ref{app:rep:cross_chain}).

\section{Reputation Market Security} 
\label{sec:rep_security}

ERC-8004 advances its Reputation Registry as the trust layer of the agent economy.
By design, the Registry provides only a minimal foundation: it makes feedback public while leaving scoring, Sybil resistance, and aggregation to off-chain reputation systems~\cite{erc8004}.
This section asks: can the Reputation  Registry, as deployed today, already serve as a \textbf{trust signal}, and what challenges must any system built on top of it overcome? We show that the Registry provides none of the guarantees a trust signal needs, and the defenses the specification points to do not yet close these gaps.

\subsection{Threat Model}

\noindent\textbf{Data model.}
A feedback record is a tuple
$f = (a, c, v, d, t_1, t_2)$, where $a$ is the rated agent, $c$ the client (reviewer), $v \in [-10^{38}, 10^{38}]$ a signed integer, $d \le 18$ a decimal precision, and $t_1, t_2$ free-form string tags.
The \emph{normalized value} is $\tilde{v} = v / 10^{d}$. Crucially, the contract does \emph{not} compute or store a canonical ``reputation''. It stores per-$(a,c)$ feedback tuples and exposes an aggregation function \texttt{getSummary(agentId,clientAddresses,tag1,tag2)}, which the \emph{caller} parameterizes by a client set $C$ and a tag filter $T$, returning the mean of $\tilde{v}$ over the selected, non-revoked records~\cite{erc8004contracts}. Therefore, any notion of ``agent $a$'s score'' is actually a consumer-side aggregation choice:
\begin{equation}
S_{T,C}(a) \;=\;
\frac{1}{|F_{a}^{T,C}|}\sum_{f \in F_{a}^{T,C}} \tilde{v}_f ,
\label{eq:score}
\end{equation}
\noindent where $F_a^{T,C}$ is the set of non-revoked feedback for $a$ restricted to clients in $C$ and tags in $T$. As a reference point, the unfiltered mean over all clients and all tags, $S(a) \equiv S_{\ast,\ast}(a)$, is the most permissive reading a relying party can take. The specification itself~\cite{erc8004} flags unfiltered aggregation as Sybil-prone and expects callers/off-chain aggregators to restrict the client set; we treat that restriction, with tag filtering, as protocol-level defenses and analyze them directly below.

\smallskip
\noindent\textbf{Necessary conditions.}
For $S(a)$ to be a trust signal under an adversary, the following must hold.

\begin{itemize}[topsep=2pt, leftmargin=*, itemsep=2pt]
\item \textit{C1: Commensurability}.
The records averaged in Equation~\ref{eq:score} measure the same quantity on the same scale, so that their mean denotes a single, well-defined property of the agent.

\item  \textit{C2: Robustness}.
$S(a)$ cannot be moved arbitrarily by a minority of adversarial inputs; influence per identity is bounded, and reaching a target requires a cost exceeding the value at stake.

\item  \textit{C3: Groundedness}.
Each feedback record reflects a verifiable, real interaction
between $c$ and $a$.

\item  \textit{C4: Economic soundness}.
Manipulating $S(a)$ should cost more than the value that can be extracted from the manipulation.
\end{itemize}

\smallskip
\noindent\textbf{Threat model.}
The adversary controls an unbounded number of EOAs,
each created and funded at a gas cost; submitting feedback requires no stake, no registration, and no prior interaction. The contract enforces exactly one \textit{integrity constraint}: an agent's
registered owner or authorized operator cannot rate its own
agent~\cite{erc8004contracts}. The adversary is otherwise unconstrained.

\noindent\textbf{The protocol's intended defenses.}
The ERC-8004 specification~\cite{erc8004} does not present the raw on-chain mean as a trust signal. It offers three defenses: \emph{(i) reviewer filtering}: \texttt{getSummary()} requires a non-empty client set, warning that unfiltered aggregation is Sybil-prone; \emph{(ii) tag filtering}: consumers may aggregate under a single tag; and \emph{(iii) off-chain aggregation}: complex scoring occurs off-chain, with reviewer-reputation systems expected to emerge. These defenses nevertheless leave gaps.
The envisioned reviewer-reputation layer does not yet exist, and reviewers remain unconstrained (\S\ref{sec:rep:wild}). Tag filtering cannot ensure comparability: tags lack standardized semantics, and identical tags can use incompatible scales (\S\ref{sec:rep:semantics}). Off-chain aggregators do not resolve these issues either: they still rely on manipulable feedback (\S\ref{sec:rep:robustness}) ungrounded in verifiable on-chain interactions (\S\ref{sec:rep:groundedness}).

\subsection{C1 --- Inconsistent Feedback Semantics and Scales}
\label{sec:rep:semantics}

\begin{figure}[htb]
\centering
\begin{minipage}[t]{0.48\textwidth}
    \centering
    \includegraphics[width=\linewidth]{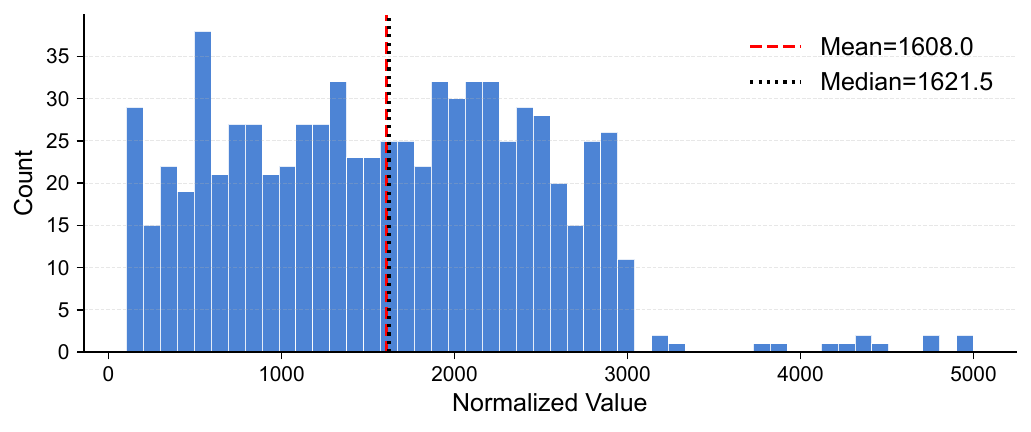}
    \caption{Value distribution for $\tilde{v}>100$ (Base chain).}
    \label{fig:base_outliers_score_distribution}
\end{minipage}
\hfill
\begin{minipage}[t]{0.48\textwidth}
    \centering
    \includegraphics[width=\linewidth]{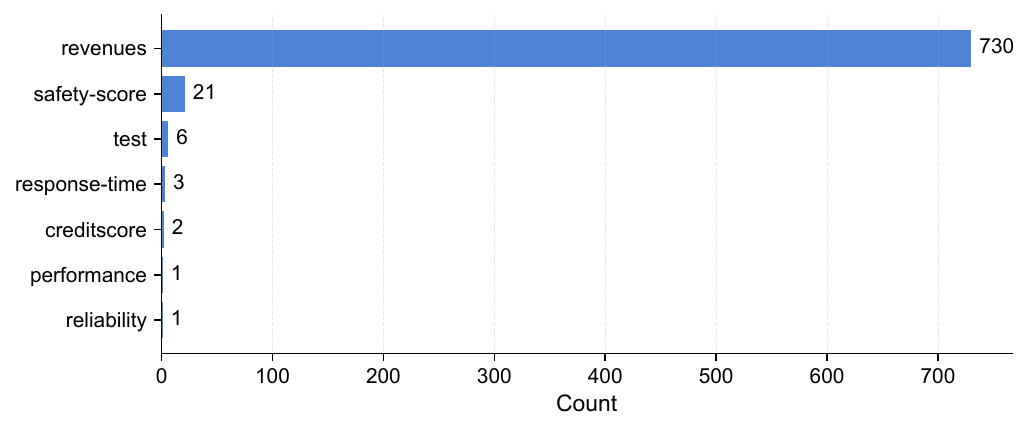}
    \caption{\texttt{tag1} distribution for $\tilde{v}>100$ (Base chain).}
    \label{fig:base_outliers_tag1_distribution}
\end{minipage}
\end{figure}

\noindent\textbf{The symptom: anomalous magnitudes.} Treating values outside any plausible rating range as off-scale (i.e., $\tilde{v}<0$ or $\tilde{v}>100$), we find $20$ negative records on Base (Figure~\ref{fig:value_distribution_BASE}) and $764$ above $100$ (Figure~\ref{fig:base_outliers_score_distribution}). Of the latter, $730$ ($95.5\%$) carry \texttt{revenues}; the rest use other unbounded metrics, including \texttt{safety-score}, \texttt{response-time}, and \texttt{creditscore} (Figure~\ref{fig:base_outliers_tag1_distribution}). These values range from roughly $100$ to $5{,}000$ (mean $1{,}608.0$, median $1{,}621.5$), submitted by $205$ clients for $225$ agents.

\begin{insightbox}
By design, the \texttt{value} field of the Reputation Registry is an untyped attestation store: the specification~\cite{erc8004} leaves tags to developers and enforces no scale, so the \texttt{value} field carries arbitrary numbers with incompatible scales and semantics. 
The flexibility is intentional; the consequence is that no single comparable rating exists for consumers.
\end{insightbox}

\noindent\textbf{The root cause.}
The value and tag fields are \textit{deliberately open}. \texttt{giveFeedback()} constrains only $|v| \le 10^{38}$ and $d \le 18$, leaving \texttt{tag1}/\texttt{tag2} to developers for composability and filtering~\cite{erc8004contracts, erc8004}.
The specification~\cite{erc8004} explicitly supports ratings (\texttt{starred}), binary flags (\texttt{reachable}, \texttt{ownerVerified}), percentages (\texttt{uptime}, \texttt{successRate}), latency (\texttt{responseTime}), USD revenue (\texttt{revenue}), and signed yields (\texttt{tradingYield}).
This heterogeneity is intended, but no comprehensive on-chain mapping defines tag units or ranges, nor is there a canonical ``reputation'' tag. Such flexibility supports diverse metrics but leaves consumers without a well-defined reputation score.

\begin{table}[htb]
\centering
\caption{\texttt{Tag1} examples. One representative tag per value-scale convention in the
feedback value field ($n \ge 10$). Each row illustrates a distinct
scale, from $0$--$100$ ratings to unbounded and signed metrics.
}
\label{tab:tag-ranges}
\resizebox{\linewidth}{!}{%
\begin{tabular}{lccccccccr}
\toprule
\multicolumn{1}{c}{\texttt{tag1}} & chain & $n$ & mean & median & min & max & $q_{25}$ & $q_{75}$ & \multicolumn{1}{c}{\textbf{apparent scale}} \\

\cmidrule(rl){1-1}\cmidrule(rl){4-5}\cmidrule(rl){6-7}\cmidrule(rl){8-9}\cmidrule(rl){10-10}

\texttt{personality}         & BSC  & 5{,}309   & 77.81 & 70   &65 &95 & 70 & 88   & $\approx$\,0--100 \\
\texttt{review}          & ETH  & 19    & 36.53	& 4  &3 & 100 & 4 & 86       & mixed 0--5 / 0--100 \\
\texttt{reachable}$^\dagger$ & ETH & 268 &  1.93 & 1 & 0 & 65 &1 &1 & mixed Boolean / 0-100\\
\texttt{health-check} & ETH & 13 & 3.77 & 4 & 3 &4 &4 & 4 & $\approx$\,0--5 \\
\texttt{miner-vouch}     & Base & 957   &1 & 1 &1 & 1     & 1 &1        & Boolean (all true) \\
\texttt{trade}           & Base & 2{,}946  & 0 & 0     & 0 & 0 & 0 & 0        & Boolean (all false)\\
\texttt{risk-assessment} & Base & 103   & 4.98& 4.95 & 4.58& 7.6  & 4.83 & 4.95  & $\approx$\,0--10 \\
\texttt{revenues}        & Base & 732 & 1639.93	& 1661.50    & 50 &4971  &961.50	&2282.50   & open metric ($\gg$100) \\
\texttt{identity} & Base & 11 & -4.55 & -10 & -10	& 20	& -10	& -10 & signed (mostly neg.)\\
\bottomrule
\end{tabular}
}

\smallskip
\footnotesize
$^\dagger$ For the \texttt{reachable} tag, the protocol presents it as a Boolean field, but in practice, some users treat it as a numerical tag.
\end{table}

\noindent\textbf{A tag space with no standard.}
The \texttt{value} field mixes unrelated conventions (Table~\ref{tab:tag-ranges}), with three layers of incommensurability.
\textit{(i)} Scales differ across tags: $0$--$100$ ratings (\texttt{personality}), $0$--$5$ assessments (\texttt{health-check}), Boolean metrics (\texttt{trade}, \texttt{miner-vouch}), unbounded metrics (\texttt{revenue}), and signed metrics (\texttt{identity}) coexist.
\textit{(ii)} Vocabulary differs \emph{across chains} (Figure~\ref{fig:tag_vocabulary}): Ethereum emphasizes \texttt{trust}/\texttt{liveness}/\texttt{quality}, BSC conversational dimensions (\texttt{personality}, \texttt{knowledge}, \texttt{timeline}, \ldots), and Base operational metrics (\texttt{contractrisk}, \texttt{counterparty}, \ldots).
\textit{(iii)} Scales differ even \emph{within} identical tags: \texttt{review} has median $4.0$ and interquartile range $4$--$86$, reflecting mixed $0$--$5$ and $0$--$100$ scales; \texttt{reachable} is used as both a Boolean and a numerical rating.
Tag filtering, the protocol's intended remedy, cannot resolve these inconsistencies: neither a canonical reputation tag nor a uniform scale within each tag is defined.

\begin{insightbox}[Insight]
The ERC-8004 reputation market lacks a canonical evaluation schema. Evaluation semantics differ across chains, rating scales vary across tags, and even identical tags are scored inconsistently, leaving raw reputation values without commensurability.
\end{insightbox}

\subsection{C2 --- Manipulation of Mean Reputation Scores}
\label{sec:rep:robustness}

\noindent\textbf{Breakdown point zero.}
We call raising the score \emph{inflation} and lowering it \emph{suppression}. Arithmetic-mean aggregation lets a single record move an agent's score to any feasible target, so the required corruption fraction approaches zero as honest ratings accumulate.\footnote{The zero breakdown point follows from arithmetic-mean aggregation (Equation~\ref{eq:score}). Median aggregation instead requires corrupting approximately half the records to arbitrarily control the output, yielding a breakdown point of $50\%$.}
With feedback values bounded only by $|v|\le10^{38}$, an adversary can move $S(a)$ from $m$ to $\tau$ for an agent with $n$ ratings by submitting
\begin{equation}
\tilde{v}^{\ast} = (n+1)\,\tau - n\,m ,
\label{eq:onefeedback}
\end{equation}
which remains within the contract bound for realistic $n$ and rating targets. The same expression enables inflation ($\tau>m$) and suppression ($\tau<m$).
Note that large $n$ offers little protection in practice. The most-rated agent has only $1{,}552$ feedback records with mean $99.9$; one crafted value $\tilde{v}^{\ast}\approx-1.55\times10^{5}$ lowers its score to $0$. Moreover, the required value grows linearly with $n(\tau-m)$ and remains within the $10^{38}$ bound until $n\gtrsim10^{37}$, far beyond any realistic volume (Appendix~\ref{app:dilution}).

\begin{figure}[htb]
\centering
\begin{minipage}[t]{0.48\textwidth}
    \centering
    \includegraphics[width=\linewidth]{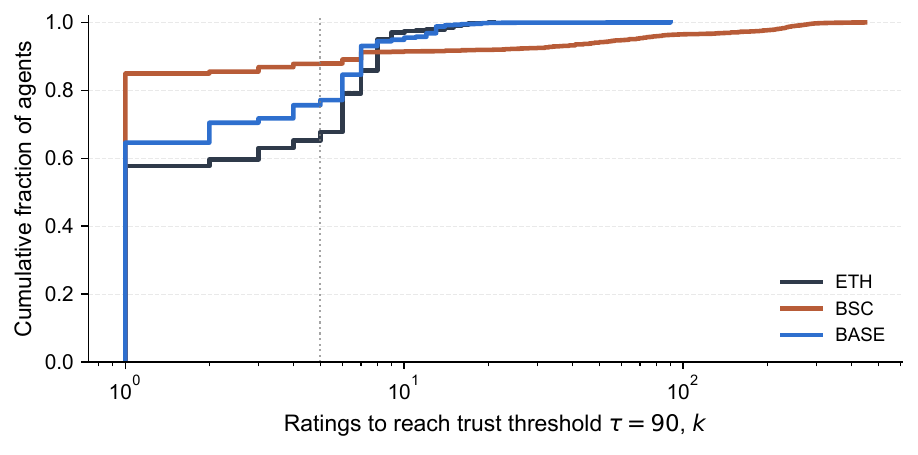}
    \caption{
    Cumulative distribution of $k$, the number of maximal ratings an adversary must add to push an agent's score past the trust threshold $\tau=90$. A smaller $k$ means the agent is cheaper to manipulate. 
    }
    \label{fig:c2_feedbacks_to_flip_cdf}
\end{minipage}
\hfill
\begin{minipage}[t]{0.48\textwidth}
    \centering
    \includegraphics[width=\linewidth]{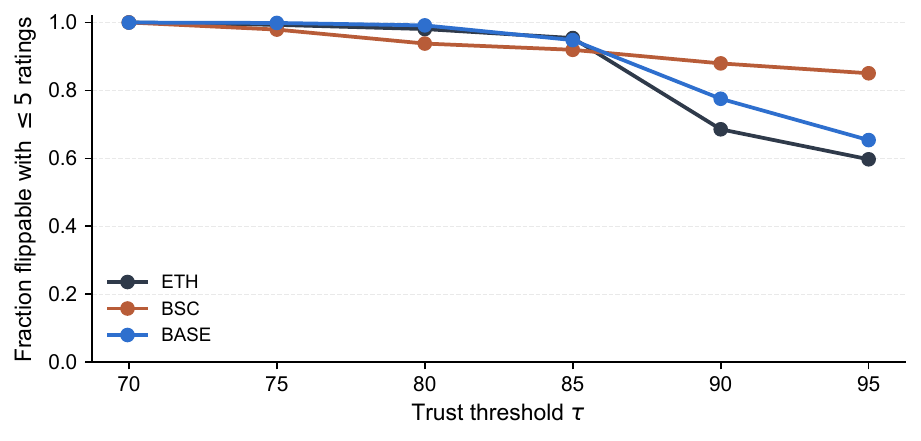}
    \caption{Fraction of agents whose mean score can be pushed past the threshold $\tau$ with $\leq5$ maximal ratings, as $\tau$ ranges from 70 to 95. Most agents can be manipulated with few maximal ratings.}
    \label{fig:c2_flip_robustness}
\end{minipage}
\end{figure}

\vspace{0.1cm}
\noindent\textbf{Bounding values does not prevent manipulation.}
Clamping each $\tilde{v}$ at $v_{\max}$ prevents the single-extreme-value attack, but an adversary can still raise the mean by submitting multiple maximal ratings: raising an agent's mean from $m$ to a threshold $\tau$, where $m<\tau<v_{\max}$, requires submitting
\begin{equation}
k \;=\; \left\lceil n\,\frac{\tau - m}{\,v_{\max} - \tau\,}\right\rceil
\label{eq:flip}
\end{equation}
ratings at $v_{\max}$.
The required number grows linearly with $n$: more existing feedback increases the attack cost but does not prevent manipulation, and each additional rating costs only gas.

Figures~\ref{fig:c2_feedbacks_to_flip_cdf} and~\ref{fig:c2_flip_robustness} quantify this vulnerability for rating-style tags,\footnote{We consider only rating-style tags that appear to use a $0$--$100$ scale and have at least $10$ samples.}
using Equation~\ref{eq:flip} with $v_{\max}=100$.
At $\tau=90$, the median agent on each chain requires only one additional maximal rating to reach the threshold, and $68$--$88\%$ require five or fewer (Figure~\ref{fig:c2_feedbacks_to_flip_cdf}).
This vulnerability persists across thresholds: as $\tau$ increases from $70$ to $95$, the proportion requiring at most five ratings decreases but remains above $60\%$ even at $\tau=95$ (Figure~\ref{fig:c2_flip_robustness}). Thus, even with bounded values, most agents can be pushed above a high rating threshold with only a few additional maximal ratings.

\begin{insightbox}[Insight]
The aggregated score~(Equation~\ref{eq:score}) has a breakdown point of zero, and the value field is bounded only by $10^{38}$.
Consequently, a single feedback record can manipulate an agent's score regardless of how many honest ratings it already has. Even under the charitable assumption that values are clamped to a ceiling, reaching any target remains possible. 
\end{insightbox}

\subsection{C3 --- Limited Evidence of Prior Interactions}
\label{sec:rep:groundedness}

\noindent\textbf{The protocol requires no proof of interaction.}
A feedback record alone does not establish a real interaction between $c$ and $a$, because any account $c$ other than the agent's owner or operator can call \texttt{giveFeedback()} without evidence of a prior interaction with $a$~\cite{erc8004contracts}.
An optional \texttt{feedbackURI} may reference payment or task proofs, but the protocol neither requires nor authenticates them. 

\begin{figure}[htb]
\centering
\begin{minipage}[t]{0.48\textwidth}
    \centering
    \includegraphics[width=\linewidth]{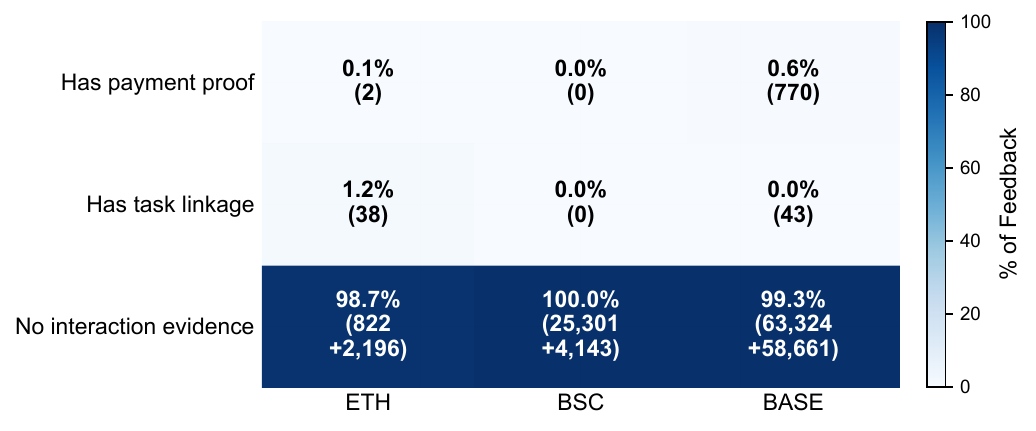}
    \caption{Feedback records classified by the strongest interaction evidence declared in their off-chain files. Most records do not provide proof of payment.}
    \label{fig:provenance}
\end{minipage}
\hfill
\begin{minipage}[t]{0.48\textwidth}
    \centering
    \includegraphics[width=\linewidth]{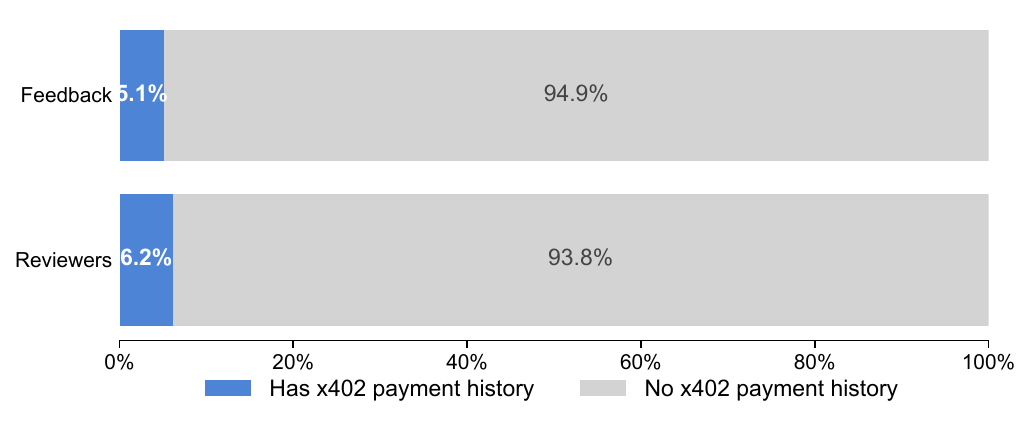}
    \caption{Share of Base chain reviewers w/wo x402 payment history (bottom bar) and their corresponding share of feedback records submitted (top bar).}
    \label{fig:usdc_overlap}
\end{minipage}
\end{figure}

\noindent\textbf{Most feedback is evidence-free.}
To assess feedback groundedness, we classify records by the strongest interaction evidence declared in their off-chain feedback documents: \emph{has payment proof} if \texttt{proofOfPayment} contains a non-null \texttt{x402Nonce} or \texttt{txHash}; \emph{has task linkage} if \texttt{proofOfPayment} is null but a non-null \texttt{a2aTaskId} or \texttt{mcpTool} declares an A2A or MCP interaction; and \emph{no interaction evidence} otherwise, including records without resolvable off-chain documents.

Figure~\ref{fig:provenance} shows that $98.7\%$ of feedback on ETH, $100.0\%$ on BSC, and $99.3\%$ on Base contains neither payment proof nor task linkage. Payment proofs account for at most $0.6\%$ (Base) and task linkages for at most $1.2\%$ (ETH). Most feedback therefore cannot be independently tied to an economic exchange or protocol-level task execution, leaving little verifiable interaction provenance.

Figure~\ref{fig:usdc_overlap} examines whether reviewers actually interacted with agents but simply did not submit evidence in their off-chain documents. Independently of self-declaration, we check whether each Base reviewer appeared as a sender in any EIP-3009 USDC transfer during our collection window, a necessary condition for USDC-based x402 payments.
Only $6.2\%$ have such payment history, accounting for $5.1\%$ of feedback. The remaining $93.8\%$ account for $94.9\%$ of feedback.\footnote{
Ideally, each feedback record would be matched to prior x402 payments between the same reviewer--agent pair. Direct attribution of x402 transactions to individual agents is, however, ambiguous in practice (Appendix~\ref{app:attribution_challenges}). We therefore use the more conservative criterion of whether the reviewer has \emph{any} x402 payment history. As a result, $93.8\%$ is a conservative lower bound. Even so, the $93.8\%$ and $94.9\%$ figures suffice to show that most feedback is evidence-free (Appendix~\ref{app:treatment}).}

\begin{insightbox}[Insight]
    The reputation contract only prohibits the agent's owner or operator from submitting feedback; any other address may rate an agent without providing proof of payment. Most feedback does not correspond to any verifiable on-chain economic interaction.
\end{insightbox}

\subsection{C4 --- Low Cost of Reputation Manipulation}
\label{sec:rep:cost}

\noindent\textbf{Per-feedback cost.} 
Because submitting feedback requires only gas, the linear cost of Equation~\ref{eq:flip} translates into a trivial dollar figure. We measure the per-feedback cost from full transaction receipts and show that moving an agent into the top decile costs pocket change on every chain. From the gas dataset, we isolate \texttt{reputation\_new\_feedback}
transactions and attribute cost per event via
$\texttt{gasCostInUSD} / \texttt{\#eventsInTx}$. Table~\ref{tab:attackcost} reports the median per-feedback cost by chain.

\begin{table}[htb]
\centering
\caption{ 
Cost to move an agent's score across the trust threshold $\tau=90$, by chain. 
The median move takes a single feedback~(\S\ref{sec:rep:robustness}), and by Equation~\ref{eq:onefeedback}, one feedback suffices in either direction (inflation or suppression).
}
\label{tab:attackcost}
\resizebox{0.55\textwidth}{!}{%
\begin{tabular}{rccc}
\toprule
 & {ETH} & {BSC} & {Base} \\
\cmidrule(rl){2-4}
Median per-feedback cost (USD) & \$0.055 & \$0.0042 & \$0.0027 \\
Median $k$ to reach $\tau$      & 1 & 1& 1 \\
Median attack cost (USD)       & \$0.055 & \$0.0042 & \$0.0027 \\
\bottomrule
\end{tabular}
}
\end{table}

On every chain, the median agent reaches $\tau$ with a single ceiling-valued feedback record (\S\ref{sec:rep:robustness}), so the median manipulation costs one feedback's gas: $\$0.0027$ on Base, $\$0.0042$ on BSC, and $\$0.055$ on Ethereum respectively. 
In fact, the same feedback works in either direction: it can lift a sub-threshold agent into the trusted tier, or, by the breakdown point of Equation~\ref{eq:onefeedback}, collapse a highly rated agent. The table thus prices the atomic manipulation, and it is pocket change everywhere.

\vspace{0.1cm}
\textit{Value at stake.}
A cheap attack matters only against value worth attacking, and these scores gate real money. 
Using settled x402 payments received by each agent's wallet as a proxy for value at stake on the Base chain, we find that the mean payment volume per agent is $\$16.74$ and the median is $\$0.7$.
On Base, the cheapest chain on which to mount the attack, the median attack cost is only $\$0.0027$, yet it can manipulate an agent with a median payment volume $259\times$ larger.\footnote{Due to the attribution challenge~(Appendix~\ref{app:attribution_challenges}), we attribute a received x402 payment to an agent only when its recipient wallet is held by a single agent at the payment's block, and we drop the ambiguous remainder (Appendix~\ref{app:treatment}).
}

\begin{insightbox}[Insight]
Manipulating an agent's score across a trust threshold costs a single feedback record. 
On Base, the median cost of manipulation is 259$\times$ lower than the median value at stake.
\end{insightbox}

\subsection{Manipulation in the Wild}
\label{sec:rep:wild}

\noindent\textbf{The one defense, and its bypass.}
\texttt{giveFeedback} rejects any caller for which  \texttt{isAuthorizedOrOwner} holds: an agent's registered owner or operator cannot rate it~\cite{erc8004contracts}. This blocks naive self-review tied to the same on-chain identity. However, the restriction applies only to the caller's address. An operator can trivially circumvent it by funding a fresh wallet and submitting feedback through it.

\begin{figure*}[htb]
\centering
\begin{minipage}[t]{0.48\textwidth}
\centering
\includegraphics[width=\linewidth]{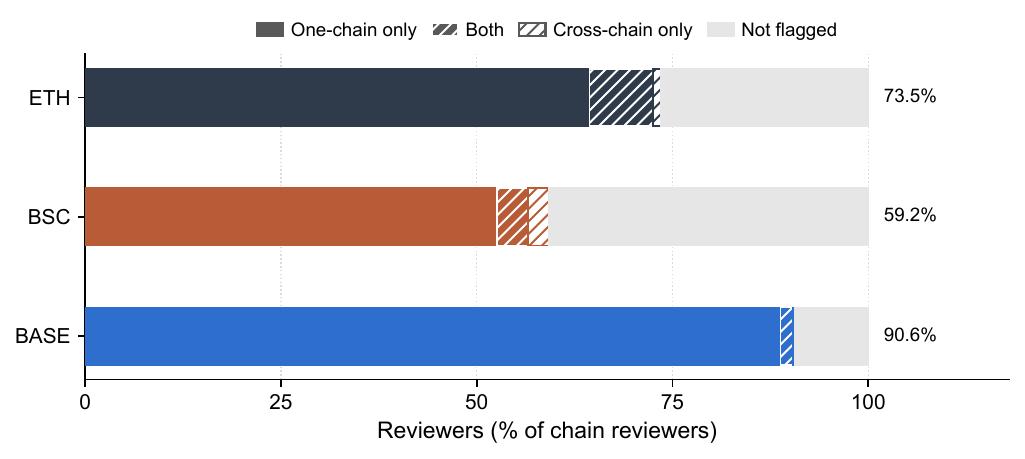}
\caption{Shared funding provenance. Bars decompose reviewer coverage in each chain into one-chain-only, cross-chain-only, and overlapping groups.}
\label{fig:eoa-first-funder-coverage}
\end{minipage}
\hfill
\begin{minipage}[t]{0.48\textwidth}
\centering
\includegraphics[width=\linewidth]{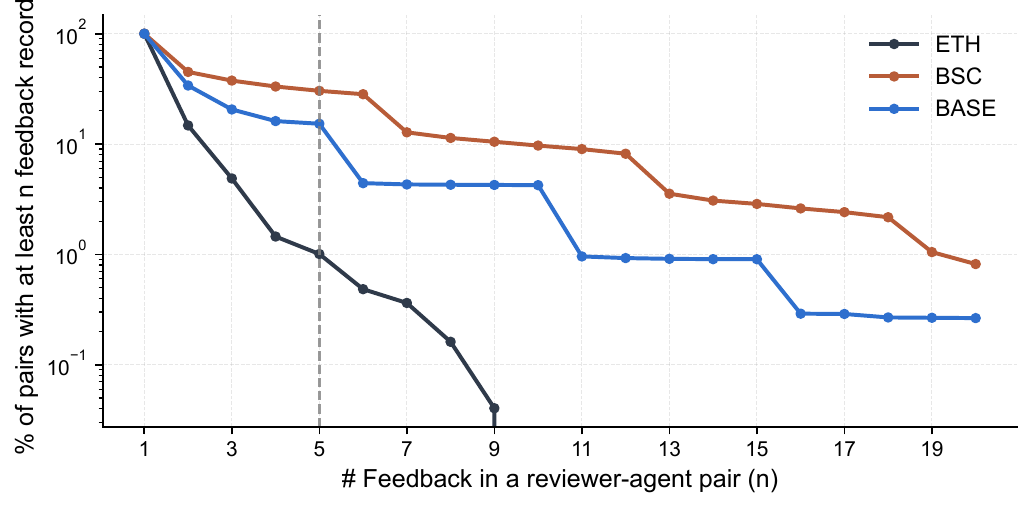}
\caption{Reviewer-agent feedback concentration. Lines show the share of reviewer-agent pairs with at least $n$ non-revoked scored feedback records.}
\label{fig:reviewer-agent-tail}
\end{minipage}
\end{figure*}

\vspace{0.2em}
\noindent\textbf{Sybil detection.} 
We apply the breadth signal to capture {\textit{shared funding provenance}}.
We construct a directed funding graph in which an edge
\(A \rightarrow B\) denotes that address \(A\) is the first native-token funder of reviewer \(B\).
Reviewer clusters are defined by tracing reviewers to a common first-funder root: if \(A \rightarrow B\) and \(B \rightarrow C\), then both \(B\) and \(C\) belong to the same funding tree rooted at \(A\).
Across chains, we merge identical addresses and extend the same rule across chains: if address \(A\) first funds reviewer \(B\) on one chain and reviewer \(C\) on another, then \(B\) and \(C\) are treated as belonging to the same candidate coordinated campaign.
We restrict first funders to EOAs and delegated EOAs, excluding smart-contract funders as infrastructure artifacts.
Figure~\ref{fig:eoa-first-funder-coverage} reports the reviewer coverage of this shared-first-funder signal across the three chains. This signal flags Sybil reviewers covering $73.5$\% of reviewers on Ethereum, $59.2$\% on BSC, and $90.6$\% on Base respectively.

\noindent\textbf{Manipulative behavior analysis.} 
Having identified candidate reviewer groups through shared funding provenance, we examine their feedback behavior. The analysis covers all reviewers, including those first funded by smart contracts; for these cases, we resolve the funding transaction to its operator EOA whenever possible and inspect the corresponding feedback, lifecycle, and nonce traces.
Figure~\ref{fig:reviewer-agent-tail} summarizes the distribution of reviewer--agent interaction counts. 

We first examine \textit{repeated-targeting} behavior. Most reviewer--agent pairs have only a few non-revoked scored feedback records, whereas the long tail contains pairs in which the same reviewer repeatedly rates the same agent, thereby increasing that reviewer's influence on the agent's aggregate reputation score.
Beyond repeatedly targeting a single agent, we observe \textit{queue-sweep} behavior: reviewers or provenance-linked groups systematically move across many agents while reusing the same score modes and tag templates. In the largest provenance-linked groups, this behavior is accompanied by multi-agent targeting, concentrated score modes, repeated tag templates, and, for contract-funded reviewers, contiguous lifecycle or nonce patterns. 
These signals are more consistent with automated reputation seeding than with independent organic reviews.
Appendix~\ref{app:sybil} further characterizes this behavior and presents additional detection heuristics.

\smallskip
\noindent\textbf{Market damage.}
To quantify the impact of Sybil-flagged feedback, we recompute each agent's aggregate reputation after removing every flagged, non-revoked scored feedback record.
After removal, an agent falls into one of two cases. 
\textit{(i)} If an agent retains at least one non-flagged, non-revoked scored feedback record, its reputation can still be recomputed. We refer to these as \emph{mixed} cases because both Sybil-flagged and non-flagged feedback originally contributed to the displayed reputation.
\textit{(ii)} Otherwise, removing the flagged feedback leaves no remaining feedback from which to recompute a reputation. We report these separately as \emph{no-baseline} cases. For mixed cases, we quantify the score-level impact by comparing the original score $S_a$ with the recomputed score
\begin{equation}
\Delta_a = S_a - S_a^{(-\mathrm{Sybil})}.
\label{eq:sybil-score-shift}
\end{equation}

\vspace{-0.3em}
\begin{table}[htb]
\centering
\footnotesize
\setlength{\tabcolsep}{3.5pt}
\caption{
Impact of removing Sybil-flagged feedback.
Sybil Feedback (\%) is computed over non-revoked scored feedback records; Affected agents (\%), No baseline (\%), and Mixed cases (\%) are over rated agents.
\#Affected agents=\#No baseline cases + \#Mixed cases. Median and mean score shifts ($\Delta$) are computed over Mixed cases.
}
\label{tab:sybil-counterfactual-damage}
\begin{tabular}{ccrrrrccc}
\toprule
Chain & \multicolumn{2}{c}{Sybil feedback} & Rated & Affected agents & No baseline cases & Mixed cases & Median $\Delta$ & Mean $\Delta$ \\

\cmidrule(lr){2-3}\cmidrule(lr){5-6}\cmidrule(lr){7-9}

ETH  & 41.4\% & \mypie{ethcolor}{41.4}  & 1,559  & 411 (26.4\%)    & 247 (15.8\%)                         & 164   & $+11.0$ & $+10.9$ \\
BSC  & 96.3\% & \mypie{bsccolor}{96.3}  & 4,310  & 3,510 (81.4\%)  & 3,356 (\textcolor{bsccolor}{77.9\%}) & 154   & $-9.1$  & $-11.5$ \\
Base & 92.6\% & \mypie{basecolor}{92.6} & 28,592 & 27,499 (96.2\%) & 24,822 (\textcolor{basecolor}{86.8\%})& 2,677 & $-0.2$  & $-5.2$ \\
\bottomrule
\end{tabular}
\end{table}

Table~\ref{tab:sybil-counterfactual-damage} shows that the dominant effect of removing Sybil-flagged feedback is not a shift in reputation scores, but the complete \textbf{disappearance} of the reputation baseline itself.  The vast majority of affected agents on BSC ($77.9\%$) and Base ($86.8\%$) do not retain any valid feedback to recompute their reputations after removal.
Among affected agents in mixed cases, the score shift is chain-specific: ETH exhibits a median (mean) inflation of
$11.0$ ($10.9$) points after removal, whereas BSC and Base exhibit median (mean) deflation of $9.1$ ($11.5$) and $0.2$ ($5.2$) points respectively. 
Detailed breakdown of market damage by agent and feedback quality is provided in Appendix~\ref{app: sybil_breakdown}.

\begin{insightbox}
ERC-8004's only on-chain anti-self-review defense is easily bypassed, enabling coordinated reviewer groups to mass-produce feedback across agents. Removing Sybil-flagged feedback leaves many agents with no remaining valid feedback to compute a reputation.
\end{insightbox}

\section{Recommendations for Protocol Designers}
\label{sec:recommendations}

This section translates the findings from our identity and reputation analyses into recommendations for the next revision of the ERC-8004 protocol and similar agent registries. 

\vspace{0.1cm}
\noindent\textbf{R1: Identity layer---separate a reserved identity from a live agent.}
Our study finds that most registered identities are placeholders rather than functional agents (\S\ref{sec:identity_analysis}). 
Reserving an identity before activation is legitimate, but the Registry does not distinguish a placeholder from a live agent. The necessary signals already exist, namely whether the URI resolves and whether the registration file is compliant, but the standard defines no canonical liveness predicate. We recommend specifying a single liveness test so that all consumers apply the same validation instead of implementing incompatible heuristics or skipping the check altogether. Reference discovery tooling should apply this predicate, hide inactive identities from discovery,
and attach no reputation to them.

\vspace{0.1cm}
\noindent\textbf{R2: Reputation layer---type the value field} (addresses C1, commensurability).
The specification deliberately leaves tags free-form, from $0$--$100$ ratings to latency in milliseconds and revenue in dollars (\S\ref{sec:rep:semantics}). It defines neither a mapping from tags to units and ranges nor a canonical reputation tag, leaving consumers unable to determine which values are comparable. We recommend a \textit{Tag Registry} that associates each tag with a unit, valid range, and direction (whether higher/lower is better), with the contract enforcing the declared range. It should define a canonical overall-rating tag on a fixed scale, and aggregate scores only within the same tag.
This is already emerging in implementations. For instance, the \href{https://github.com/QuantuLabs/8004-solana}{QuantuLabs} SDK encodes \texttt{revenues} of \$$150.00$ as \texttt{value}$=15000$ with \texttt{decimals}$=2$~\cite{quantulabs8004solana}; the \href{https://github.com/nuwa-protocol/nuwa-8004}{Nuwa} fork clamps feedback values to a $0$--$100$~\cite{nuwa8004}. Both restore comparable scales locally. Standardizing tag typing would make this portable across implementations.

\vspace{0.1cm}
\noindent\textbf{R3: Reputation layer---aggregate with a robust, bounded-influence rule} (addresses C2, robustness).
The specification leaves sophisticated aggregation to off-chain systems. The on-chain \texttt{getSummary()} returns a plain mean, which allows a single extreme value or repeated feedback from one wallet to influence reputation (\S\ref{sec:rep:robustness}). The Registry should therefore provide a safe aggregator with four properties:
\textit{(i)} Use a median or trimmed mean so no single outlier dominates.
\textit{(ii)} Clamp values to each tag's declared range.
\textit{(iii)} Limit each reviewer's contribution; we observe up to $1{,}181$ feedback records from a single reviewer--agent pair on Base.
\textit{(iv)} Weight scores by the number of distinct, evidence-backed reviewers.
Some of these aggregation design points already exist in implementation. For example, \href{https://helixa.xyz}{Helixa} ignores the raw on-chain mean and computes its own $0$--$100$ Cred Score for Base agents~\cite{helixa} (like an aggregator), but only consumers using Helixa benefit.

\vspace{0.1cm}
\noindent\textbf{R4: Reputation layer---tie feedback to a verifiable interaction} (addresses C3, groundedness).
We find that most feedback is not backed by a verifiable on-chain interaction (\S\ref{sec:rep:groundedness}).
Across the three chains, $98.7\%$--$100\%$ of records contain neither proof of payment nor a task link, and only $0.6\%$ of Base chain feedback includes a \texttt{proofOfPayment}. Although the specification defines an optional payment proof field, it neither requires nor verifies it. We recommend requiring each feedback record to reference a verifiable interaction, such as a settled x402 payment or an attested task. If mandatory evidence is impractical initially, the contract should at least record whether feedback is evidence-backed, and the default summary should report backed and unbacked counts for filtering.

\vspace{0.1cm}
\noindent\textbf{R5: Reputation layer---make influence cost scale with stakes} (addresses C4, economic soundness).
We find that fabricating an agent's reputation takes a single feedback record: $\$0.0027$ on Base, $\$0.0042$ on BSC, and $\$0.055$ on Ethereum respectively (\S\ref{sec:rep:cost}). That cost is far below the real payment values the agent guards.
We recommend attaching a cost to feedback that rises with its influence. This could be a stake that is slashed after a successful dispute, or a weight tied to settled payment volume. The cost of changing a score should grow with the value that the score controls.

\vspace{0.2em}
\noindent\textbf{R6: Reputation layer---provide a default Sybil defense} (addresses C2 \& C4).
Our study reveals widespread Sybil-style reputation manipulation across all three chains (\S\ref{sec:rep:wild}). Yet the specification leaves Sybil resistance to callers and to off-chain systems that have yet to emerge. On-chain, the only enforced rule is that an owner cannot rate its own agent, a restriction easily bypassed by funding a fresh wallet. 
We recommend that influence be bounded per \emph{principal} rather than per address. Concretely, the protocol should provide a reference Sybil filter that groups reviewers by shared funding provenance or by an attested identity, extends the self-review rule from the owner's address to the owner's funding cluster, and caps the contribution of each funder or cluster.

\section{Limitations and Future Work}
\label{sec:limitations}

\noindent\textbf{Data scope.}
We study the Identity and Reputation Registries on Ethereum, BSC, and Base, the chains with the highest registration and feedback volumes. No mainnet Validation Registry deployment was observed during our collection window. Future work should extend coverage to other chains and deployed Validation Registries to assess how attestations affect trust and reputation robustness.

\noindent\textbf{Downstream impact of reputation manipulation.}
\S\ref{sec:rep:wild} documents score manipulation in the wild, but establishing its causal effects on downstream outcomes is beyond this study's scope. Future work should investigate whether such manipulation causes financial losses or other harms.

\noindent\textbf{Design of the identity and reputation layers.}
Our study characterizes the deployed ERC-8004 ecosystem and identifies potential weaknesses. We propose recommendations in \S\ref{sec:recommendations}, but implementing and evaluating them is beyond this study's scope. Future work could develop a standardized identity-liveness check for the identity layer, and build a reputation mechanism combining typed values, verified interactions, Sybil defenses, and robust aggregation. Our dataset could help assess these designs against adaptive adversaries and their effects on new agents.

\section{Related Work}
\label{sec:related-work}

\noindent\textbf{Agent economies.} 
Autonomous agents are increasingly framed not only as software assistants, but as economic actors that discover services, delegate work, and settle payments with minimal human intervention~\cite{park2023generative, tomasev2025virtual}. Recent work on the emerging agent economy argues that such systems require machine-readable identity, programmable settlement, and portable trust signals before agents can transact safely across organizational boundaries~\cite{mao2026sok,zhang2026sok}. This has motivated a stack of agent interoperability and payment protocols: A2A and MCP standardize communication and tool use~\cite{google2025a2a, anthropic2024mcp, a2amcp2025, sun2025vision, jiang2026sok}, while x402, A402, and related systems target web-native or blockchain-mediated machine-to-machine payments~\cite{coinbase2025x402, li2026a402, l402, zhang2026sok}. Security work on agentic commerce further identifies risks around authorization, execution evidence, replay protection, and accountability~\cite{goenka2026tesspay, mao2026sok, li2026five,ling2026free}. 

\vspace{0.2em}
\noindent\textbf{Agent registries and ERC-8004.}
Agent registries make autonomous agents discoverable by binding persistent identifiers to operational metadata. Their closest precursor is decentralized identity: W3C DIDs ~\cite{sporny2022did} define controller-managed identifiers that resolve to DID documents containing verification material and service endpoints, and subsequent studies ~\cite{mazzocca2025survey,fathalla2026self,orru2025revisiting,liu2025fully} place DIDs within broader self-sovereign identity and verifiable credential ecosystems.
ERC-8004 extends it to agent markets by combining an Identity Registry with a Reputation Registry for trust decisions. Recent work~\cite{hu2025interagenttrust} positions ERC-8004 within the broader inter-agent trust landscape, while existing empirical studies remain limited to Ethereum~\cite{liu2026ethagentdataset,mafrur2026agent}.
Our work is the first multi-chain study of ERC-8004, focusing on Identity and Reputation Registries across Ethereum, BNB, and Base.

\vspace{0.2em}
\noindent\textbf{Reputation integrity in agent registries.}
Reputation is a crucial mechanism for open multi-agent systems. Classical trust models~\cite{huynh2006integrated,josang2007survey} and recent LLM-agent work~\cite{lou2025drf,chishti2026agentreputation} emphasize that reputation is meaningful only when feedback is contextualized and the reviewer is credible. Blockchain-based reputation systems~\cite{hasan2022privacy,yu2019repucoin} improve transparency by recording feedback on public ledgers, but permissionless participation makes Sybil resistance
essential~\cite{budish2024economic,crites2025syra,rajabi2023feasibility}. Recent
defenses therefore tie influence to graph structure~\cite{sybil2025airdrop,ren2025beyond} or stronger identity evidence~\cite{choudhuri2026cryptographic}. ERC-8004 adopts a thinner design, recording feedback while leaving reputation aggregation to downstream consumers. We provide the first empirical evaluation of whether this deployed
feedback layer constitutes a trustworthy basis for agent trust decisions, a question unexplored in prior work.

\section{Conclusion}
\label{sec:conclusion}

When an autonomous agent consults ERC-8004 about an unknown counterpart, can it rely on the resulting reputation to make a trust decision? Our measurements suggest not. The ERC-8004 protocol succeeds at what it explicitly provides---public identities and feedback---but they do not necessarily establish \textit{trust}. 
The root cause lies in the protocol's design. It deliberately anchors only identities and feedback on-chain while leaving semantics, evidence, aggregation, and Sybil resistance to off-chain systems that have yet to emerge.
This flexibility makes the protocol easy to adopt, but also leaves reputation signals incomparable, weakly grounded, and cheap to manipulate. 

Our findings also have implications for the emerging agent economy.
Registration counts and feedback volume are poor indicators of maturity when many identities are placeholders and much of the reputation is coordinated. Similar risks are likely to arise in any open agent registry that binds identity to behavioral reputation without requiring verifiable evidence
or robust aggregation.

As a young protocol, ERC-8004 still has an opportunity to address these weaknesses before adoption locks in the design.
We hope the empirical baseline provided by this work will inform future revisions of the protocol and help close the gap between recording trust and earning it.

\bibliographystyle{ACM-Reference-Format}
\bibliography{references}

\appendix

\section{ERC-8004 Background: Supplementary Diagram}

\begin{figure}[H]
\centering
    \includegraphics[width=0.9\linewidth]{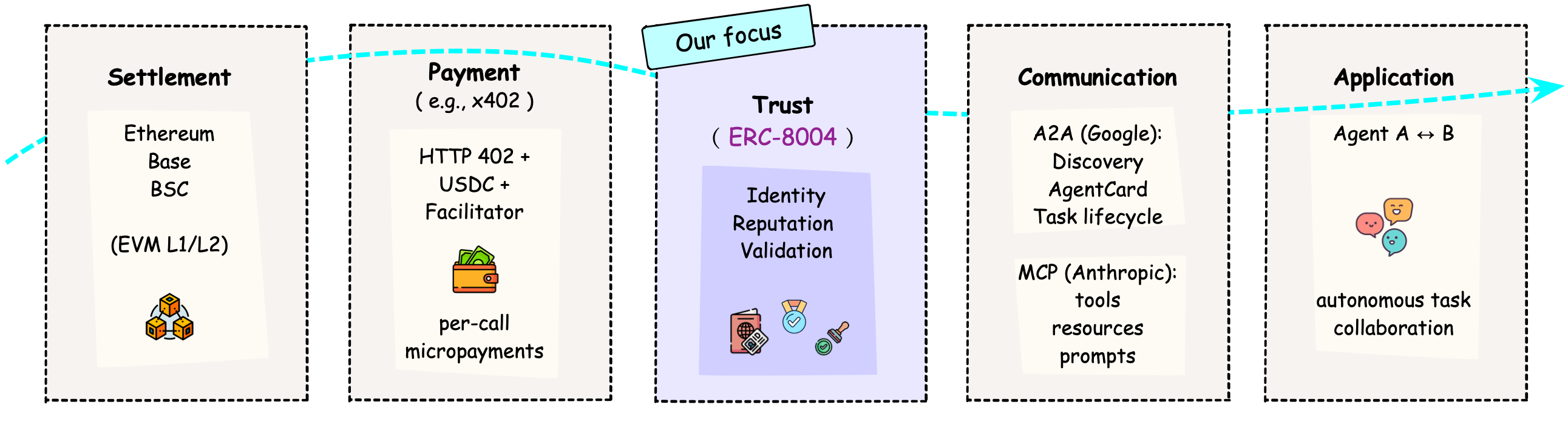}
    \caption{The agentic-internet stack. ERC-8004 supplies the missing \emph{trust layer} between agent communication protocols (A2A/MCP) and value settlement (x402 over EVM chains), so that each layer makes the others usable
for open, cross-organizational agent economies. }
    \label{fig:stack}
\end{figure}

Figure~\ref{fig:stack} illustrates the agentic-internet stack and the position of ERC-8004 as the trust layer between agent communication protocols and value settlement infrastructure.

\section{Supplementary Details on Identity and Reputation Data Collection}
\label{app:data_more}

\begin{table*}[htb]
  \centering
  \caption{On-chain events crawled and off-chain files fetched for the Identity and Reputation registries.}
  \label{tab:datasets}
  \resizebox{\linewidth}{!}{
  \begin{tabular}{crp{13cm}}
    \toprule
     & {Type} & \multicolumn{1}{c}{ {File / Content}} \\
    \cmidrule(rl){2-3}
    \multirow{9}{*}{\rotatebox{90}{Identity}}
      & Event & \texttt{Registered} --- one record per agent, storing
                 \texttt{agentId}, wallet, and mint URI.\\
      & Event & \texttt{Transfer} (mint, \texttt{from=0x0}) --- ERC-721 mint
                signal; row count mirrors \texttt{Registered}. \\
      & Event & \texttt{Transfer} (secondary, \texttt{from$\neq$0x0}) ---
                post-mint ownership transfers. \\
      & Event & \texttt{URIUpdated} --- records every \texttt{setAgentURI()}
                call after minting; block-ordered replay over
                \texttt{Registered} yields the current URI per agent. \\
      & Event & \texttt{MetadataSet} (\texttt{agentWallet}) --- payment
                wallet set by the owner; ABI-encoded address. \\
      & Event & \texttt{MetadataSet} (other keys) --- arbitrary key--value
                metadata (e.g.\ \texttt{version}, \texttt{framework}). \\
      & Derived & Agent URI and
                  ownership at \texttt{END\_BLOCK}, computed from event replay. \\
      & Off-chain & Registration file per agent: parsed JSON at the agent's current URI. \\
     \cmidrule(rl){2-2}\cmidrule(rl){3-3}
    \multirow{8}{*}{\rotatebox{90}{Reputation}}
      & Event & \texttt{NewFeedback} --- one record per feedback submission;
                includes raw signed value, decimal precision, and semantic tags
                (\texttt{tag1}, \texttt{tag2}), rated endpoint,
                and \texttt{feedbackURI}.\\
      & Event & \texttt{FeedbackRevoked} --- \texttt{NewFeedback} revocations by
              $(\textit{agentId}, \textit{clientAddress}, \textit{feedbackIndex})$. \\
      & Event & \texttt{ResponseAppended} --- responses to feedback;
                includes \texttt{responseURI}. \\
      & Derived & Rolling reputation timeseries: mean score over active feedback
              at each event block. \\
      & Derived & Client snapshot: \texttt{getClients()} per agent at \texttt{END\_BLOCK} \\
      & Off-chain & Feedback file per submission: parsed JSON at \texttt{feedbackURI}. \\
      & Off-chain & Response file per response: parsed JSON at
                    \texttt{responseURI}. \\
    \bottomrule
  \end{tabular}
  }
\end{table*}

Table~\ref{tab:datasets} summarizes the collected on-chain events for the Identity and Reputation registries, with off-chain files referenced by those events and derived metrics. 

\vspace{0.2cm}
\noindent\textbf{Off-chain data collection and temporal scope}.
We retain each agent’s initial URI and all subsequent \texttt{URIUpdated} events throughout the observation window, as described in \S\ref{sec:data}. Replaying these events reconstructs its URI history for the activation-status and activation-lag analyses and determines its final URI at \texttt{END\_BLOCK}. At the end of the observation window, we fetch the registration file at this final URI to assess availability, compliance, and declared services. Thus, URI activation (\S\ref{sec:Identity-Activity Gap}) is analyzed longitudinally, while off-chain content is assessed as an end-of-window snapshot, which does not establish historical file or service availability.

\section{x402: Data Collection and Attribution Challenges}
\label{app:x402}

\subsection{The x402 Protocol}
x402~\cite{coinbase2025x402,li2026five} is an open payment protocol built on the HTTP 402 ``Payment Required'' status code.
It is designed for programmatic payments, including payments made by autonomous agents, where a client pays for a resource such as an API call on a per-request basis. The flow has three steps.
A client first requests a resource. 
The server replies with a 402 status and a set of payment requirements.
The client then resends the request together with a cryptographically signed payment payload.
A third party called a facilitator verifies the payment and settles it on-chain.
Payments are generally denominated in USDC and are intended to be small and frequent.

\smallskip
\noindent\textbf{On-chain form.}
An x402 payment settles as an EIP-3009~\cite{kim2020eip3009} authorized transfer of a stablecoin, in practice USDC.
The payer signs a transfer authorization off-chain and therefore pays no gas. A facilitator then submits the corresponding \texttt{transferWithAuthorization} call, or its \texttt{receiveWithAuthorization} variant, to the token contract.
The authorization specifies the payer, the recipient, the amount, a validity window, and a unique nonce that prevents replay.
On execution, the token contract emits two events.
The first is \texttt{AuthorizationUsed}, which records the authorizing payer and the nonce.
The second is a standard ERC-20 \texttt{Transfer}, which records the payer, the recipient, and the amount.
An on-chain x402 settlement therefore takes the form of a USDC authorized transfer that emits this paired \texttt{AuthorizationUsed} and \texttt{Transfer}.

The x402 protocol layer itself is off-chain.
The HTTP 402 exchange and the facilitator routing leave no on-chain marker beyond the EIP-3009 settlement.
On-chain, an x402 payment is therefore indistinguishable from any other EIP-3009 authorized USDC transfer.
We adopt this settlement form as our operational definition of an on-chain x402 payment, and we treat it as an upper bound on protocol-level x402.
Every x402 settlement in USDC satisfies the definition, while a transaction that satisfies it may also be an EIP-3009 transfer that did not originate from the x402 protocol.

A transaction $tx$ matches this form when, on the USDC contract, it emits an \texttt{AuthorizationUsed} from a payer together with a \texttt{Transfer} of the same payer. Formally,
\begin{equation}
\label{eq:x402}
\begin{aligned}
\mathrm{x402}(tx)\ \Longleftrightarrow\ &
\exists\, a, r, v, \nu:\; \\
&\mathsf{AuthorizationUsed}(a,\nu)
   \in \mathrm{logs}_{\mathrm{USDC}}(tx)
   \;\wedge \mathsf{Transfer}(a,r,v)
   \in \mathrm{logs}_{\mathrm{USDC}}(tx)
\end{aligned}
\end{equation}
where $a$ is the paying address (the authorizer), $r$ the recipient, $v$ the transferred amount, $\nu$ the authorization nonce, and $\mathrm{logs}_{\mathrm{USDC}}(tx)$ the events emitted by the USDC contract in $tx$. The shared symbol $a$ encodes that the payer of the \texttt{Transfer} is the authorizer of the \texttt{AuthorizationUsed}.

\subsection{Two Forms of x402 Settlement}
\label{sec:x402-forms}

An x402 payment reaches an agent in one of two forms (see Figure~\ref{fig:x402-forms}):

\begin{figure}[htb]
\centering
\begin{minipage}[c]{0.38\linewidth}
\centering
\resizebox{\linewidth}{!}{%
\begin{tikzpicture}[
  >={Stealth[length=2mm]},
  box/.style={draw, rounded corners=2pt, align=center,
              text width=22mm, minimum height=11mm,
              font=\small, inner sep=3pt},
  x402/.style={->, thick},
  release/.style={->, thick, dashed},
  edgelbl/.style={font=\scriptsize, align=center, inner sep=2pt}
]
\node[box, fill=white]   (p) at (0,0)       {Payer\\[-1pt]\scriptsize client};
\node[box, fill=white]   (a) at (3.6,0)     {Agent wallet\\[-1pt]\scriptsize ERC-8004};
\node[box, fill=black!4] (e) at (0,-1.8)    {Escrow\\[-1pt]\scriptsize ACP escrow};
\node[box, fill=white]   (v) at (3.6,-1.8)  {Provider\\[-1pt]\scriptsize Virtuals / operator};

\draw[x402] (p.east) -- node[edgelbl, above]{x402\\(direct)} (a.west);
\draw[x402] (p.south) -- node[edgelbl, right]{x402} (e.north);
\draw[release] (e.east) -- node[edgelbl, above]{release} (v.west);
\end{tikzpicture}%
}
\end{minipage}\hfill
\begin{minipage}[c]{0.59\linewidth}
\captionsetup{skip=0pt}
\caption{Two on-chain forms of an x402 payment. \emph{Solid} arrows denote EIP-3009 settlements emitting \texttt{AuthorizationUsed} and \texttt{Transfer}. In the \emph{direct} form (top row), USDC reaches the agent's declared ERC-8004 wallet. In the \emph{escrow} form (down, then right), funds enter the ACP escrow before a plain \texttt{Transfer} releases them to the provider (\emph{dashed} arrow). The provider address is typically not an ERC-8004 wallet.}
\label{fig:x402-forms}
\end{minipage}
\Description{Four boxes in a two-by-two grid: Payer and Agent wallet above Escrow and Provider. A solid arrow connects Payer to Agent wallet for direct x402 settlement; another connects Payer to Escrow. A dashed release arrow connects Escrow to Provider.}
\end{figure}

\smallskip
\noindent\ding{192}~\textit{\textbf{Direct settlement.}}
The payer settles USDC straight to the recipient.
The payer signs an EIP-3009 authorization off-chain.
A facilitator submits the corresponding \texttt{transferWithAuthorization} call.
The USDC contract then emits \texttt{AuthorizationUsed} and a \texttt{Transfer}
whose recipient is the agent wallet.
The whole payment is a single on-chain settlement.
The agent is paid immediately.

\smallskip
\noindent\ding{193}~\textit{\textbf{Escrow settlement.}}
The client does not pay the agent directly.
The agent acts as a provider on the Agent Commerce Protocol (ACP)~\cite{acp2026github-virtual}, and it is paid at the provider address recorded for the job, not at the wallet it declares for direct payment.
The client's x402 payment funds a shared ACP escrow contract, and the escrow later releases the funds to the provider address through a separate transfer.
In our dataset, 669 feedback records on the Base chain carry an x402 authorization nonce in their \texttt{proofOfPayment} field.
Each nonce is the indexed parameter of an \texttt{AuthorizationUsed} event, so it resolves to a unique settlement transaction.
For 668 of these 669 records, the USDC settles to a single address, the ACP~\cite{acp2026github-virtual} \texttt{PaymentManager}.\footnote{\href{https://basescan.org/address/0xEF4364Fe4487353dF46eb7c811D4FAc78b856c7F}{0xEF4364Fe4487353dF46eb7c811D4FAc78b856c7F}, an \texttt{ERC-1967} escrow proxy shared by all ACP jobs on Base.}
This concentration is structural, since every ACP job settles through that single escrow.
The escrow lifecycle has the following stages.
\begin{enumerate}[leftmargin=*]
  \item \textbf{Account.} Every job is created under an account that records the client--provider pair, emitted as \texttt{AccountCreated} and referenced by the job's \texttt{accountId}. The account is created automatically in the same
  transaction as the job, so it is not a separate client action. A new account is minted per job on the default path, while a pre-existing account may instead be reused for repeated jobs between the same pair. The account groups jobs and tracks interaction counts.

    \item \textbf{Job creation.} The client opens a job through the router entry \texttt{createX402Job}, which routes to \texttt{createJobWithX402} on the job module. The contract emits \texttt{JobCreated}, which records the job identifier (\texttt{jobId}), the account, the client, the provider, an optional evaluator, and the expiry. A companion \texttt{BudgetSet} event records the budget. The job is marked as x402-funded.
    
  \item \textbf{Funding (the x402 leg).} The payer signs an EIP-3009 authorization and a facilitator submits it.
  The USDC moves into the escrow contract.
  This is the only stage that is an x402 settlement.
  It emits \texttt{AuthorizationUsed} and a \texttt{Transfer} whose recipient is the escrow contract, not the agent.
  The transfer itself carries no job identifier.

  \item \textbf{Receipt confirmation.} The protocol uses \texttt{confirmX402PaymentReceived} to mark the budget as received and emits \texttt{X402PaymentReceived}.
  This step links the deposited funds to a \texttt{jobId}.

  \item \textbf{Execution and evaluation.} The provider performs the work. The job advances through its phases, and an evaluator may validate the deliverable.

\item \textbf{Settlement.}
  On success, the contract calls \texttt{releasePayment} and emits
  \texttt{PaymentReleased}, which records the \texttt{jobId}, the recipient, and the released amount. The escrow sends the budget, minus platform and evaluator fees, to the provider. The recipient passed to \texttt{releasePayment} is the provider, 
  and it is the address recorded in \texttt{PaymentReleased}.
  On failure or expiry, the contract calls \texttt{refundBudget} and emits \texttt{PaymentRefunded}, returning the funds to the client.
\end{enumerate}

\subsection{Attribution Challenges}
\label{app:attribution_challenges}

In practice, linking a given x402 payment transaction to an ERC-8004 \texttt{agentId} is challenging.

In the direct form, the recipient of the x402 \texttt{Transfer} is an agent wallet, which may be shared by multiple agents. 
For each block, we can build the map from a wallet to the agents that declare it as their payment wallet at that block.
The map is block-dependent, since an agent can reassign its wallet over time. When the wallet is declared by exactly one agent at the payment's block, we attribute the payment to that \texttt{agentId}. However, when several agents \textbf{share} the same wallet at that block, the transfer does not reveal which agent the payment is for, making attribution \textbf{ambiguous}.

In the escrow form, the challenge is deeper, and it has three layers. First, the x402 \texttt{Transfer} is sent to the \textbf{shared} ACP \texttt{PaymentManager}, not to an agent wallet, so the recipient is identical across agents and reveals nothing. 
Second, the deposit (x402 payment in step (3)) carries no \texttt{jobId}, and the \texttt{X402PaymentReceived} event marks a job as funded without referencing the deposit, so the transaction itself \textbf{does not link} a specific deposit to a \texttt{jobId}. 
Third, even once a \texttt{jobId} is known, the provider is the agent's ACP-side payout address that need not equal its declared ERC-8004 wallet or owner. In our data, the provider set and the declared-wallet set do not overlap. 

\vspace{0.2em}
In summary, these challenges make attribution unreliable. \textbf{Only one case provides a clean link}: a direct payment whose recipient wallet is declared by a single agent in the payment's block. 
This payment can be mapped to that \texttt{agentId}. 
In every other case, a direct payment into a shared wallet or any escrow payment, attributing the payment to an \texttt{agentId} has no on-chain anchor.

\subsection{Our Treatment}
\label{app:treatment}
These challenges shape how we report x402 payments in the paper's main text.
Rather than force a single attribution, we make a few deliberate choices about what we attribute, what we aggregate, and what we only bound.
We describe these choices in detail below.

\smallskip
\noindent{\textbf{\S\ref{sec:rep:groundedness} (reviewer-level test)}}.
To determine whether each feedback record is supported by a genuine interaction, the ideal test is to check whether reviewer $c$ made an x402 payment to agent $a$ before the feedback was submitted. This \textbf{pair-level} test would
confirm the economic interaction behind that feedback.
Formally, for a feedback record $(c,a,b)$ submitted at block $b$, define
\begin{equation}
P(c,a,b)=
\mathbf{1}\!\left[
\exists\, t<b:\;
c\xrightarrow{\mathrm{x402}} a
\right],
\label{eq:pair-test}
\end{equation}
where $P(c,a,b)=1$ indicates that reviewer $c$ made an x402 payment to agent $a$ before the feedback.
This test requires that each x402 payment be attributed to the specific agent $a$. However, as established above, that attribution is unreliable.
In the direct form, the recipient wallet may be declared by several agents at the payment's block, so a payment into it cannot be assigned to a unique \texttt{agentId}. In the escrow form, the payment enters a shared escrow, carries no \texttt{jobId}, and the eventual provider is not the agent's declared wallet, so we cannot link the payment to an agent.
We therefore cannot, in general, evaluate $P(c,a,b)$.
Instead, we evaluate the \textbf{reviewer-level} test:
\begin{equation}
H(c)=
\mathbf{1}\!\left[
\exists\, t:\;
c\xrightarrow{\mathrm{x402}} *
\right],
\label{eq:reviewer-test}
\end{equation}

where $*$ denotes any agent.
Concretely, we check whether $c$ appears as the authorizing payer of an EIP-3009 USDC transfer, namely as the \texttt{authorizer} of an \texttt{AuthorizationUsed} event and the value sender of the paired \texttt{Transfer}. We match on the authorizer because x402 is gasless and each
transfer is submitted by a facilitator. Accepting any EIP-3009 transfer makes the criterion even more conservative.
The reviewer-level test is a necessary condition for the pair-level test:
\begin{equation}
P(c,a,b)
\Longrightarrow
H(c).
\label{eq:necessary}
\end{equation}
Equivalently, by contraposition,
\begin{equation}
H(c)=0
\Longrightarrow
P(c,a,b)=0.
\label{eq:contrapositive}
\end{equation}

Thus, reviewers who fail our test---those with no x402 payment history at all---necessarily submit feedback unsupported by any x402 transaction.
The converse does not hold. A reviewer may have made prior x402 payments to another agent $a'$, yet submitted feedback for agent $a$. Our criterion may therefore over-credit a feedback record, but it can never wrongly reject one.

As a result, the reported share in
Figure~\ref{fig:usdc_overlap} is a conservative bound.
The $93.8\%$ of reviewers and $94.9\%$ of feedback that we report as lacking on-chain payment backing can only increase under the stricter pair-level test, which would also reject the over-credited reviewers. Equivalently, the grounded
share is at most $6.2\%$ of reviewers and $5.1\%$ of feedback.

Our conclusion is therefore robust to the attribution gap. Since Equation~\eqref{eq:contrapositive} guarantees that every reviewer counted as unsupported is truly unsupported, these conservative lower bounds already establish that the overwhelming majority of feedback lacks transaction evidence.

\medskip
\noindent{\textbf{\S\ref{sec:rep:cost} (value at stake)}}.
The value-at-stake proxy needs a per-agent quantity, the
x402 volume an agent has received at its own wallet.  Building the proxy therefore requires attributing received payments to specific agents, which is exactly the step that
is not always possible. We resolve this by attributing only the unambiguous payments and by stating precisely what we discard.

We attribute through a block-dependent wallet ownership map. For a wallet $w$ and a block $b$, let $\mathrm{own}(w,b)$ be the set of agents that declare $w$ as their payment wallet at $b$. This map is block-dependent because an agent can reassign its wallet over time, so we build it from the wallet metadata history and evaluate it at the block of each payment. We attribute a received payment $p$,
with recipient wallet $w(p)$, block $b(p)$, and amount $v(p)$, to an agent $a$
only when
\begin{equation}
\label{eq:ownership}
\mathrm{own}\!\left(w(p),\,b(p)\right)=\{a\}.
\end{equation}
That is, the recipient wallet is held by exactly one agent at the payment's block.
When the wallet is shared by several agents at that block, we discard the payment
rather than split or guess it. This rule only ever drops payments, and it never
moves a payment to the wrong agent. Each agent's attributed volume is thus a lower bound on the volume that agent truly received.
\begin{equation}
\label{eq:agent-volume}
\widehat{\mathrm{vol}}(a)
=
\sum_{p:\,\mathrm{own}(w(p),\,b(p))=\{a\}}
v(p).
\end{equation}

\textit{What the proxy omits.}
The proxy understates the value at stake on two independent counts. First, it drops every direct payment into a wallet that is shared at the payment's block, as above. Second, it counts only the direct form. Escrow-routed x402 never reaches
the declared wallet, since it is sent to the shared ACP \texttt{PaymentManager}
and later released to a provider address that is not the agent's declared wallet. Such volume is absent from
$\widehat{\mathrm{vol}}(a)$ by construction. Both omissions remove non-negative amounts, so for every agent, the counted volume cannot exceed the true volume.

\textit{Population, estimator, and robustness.}
We report the mean and median of $\widehat{\mathrm{vol}}(a)$ over agents with at least one attributable payment. We condition on this set because the unconditional median over all registered agents is dominated by agents with no recorded payment. In contrast, conditional figures describe a typical paid agent, the kind of agent an attacker would target.
Importantly, we do not read them as bounds, since closing the attribution gap would also change which agents enter the conditioning set, so the conditional median need not move in a fixed direction.

The argument in \S\ref{sec:rep:cost} does not rest on the conditional median.
It rests on three quantities that the attribution gap cannot turn against us.
First, a per-agent bound. For any agent we price, $\widehat{\mathrm{vol}}(a)\le
\mathrm{vol}_{\mathrm{true}}(a)$, so the value we attribute is a floor on its true
value at stake, and recovering dropped or escrow-routed volume can only raise it.
Second, a count that is a floor by construction. Since
$\widehat{\mathrm{vol}}(a)\le\mathrm{vol}_{\mathrm{true}}(a)$ holds pointwise,
every agent we observe above the attack cost $\tau_{\mathrm{atk}}=\$0.0027$ is
genuinely above it, and the true number of such agents can only be larger:
\begin{equation}
\label{eq:attack-threshold}
\{a:\widehat{\mathrm{vol}}(a)\ge\tau_{\mathrm{atk}}\}
\subseteq
\{a:\mathrm{vol}_{\mathrm{true}}(a)\ge\tau_{\mathrm{atk}}\}.
\end{equation}
Finally, the aggregate attributed volume, $\sum_a\widehat{\mathrm{vol}}(a)$, is also a lower estimate of the total value at stake. Filling in the missing attribution can only increase this total. We therefore use the conditional mean and median mainly to give a sense of scale; the substantive comparison between attack cost and value at stake relies on quantities that remain conservative under the attribution gap.


\section{Supplementary Analysis of Agent Quality}
\label{app:quality}

\S\ref{sec: service_and_quality} provides a coarse-grained view of agent quality. This Appendix aims to present a finer-grained assessment.
We measure quality along two owner-side dimensions: the genuineness of an agent's registration and the maintenance provided by its owner afterward, captured by registration content and post-registration owner activity respectively. We deliberately exclude reputation signals.\footnote{As shown in \S\ref{sec:rep_security}, the reported reputation score cannot serve as a reliable trust signal.}
The analysis covers the full registration population.
Table~\ref{tab:qa-inputs} summarizes the input indicators and their satisfaction rates by chain.
We then combine the two unit-interval dimensions into a single relative score to facilitate cross-agent and cross-chain comparison.

\begin{table}[htb]
\centering
\caption{Descriptive satisfaction rates of the construction inputs, by chain. All inputs are binary indicators; the last three columns report the percentage of agents satisfying each indicator on each chain.}
\label{tab:qa-inputs}
\resizebox{0.91\textwidth}{!}{%
\begin{tabular}{lllc c ccc}
\toprule
& \multicolumn{1}{c}{Indicator} & \multicolumn{1}{c}{Description} & \multicolumn{1}{c}{Type} &   & BASE & BSC & ETH \\

\cmidrule(rl){2-4}\cmidrule(rl){6-6}\cmidrule(rl){7-7}\cmidrule(rl){8-8}
\multirow{7}{*}{\rotatebox{90}{\textit{Content}}}
& ERC8004        & valid ERC-8004 registration   & Binary $\{0,1\}$ & \multirow{12}{*}{\rotatebox{90}{\textit{Share Satisfied} (\%)}} & 26.9~\mypie{basecolor}{26.9}  & 83.3~\mypie{bsccolor}{83.3} & 29.1~\mypie{ethcolor}{29.1} \\
& Description    & non-empty description         & Binary $\{0,1\}$ & & 52.2~\mypie{basecolor}{52.2} & 85.3~\mypie{bsccolor}{85.3} & 39.2~\mypie{ethcolor}{39.2} \\
& Service        & service declared$^{\dagger}$              & Binary $\{0,1\}$ & & 15.9~\mypie{basecolor}{15.9} &  5.2~\mypie{bsccolor}{5.2} &  3.5~\mypie{ethcolor}{3.5} \\
& Trust Model    & trust model declared          & Binary $\{0,1\}$ & & 10.6~\mypie{basecolor}{10.6} & 73.4~\mypie{bsccolor}{73.4} &  4.2~\mypie{ethcolor}{4.2} \\
& x402           & x402 payment support          & Binary $\{0,1\}$ & & 3.2~\mypie{basecolor}{3.2} &  0.8~\mypie{bsccolor}{0.8} & 14.0~\mypie{ethcolor}{14.0} \\
& Name           & valid agent name              & Binary $\{0,1\}$ & & 50.0~\mypie{basecolor}{50.0} & 84.3~\mypie{bsccolor}{84.3} & 38.9~\mypie{ethcolor}{38.9} \\
& Well-known     & .well-known resolution        & Binary $\{0,1\}$ & & 1.9~\mypie{basecolor}{1.9} &  0.0~\mypie{bsccolor}{0.0} &  1.2~\mypie{ethcolor}{1.2} \\
\cmidrule(rl){2-4}\cmidrule(rl){6-6}\cmidrule(rl){7-7}\cmidrule(rl){8-8}
\multirow{4}{*}{\rotatebox{90}{\textit{Engaged}}}
& URI Update      & URI updated post-mint$^{\ddagger}$            & Binary $\{0,1\}$ & & 14.9~\mypie{basecolor}{14.9} & 0.8~\mypie{bsccolor}{0.8} & 1.4~\mypie{ethcolor}{1.4} \\
& Metadata Update & metadata updated post-mint       & Binary $\{0,1\}$ & & 5.0~\mypie{basecolor}{5.0} & 0.0~\mypie{bsccolor}{0.0} & 0.0~\mypie{ethcolor}{0.0} \\
& Review Response & owner replied to reviews         & Binary $\{0,1\}$ & & 8.5~\mypie{basecolor}{8.5} & 0.0~\mypie{bsccolor}{0.0} & 0.0~\mypie{ethcolor}{0.0} \\
& Delayed Update  & last update $>$ 1 day after mint & Binary $\{0,1\}$ & & 3.1~\mypie{basecolor}{3.1} & 0.0~\mypie{bsccolor}{0.0} & 0.2~\mypie{ethcolor}{0.2} \\

\bottomrule
\end{tabular}%
}
\vspace{2pt}
\begin{minipage}{0.91\textwidth}
\footnotesize
$^{\dagger}$ Includes services declared in non-compliant registration files, so values are slightly higher than in Figure~\ref{fig:quality_taxonomy}.\\
$^{\ddagger}$ Also includes agents that already had a URI at mint and later changed it, so values differ slightly from ``Activated Later'' in Figure~\ref{fig:activation_grouped}.
\end{minipage}
\end{table}

\noindent\textbf{Content axis.} We score each agent using seven binary indicators capturing registration completeness and compliance. Each indicator is binary: a valid ERC-8004 registration, a non-empty description, a declared service, a declared trust model, x402 support, a valid name, and successful \texttt{.well-known} resolution. The unretrievable registrations score zero on all seven. The content score is defined as the first dimension of a Multiple Correspondence Analysis (MCA) on these indicators and is rescaled to $[0,1]$, with higher values indicating greater compliance and completeness. The first dimension explains $47.6\%$ of inertia and closely matches the raw share of satisfied indicators (Spearman $\rho = 0.982$), indicating a robust aggregation. Mean scores are $0.565$ on BSC, $0.268$ on Base, and $0.214$ on Ethereum, reflecting the greater prevalence of standardized registrations on BSC.

\begin{figure}[htb]
\begin{minipage}[t]{0.48\textwidth}
\centering
\includegraphics[width=\linewidth]{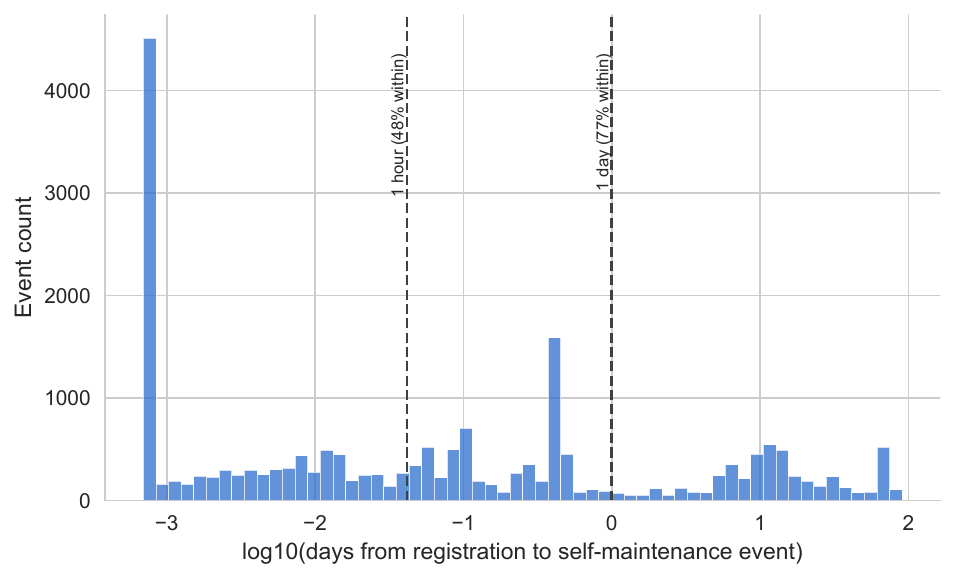}
\caption{Distribution of the time between registration and each post-registration self-maintenance event.
}
\label{fig:qa-setup}
\end{minipage}
\hfill
\begin{minipage}[t]{0.48\textwidth}
\centering
\includegraphics[width=\linewidth]{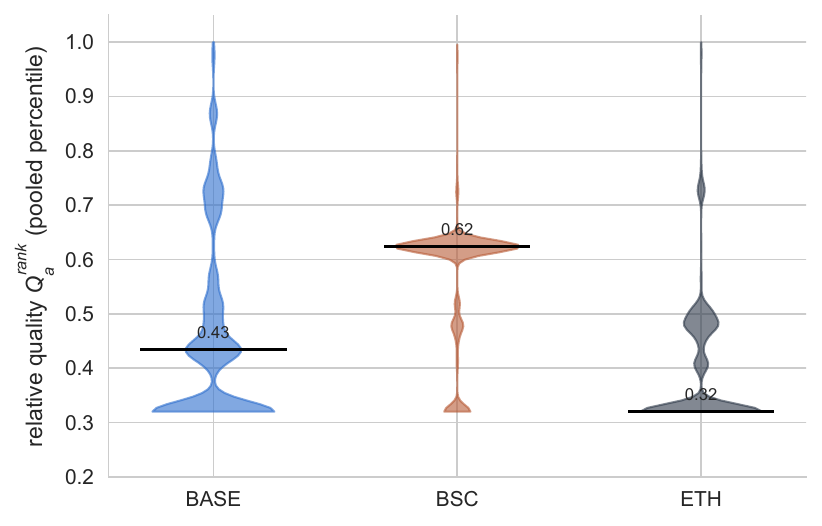}
\caption{Distribution of relative quality score $Q_a^{\text{rank}}$ by chain (with the median marked).
}
\label{fig:qa-relative}
\end{minipage}
\end{figure}

\smallskip
\noindent\textbf{Engagement axis.} This axis measures post-mint owner engagement, serving as a proxy for agent liveness. Because ERC-8004 agents operate off-chain and engagement is largely unobservable on-chain, we use post-registration updates as a proxy for owner engagement.
We set four binary indicators: any URI update, any metadata update, any appended client review response, and whether these actions occur more than one day after registration. The final indicator distinguishes sustained engagement from setup-time activity, as $48.5\%$ of updates occur within one hour and $77.3\%$ within one day of minting (Figure~\ref{fig:qa-setup}). The engagement score is the first MCA dimension of these indicators, explaining $42.0\%$ of inertia and with high correlation (Spearman $\rho = 0.824$) with the raw share of satisfied indicators. Engagement levels are low across all three chains, but remain higher on Base, with mean scores of $0.069$, $0.005$, and $0.002$ for Base, Ethereum, and BSC respectively.

\smallskip
\noindent\textbf{Relative quality score.} Content and engagement capture distinct aspects of quality and are essentially uncorrelated (Pearson $0.013$, Spearman $-0.008$). As a result, we construct a relative quality score $Q_a^{\text{rank}}$ by averaging them to obtain a single score for comparing platforms on a common scale. Pooling ranks over the full population ensures comparability across chains and yields a purely relative measure of position within the overall distribution.
As shown in Figure~\ref{fig:qa-relative}, BSC attains the highest relative score (median $0.62$), followed by Base ($0.43$) and Ethereum ($0.32$). This reflects BSC’s high registration-file compliance rather than operational activity: its mean content score is $0.565$, whereas its mean engagement score is only $0.002$.

\begin{figure}[htb]
\begin{minipage}[t]{0.48\textwidth}
\centering
\includegraphics[width=\linewidth]{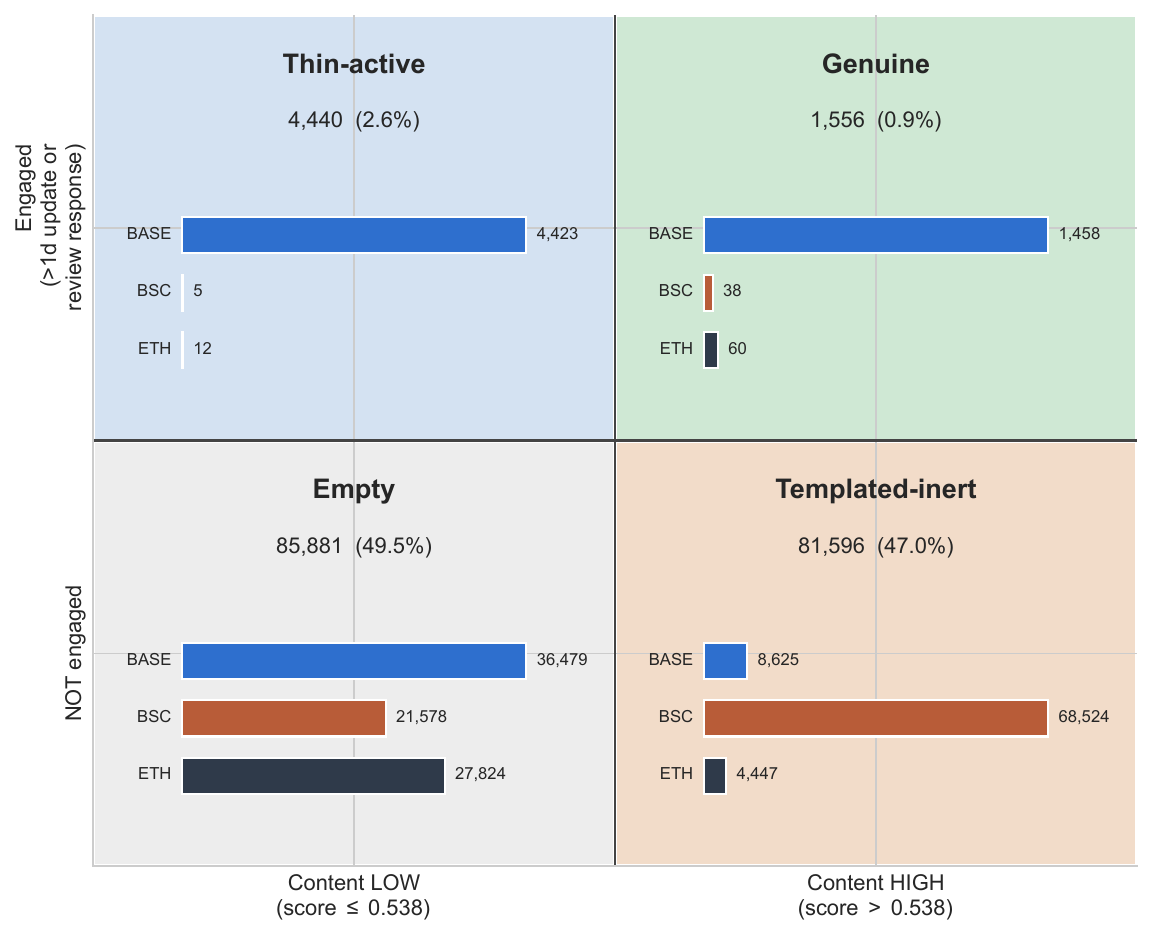}
\caption{Two-dimensional quality typology.}
\label{fig:qa-typology}
\end{minipage}
\hfill
\begin{minipage}[t]{0.48\textwidth}
\centering
\includegraphics[width=\linewidth]{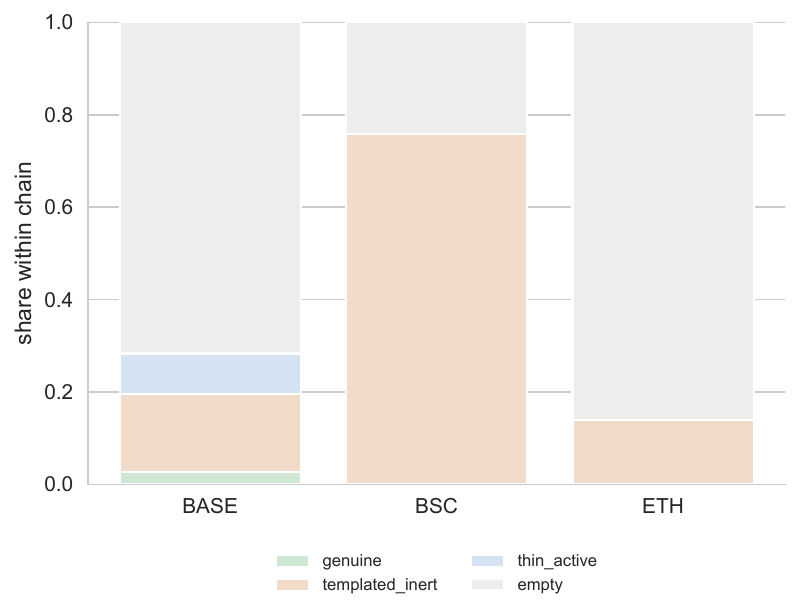}
\caption{Quality type composition by chain.}
\label{fig:qa-composition}
\end{minipage}
\end{figure}

\smallskip
\noindent\textbf{Quality typology.} 
To retain both dimensions and classify agents jointly along them, content is dichotomized at the pooled median score ($0.538$), and engagement is classified as engaged if an agent exhibits post-setup activity. Crossing the two binary dimensions yields four types (Figure~\ref{fig:qa-typology}).

\textit{Empty} agents form a large placeholder population, characterized by frequent unreachability and batch minting. \textit{Thin-active} agents exhibit post-registration activity despite poor registration quality, suggesting owner engagement without a functioning off-chain endpoint. \textit{Templated-inert} agents are content-complete and reachable but rarely revisited after registration, consistent with large-scale templated deployments. In contrast, \textit{genuine} agents constitute the engaged core of the ecosystem, exhibiting richer declared functionality and little evidence of batch minting.

These types also exhibit sharply different cross-chain distributions (Figure~\ref{fig:qa-composition}). Overall, registrations are dominated by the \textit{empty} ($49.5\%$) and \textit{templated-inert} ($47.0\%$) types, while \textit{genuine} agents remain rare ($0.9\%$). Base hosts nearly all \textit{genuine} agents and most \textit{thin-active} agents, whereas BSC is dominated by \textit{templated-inert} registrations and Ethereum by \textit{empty} registrations. Together, these patterns suggest that Base contains most of the ecosystem's active agents, while BSC and Ethereum are largely characterized by inactive registrations.

\section{Cross-Chain Reputation Analysis}
\label{app:rep:cross_chain}

We examine agents that declare multi-chain deployment in their off-chain registration files, and ask whether reputation is \emph{portable} across chains, i.e.,  whether an agent's standing on one chain is reflected in its standing on another; or whether each chain constitutes an isolated reputation silo.

\begin{table}[htb]
\centering
\caption{Cross-chain registration combinations declared in registration files, excluding testnet registrations and invalid \texttt{agentID}s.
Of $173{,}473$ registered agents, $629$ ($0.4$\%) declare registration on two or more mainnets.}
\label{tab:crosschain_combos}
\resizebox{\textwidth}{!}{%
\begin{tabular}{cc cc cc cc cc cc cc cc cc}
\toprule
\multicolumn{2}{c}{BSC+Base}
 & \multicolumn{2}{c}{BSC+Base+ETH}
 & \multicolumn{2}{c}{Base+ETH}
 & \multicolumn{2}{c}{Base+MegaETH}
 & \multicolumn{2}{c}{Base+ETH+MegaETH}
 & \multicolumn{2}{c}{ARB+BSC+Base+ETH}
 & \multicolumn{2}{c}{Billions+ETH}
 & \multicolumn{2}{c}{$\geq$10 chains$^\dagger$}
 & \multicolumn{2}{c}{Other} \\
\cmidrule(lr){3-4}
\cmidrule(lr){7-8}
\cmidrule(lr){9-10}
\cmidrule(lr){11-12}
\cmidrule(lr){15-16}
$n$ & \%
 & $n$ & \%
 & $n$ & \%
 & $n$ & \%
 & $n$ & \%
 & $n$ & \%
 & $n$ & \%
 & $n$ & \%
 & $n$ & \% \\
\cmidrule(lr){1-18}
 266 & (42.3~\mypie{combocolor}{42.3})
 & 197 &  (31.3~\mypie{combocolor}{31.3}) 
 &  59 &  (9.4~\mypie{combocolor}{9.4})
 &  21 &  (3.3~\mypie{combocolor}{3.3})
 &   8 &  (1.3~\mypie{combocolor}{1.3})
 &   7 &  (1.1~\mypie{combocolor}{1.1})
 &   6 &  (1.0~\mypie{combocolor}{1.0})
 &   6 &  (1.0~\mypie{combocolor}{1.0})
 &  59 &  (9.4~\mypie{combocolor}{9.4})  \\
\bottomrule
\end{tabular}}

\smallskip
\footnotesize
$^\dagger$Two agent groups declared $10$--$22$ mainnets each, likely via templated batch registration scripts.
\end{table}

\noindent\textbf{Cross-chain registration.}
Table~\ref{tab:crosschain_combos} reports the declared chain combinations of the $629$ agents deployed on at least two mainnets. The vast majority concentrate on the three chains in our study: BSC$+$Base and BSC$+$Base$+$ETH account for $73.6\%$ of all multi-chain declarations.

\begin{figure}[htb]
    \centering
    \begin{subfigure}{0.48\linewidth}
        \centering
        \includegraphics[width=\linewidth]{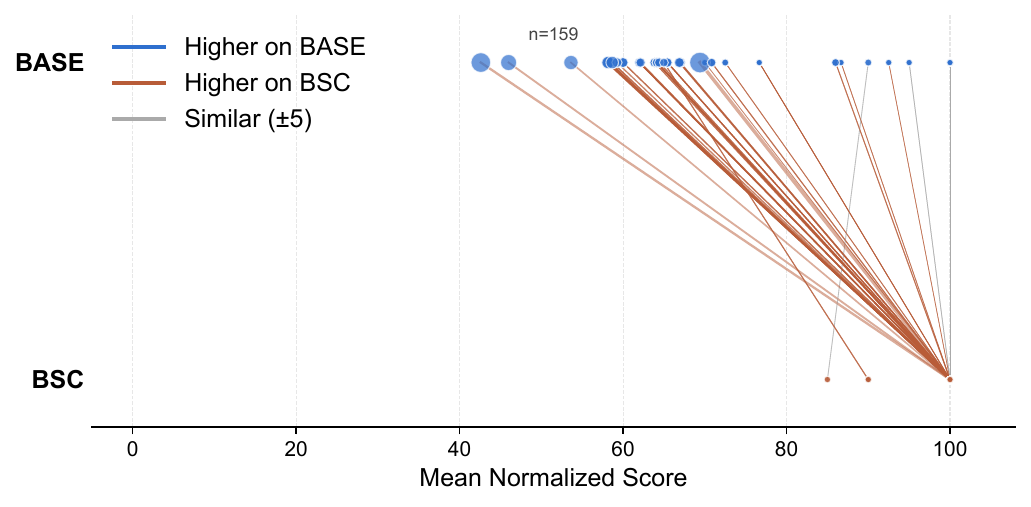}
        \caption{BSC--Base}
        \label{fig:crosschain_BSC_BASE.pdf}
    \end{subfigure}
    \hfill
    \begin{subfigure}{0.48\linewidth}
        \centering
        \includegraphics[width=\linewidth]{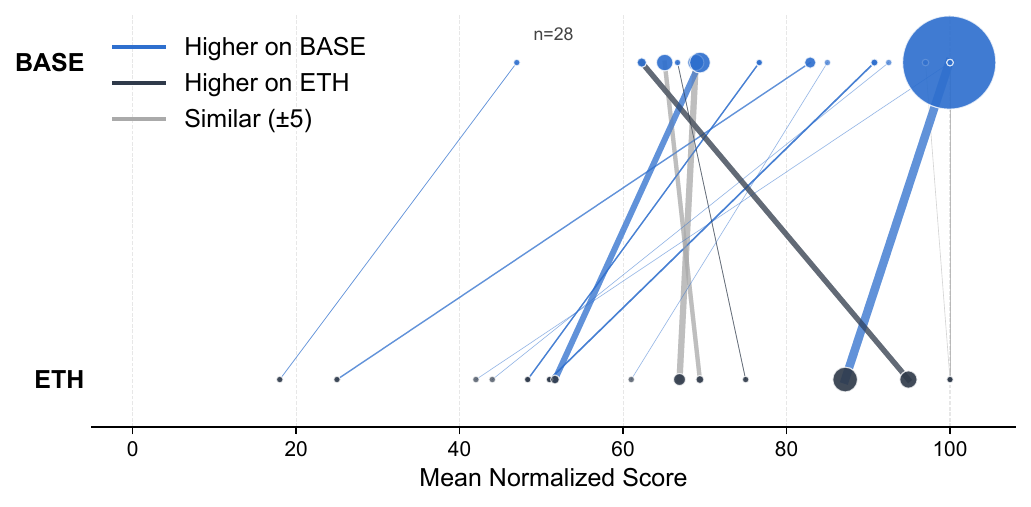}
        \caption{ETH--Base}
        \label{fig:crosschain_ETH_BASE}
    \end{subfigure}
   \caption{Mean normalized reputation scores for agents declaring registration on both BSC and Base and on both ETH and Base, restricted to agents with at least one feedback record on each chain. Line color indicates which chain carries the higher score. Bubble size reflects the number of feedback records on that chain.
   }
    \label{fig:cross_chain_score}
\end{figure}

\noindent\textbf{Cross-chain reputation consistency.}
Figure~\ref{fig:cross_chain_score} compares the mean normalized reputation scores of agents registered on both chains with at least one feedback record on each. We analyze the BSC--Base ($n=159$) and ETH--Base ($n=28$) pairs; the ETH--BSC pair is omitted because only two agents overlap.
For BSC--Base, nearly all agents cluster at the top of the BSC score range (90--100), reflecting score inflation, whereas the same agents span 40--100 on Base. The two score distributions are uncorrelated (Spearman
$\rho=0.05$, $p=0.56$).
For ETH--Base, both chains exhibit broader score ranges, yet scores again show
no significant correlation ($\rho=0.14$, $p=0.48$).
These results indicate that ERC-8004 reputations are chain-specific and do not transfer across deployments.

This divergence is consistent with ERC-8004’s per-chain design: feedback is recorded independently, while cross-chain identity linking and reputation aggregation require additional infrastructure.   
Addressing this fragmentation motivates emerging efforts such as Trustless Agents Plus\footnote{See \url{https://www.decipherclub.com/trustless-agents-plus/}.}, which binds per-chain identities and aggregates their reputation to make an agent’s track record usable across chains. Our results highlight the practical relevance of this gap.

\smallskip
\noindent\textbf{Future development}. Today, reputation is chain-specific: scores do not transfer across chains. Cross-chain portability is a desirable long-term goal, but it should follow, not precede, stronger per-chain guarantees. Before reputation is shared across chains, the protocol needs \textit{(i)} a verifiable cross-chain identity binding and \textit{(ii)} robust per-chain reputation integrity. Otherwise, portability would merely propagate manipulation from the weakest chain to others.

\section{Why Honest Volume Does Not Dilute a Single Crafted Feedback Record}
\label{app:dilution}

The intuition that a large number of honest ratings should ``average away'' a single adversarial input is correct for a bounded score, but fails for the ERC-8004 mean score because the value field is almost unbounded. This appendix makes that precise, derives the crafted value of Equation~\ref{eq:onefeedback}, and shows why the one-feedback attack remains admissible at every volume a real agent can attain.

\smallskip
\noindent\textbf{The crafted value (\S\ref{sec:rep:robustness})}.
An agent $a$ holds $n$ honest feedback records with mean $m$, so its score is
$S(a)=m$. The adversary appends one record of normalized value $\tilde v^{\ast}$.
Because the Registry aggregates by the arithmetic mean (Equation~\ref{eq:score}),
the new score over $n+1$ records is
\begin{equation}
S'(a) \;=\; \frac{n\,m + \tilde v^{\ast}}{\,n+1\,}.
\end{equation}
Setting $S'(a)$ equal to any chosen target $\tau$ and solving for the single
value gives
\begin{equation}
\frac{n\,m + \tilde v^{\ast}}{n+1} = \tau
\;\;\Longrightarrow\;\;
\tilde v^{\ast} = (n+1)\,\tau - n\,m ,
\label{eq:app-onefeedback}
\end{equation}
which is exactly Equation~\ref{eq:onefeedback}. The target $\tau$ is free, so the same expression covers both directions: $\tau>m$ inflates the score and $\tau<m$ suppresses it. One record is always enough in principle; the only question is whether the required $\tilde v^{\ast}$ is admissible on-chain.

\smallskip
\noindent\textbf{Admissibility and the growth in $n$}.
The contract permits any value with $|v|\le 10^{38}$. For large $n$,
Equation~\ref{eq:app-onefeedback} simplifies to
\begin{equation}
\tilde v^{\ast} = (n+1)\tau - n m = n(\tau-m) + \tau \;\approx\; n\,(\tau-m).
\end{equation}
The magnitude of the crafted value grows \emph{linearly} in $n$, while the contract ceiling $10^{38}$ is \emph{fixed}. The one-feedback attack stops being admissible only when the required value exceeds that ceiling:
\begin{equation}
\bigl|\,n(\tau-m)\,\bigr| > 10^{38}
\;\;\Longrightarrow\;\;
n > \frac{10^{38}}{\,|\tau-m|\,}.
\label{eq:app-threshold}
\end{equation}

\smallskip
\noindent\textbf{Why the threshold is never reached}.
The bound in Equation~\ref{eq:app-threshold} depends on the gap $|\tau-m|$ the adversary wants to traverse. Across the full $0$--$100$ rating range this gap is at most $100$, so the threshold spans only a narrow band:
\begin{equation}
\text{gap }1:\; n>10^{38},
\qquad
\text{gap }10:\; n>10^{37},
\qquad
\text{gap }100:\; n>10^{36}.
\end{equation}
Taking the representative gap $|\tau-m|=10$ used in the main text gives $n>10^{38}/10 = 10^{37}$. In every case, the threshold lies between $10^{36}$ and
$10^{38}$.
This number is unreachable by orders of magnitude. The most-rated agent in our entire dataset carries $1{,}552$ feedback records, i.e., $n\approx10^{3}$, leaving
a gap of more than thirty orders of magnitude to $10^{37}$. For further scale, issuing one feedback every second record for the current age of the universe ($\approx 4\times10^{17}$ seconds) would produce only about $10^{17}$ records, still twenty orders of magnitude short. No agent will ever accumulate $10^{37}$ honest ratings, so the dilution effect is irrelevant and a single crafted feedback entry remains admissible.

\smallskip
\noindent\textbf{A worked suppression example}.
The most-rated agent above illustrates the point concretely. It holds
$n=1{,}552$ records at mean $m=99.9$. To suppress its score to $\tau=0$, the
adversary submits
\begin{equation}
\tilde v^{\ast} = (1{,}552+1)\times 0 \;-\; 1{,}552\times 99.9
\;\approx\; -1.55\times 10^{5}.
\end{equation}
This value satisfies $|\tilde v^{\ast}|\approx 10^{5} \ll 10^{38}$, so a single
record collapses an agent with over fifteen hundred honest ratings from near-perfect
to zero. The same construction with $\tau>m$ inflates an unknown agent to the top
of the scale at equally negligible cost.

\section{Supplementary Manipulative Behavior Analysis}
\label{app:sybil}

This appendix expands the manipulative-behavior analysis in
\S\ref{sec:rep:wild}. We take the provenance-linked reviewer groups defined in \S\ref{sec:rep:wild} as input and characterize how they write feedback.
We provide operational detail for repeated targeting, queue-sweep behavior, and template concentration.

\subsection{Repeated Targeting and Queue Sweeps}

For a provenance-linked group \(g\), let \(\mathcal{F}_g\) denote its feedback
set, \(N_g = |\mathcal{F}_g|\), and \(A_g\) the number of distinct rated agents.
We define fan-out as
\begin{equation}
\label{eq:fanout}
\mathrm{Fanout}(g)
=
\frac{A_g}{N_g}
\end{equation}

For each reviewer--agent pair \((r,a)\), let \(m_{r,a}\) denote the number of
feedback records submitted by reviewer \(r\) to agent \(a\), and define the observed pair set
\begin{equation}
\label{eq:pair-set}
P_g
=
\{(r,a): m_{r,a} > 0\}
\end{equation}

We define the repeated-feedback share as
\begin{equation}
\label{eq:rs}
\mathrm{RS}(g)
=
1-
\frac{|P_g|}{N_g}
\end{equation}
This quantity is zero when every feedback record belongs to a distinct reviewer--agent pair and increases as repeated ratings of the same targets account for a larger share of the group's activity.

We use \emph{queue-sweep} to denote broad traversal across targets: a group rates many distinct agents while contributing relatively few repeated records to
each reviewer--agent pair. Operationally, this corresponds to high
\(\mathrm{Fanout}(g)\) and low \(\mathrm{RS}(g)\). Conversely, low
\(\mathrm{Fanout}(g)\) and high \(\mathrm{RS}(g)\) indicate repeated targeting,
where a group concentrates influence on a small set of targets. We use these metrics jointly with score-value and tag-template concentration to characterize the groups below.

\subsection{Entropy-Based Template Concentration}

We measure template concentration using Shannon entropy over empirical
distributions of score values and tag pairs. For a provenance-linked group
\(g\), let \(\mathcal{F}_g\) be its non-revoked scored feedback set. For an
attribute \(X\), such as clipped score or the pair
\((\texttt{tag1},\texttt{tag2})\), define
\begin{equation}
\label{eq:entropy}
\begin{aligned}
p_g(x)
&=
\frac{
\left|
\{f\in\mathcal{F}_g : X(f)=x\}
\right|
}{
|\mathcal{F}_g|
}, \\
H_X(g)
&=
-\sum_x p_g(x)\log_2 p_g(x)
\end{aligned}
\end{equation}
We report the effective number of values or templates
\begin{equation}
\label{eq:effective-diversity}
D_X(g)
=
2^{H_X(g)}
\end{equation}
Lower \(D_X(g)\) indicates stronger reuse of score values or tag templates.
Score entropy captures literal score-value reuse within a group; it is not
used to assert semantic comparability across tags.

\subsection{Case Studies}

We present one reviewer group or operator-resolved group per chain to
illustrate how the measurements separate repeated targeting from queue-sweep
behavior.

\vspace{0.2em}
\noindent\textbf{Ethereum.} The EOA funder
{0xcc28...4821} links nine reviewers that write $90$ non-revoked scored feedback records over three agents. The group has low fanout and high repetition (\(\mathrm{Fanout}=0.033\), \(\mathrm{RS}=0.700\)), consistent with repeated targeting. Its score values are concentrated (\(D_{\mathrm{score}}=1.7\); score
100 accounts for $88.9\%$ of records), while tag-pair use is diffuse (\(D_{\mathrm{tag}}=17.9\)), suggesting that reinforcement is driven primarily by target reuse rather than a single tag template.

\vspace{0.2em}
\noindent\textbf{BSC.} The delegated-EOA funder
{0xaef8...d18d}
links $20$ reviewers that write $21{,}863$ non-revoked scored feedback records over $312$ agents. 
This large group exhibits repeated-targeting behavior:
\(\mathrm{Fanout}=0.014\) and \(\mathrm{RS}=0.873\). Scores are
concentrated (\(D_{\mathrm{score}}=3.6\)), with score $70$ accounting for $73.2\%$
of records. Tag-pair diversity is higher than in the Base sweep (\(D_{\mathrm{tag}}=6.0\)), but the group still most often reuses \texttt{personality|fragment} and emits $6{,}295$ records in its busiest
24-hour window.

\vspace{0.2em}
\noindent\textbf{Base.} The Base case shows a contract-resolved batch sweep. When a
reviewer's first funder is a smart contract, we resolve the funding transaction
to its operator EOA when trace data is available; this succeeds for $155$ of $157$ contract-funded reviewers. The largest resolved operator, {0x4055...5cc5}, funds $80$ reviewer wallets through the same contract, and these wallets write $800$ non-revoked scored feedback records. The footprint is highly regular: all $80$ wallets issue ten feedback transactions with score $100$ to ten agents, and $79$ have no transactions after the feedback block. The modal lifecycle/nonce template covers $56$ wallets: each has eight pre-feedback transactions, then a contiguous block of ten feedback nonces to ten agents, and then no post-feedback activity.
This repeated template is consistent with a clone-like queue sweep, where short-lived reviewer wallets attributable to the same operator execute the same fixed-length feedback routine.

\section{Reputation Score Manipulation: Damage Breakdown}
\label{app: sybil_breakdown}

\subsection{Breakdown by Agent Quality}
\label{app:sybil_breakdown_by_quality}
\begin{table}[H]
\centering
\scriptsize
\setlength{\tabcolsep}{2.6pt}
\caption{
Impact of removing Sybil-flagged feedback, broken down by recipient-agent quality.
Sybil Feedback (\%) is computed over non-revoked scored feedback records in each row; Affected agents (\%), No baseline cases (\%), and Mixed cases (\%) are over rated agents in the same row.
\#Affected agents=\#No baseline cases + \#Mixed cases. Median and mean score shifts ($\Delta$) are computed over Mixed cases.
}
\label{tab:sybil-counterfactual-damage-breakdown}
\resizebox{0.98\textwidth}{!}{%
\begin{tabular}{cl@{\hspace{6pt}}ccrrr@{\hspace{8pt}}@{\hspace{8pt}}ccc}
\toprule
Chain & \multicolumn{1}{c}{Agent quality} & \multicolumn{2}{c}{Sybil feedback} & Rated & Affected agents & No baseline cases & Mixed & Median $\Delta$ & Mean $\Delta$ \\

\cmidrule(lr){2-2}\cmidrule(lr){3-4}\cmidrule(lr){6-7}\cmidrule(lr){8-10}

ETH  & All & 41.4\% & \mypie{ethcolor}{41.4}  & 1,559  & 411 (26.4\%)     & 247 (15.8\%)                         & 164   & $+11.0$ & $+10.9$ \\
ETH  & Valid ERC-8004, with service      & 61.1\% & \mypie{ethcolor}{61.1}  & 253    & 205 (81.0\%)     & 61 (24.1\%)                          & 144   & $+15.9$ & $+14.8$ \\
ETH  & Valid ERC-8004, no service & 38.9\% & \mypie{ethcolor}{38.9}  & 375    & 141 (37.6\%)     & 128 (34.1\%)                         & 13    & $-33.5$ & $-24.2$ \\
ETH  & File retrieved but not compliant & 15.7\% & \mypie{ethcolor}{15.7}  & 54     & 8 (14.8\%)       & 5 (9.3\%)                            & 3     & $-9.3$  & $-11.2$ \\
ETH  & URI unresolvable  & 27.5\% & \mypie{ethcolor}{27.5}  & 421    & 17 (4.0\%)       & 15 (3.6\%)                           & 2     & $+4.2$  & $+4.2$ \\
ETH  & No URI            & 10.1\% & \mypie{ethcolor}{10.1}  & 456    & 40 (8.8\%)       & 38 (8.3\%)                           & 2     & $-4.2$  & $-4.2$ \\

\cmidrule(lr){2-2}\cmidrule(lr){3-4}\cmidrule(lr){6-7}\cmidrule(lr){8-10}

BSC  & All               & 96.3\% & \mypie{bsccolor}{96.3}  & 4,310  & 3,510 (81.4\%)   & 3,356 (\textcolor{bsccolor}{77.9\%}) & 154   & $-9.1$  & $-11.5$ \\
BSC  & Valid ERC-8004, with service     & 99.3\% & \mypie{bsccolor}{99.3}  & 519    & 470 (90.6\%)     & 447 (86.1\%)                         & 23    & $-7.1$  & $-3.3$ \\
BSC  &  Valid ERC-8004, no service & 78.5\% & \mypie{bsccolor}{78.5}  & 2,896  & 2,347 (81.0\%)   & 2,229 (77.0\%)                       & 118   & $-10.0$ & $-12.6$ \\
BSC  & File retrieved but not compliant & 92.5\% & \mypie{bsccolor}{92.5}  & 153    & 141 (92.2\%)     & 141 (92.2\%)                         & 0     & --      & -- \\
BSC  & URI unresolvable  & 64.1\% & \mypie{bsccolor}{64.1}  & 373    & 248 (66.5\%)     & 244 (65.4\%)                         & 4     & $-27.5$ & $-21.2$ \\
BSC  & No URI            & 81.3\% & \mypie{bsccolor}{81.3}  & 369    & 304 (82.4\%)     & 295 (79.9\%)                         & 9     & $-12.5$ & $-14.4$ \\

\cmidrule(lr){2-2}\cmidrule(lr){3-4}\cmidrule(lr){6-7}\cmidrule(lr){8-10}

Base & All               & 92.6\% & \mypie{basecolor}{92.6} & 28,592 & 27,499 (96.2\%)  & 24,822 (\textcolor{basecolor}{86.8\%})& 2,677 & $-0.2$  & $-5.2$ \\
Base & Valid ERC-8004, with service      & 89.2\% & \mypie{basecolor}{89.2} & 4,495  & 4,245 (94.4\%)   & 3,528 (78.5\%)                       & 717   & $+0.0$  & $-0.0$ \\
Base &  Valid ERC-8004, no service & 92.9\% & \mypie{basecolor}{92.9} & 3,458  & 3,255 (94.1\%)   & 2,780 (80.4\%)                       & 475   & $-14.6$ & $-14.3$ \\
Base & File retrieved but not compliant & 91.6\% & \mypie{basecolor}{91.6} & 4,997  & 4,698 (94.0\%)   & 4,345 (87.0\%)                       & 353   & $+0.0$  & $-5.1$ \\
Base & URI unresolvable  & 93.5\% & \mypie{basecolor}{93.5} & 3,842  & 3,683 (95.9\%)   & 3,268 (85.1\%)                       & 415   & $+0.0$  & $-5.6$ \\
Base & No URI            & 94.6\% & \mypie{basecolor}{94.6} & 11,800 & 11,618 (98.5\%)  & 10,901 (92.4\%)                      & 717   & $-0.4$  & $-4.2$ \\
\bottomrule
\end{tabular}
}
\end{table}

\begin{figure}[htb]
    \centering
    \includegraphics[width=0.92\linewidth]{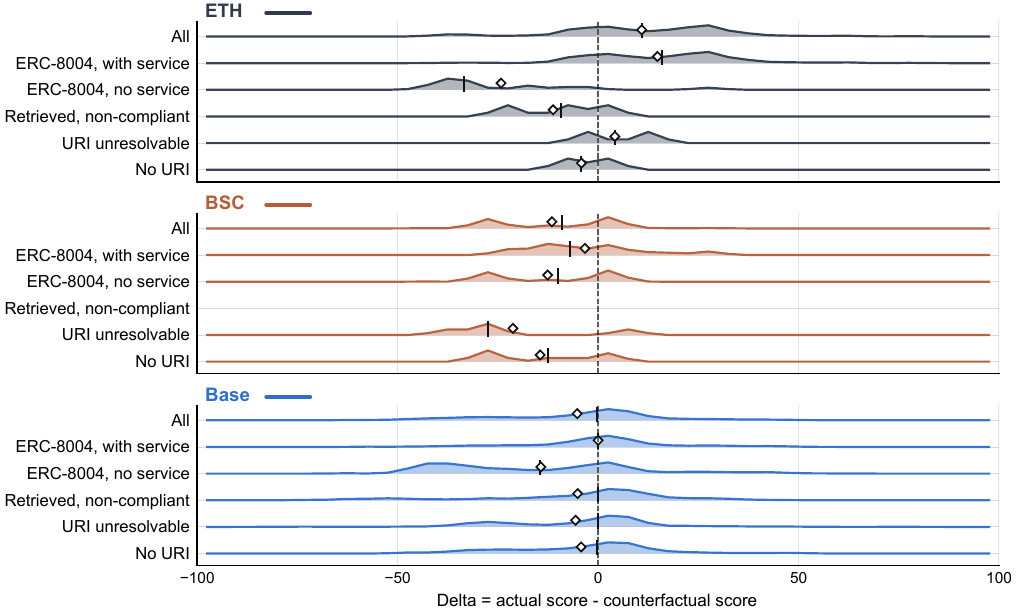}
    \caption{
    Reputation score change after removing Sybil-flagged feedback, by recipient-agent quality (mixed-case agents only). $\Delta = S_a - S_a^{(-\mathrm{Sybil})}$. Bars mark medians, diamonds mark means.
    }
    \label{fig:sybil-shift-distributions}
\end{figure}

\S\ref{sec:rep:wild} quantifies the aggregate effect of removing Sybil-flagged feedback. Here we break down that market damage by the quality of the \emph{recipient} agent, using the five-category taxonomy of \S\ref{sec: service_and_quality} (\emph{No URI}, \emph{URI unresolvable}, \emph{file retrieved but non-compliant}, \emph{Valid ERC-8004 agent but no service}, and \emph{Valid ERC-8004 agent with service}). 
For each category, we recompute agents' reputation scores after removing all Sybil-flagged feedback, and report the same quantities as in the main text: the share of Sybil feedback, the share of \emph{affected} agents, the share left with \emph{no baseline}, the number of \emph{mixed} cases, and the median and mean score shift $\Delta_a$ over mixed cases
(Table~\ref{tab:sybil-counterfactual-damage-breakdown}, Figure~\ref{fig:sybil-shift-distributions}).

\noindent\textbf{On Ethereum, inflation concentrates on service-bearing agents.}
Reputation manipulation appears to be selective: the agents that appear most trustworthy also experience the greatest reputation inflation. 
Among \emph{Valid ERC-8004 with service} agents, $81\%$ are affected by Sybil-flagged feedback, the highest proportion of any agent category. For those in mixed cases, Sybil attacks bring a significant score inflation: removing Sybil-flagged feedback decreases the median reputation score by $15.9$ points. Rather than being spread uniformly across the ecosystem, Sybil manipulation is concentrated on the agents most likely to be chosen by relying parties, suggesting that attackers strategically inflate the reputation of high-value service-bearing agents.

\vspace{0.2em}
\noindent\textbf{On Base, the manipulation is indiscriminate.}
In contrast, Base exhibits little evidence of selective targeting. The share of affected agents remains consistently high across all agent quality categories, ranging from $94.0\%$ for service-bearing agents to $98.5\%$ for \emph{No URI} agents. Likewise, the proportion of Sybil-flagged feedback varies little with recipient quality, staying between approximately $89\%$ and $95\%$ across all categories. Rather than concentrating on the most credible agents, Sybil activity appears to blanket the entire registered agent population, consistent with the large-scale automated seeding and queue-sweep behavior described in \S\ref{sec:rep:wild} and Appendix~\ref{app:sybil}.

\vspace{0.2em}
\noindent\textbf{On BSC, contamination is pervasive but mildly selective.}
BSC is the most heavily contaminated chain: $96.3\%$ of its feedback is Sybil-flagged, and the proportion of affected agents remains relatively high across quality categories, ranging from roughly $66.5\%$ to $92.2\%$. Even under such pervasive contamination, some degree of selectivity remains. Service-bearing agents are more likely to have no baseline ($86.1\%$) than agents without services ($77.0\%$), echoing the Ethereum pattern but much more weakly. Overall, BSC represents an intermediate pattern, lying between Ethereum's targeted reputation inflation and Base's indiscriminate Sybil seeding.

\subsection{Breakdown by Interaction Evidence}

Figure~\ref{fig:provenance} groups feedback records according to the strongest interaction evidence declared in their off-chain documents: a payment proof (\emph{has payment proof}), a task linkage to an A2A or MCP call (\emph{has task linkage}), or no verifiable interaction artifact (\emph{no interaction evidence}).

This naturally raises the question: are records that claim payment proof or task linkage less susceptible to Sybil manipulation than those with no evidence of interaction?
Table~\ref{tab:feedback-evidence-breakdown} examines this relationship by cross-tabulating each evidence category with its Sybil contamination rate.

\begin{table}[htb]
\centering
\small
\setlength{\tabcolsep}{2pt}
\caption{
Sybil contamination rate by feedback evidence class and chain.
}
\label{tab:feedback-evidence-breakdown}
\begin{tabular}{c crccr}
\toprule
Chain & Feedback quality & \#Feedback & \multicolumn{2}{c}{Sybil fb rate} & \# Sybil fb \\

\cmidrule(lr){2-3}\cmidrule(lr){4-5}

ETH  & Payment proof           & 2 (0.07\%)       & 100.0\% & \mypie{ethcolor}{100.0}  & 2  \\
ETH  & Task linkage            & 38 (1.2\%)       & 100.0\% & \mypie{ethcolor}{100.0}  & 38  \\
ETH  & No interaction evidence & 3,018 (98.7\%)   & 40.6\%  & \mypie{ethcolor}{40.6}   & 1,225  \\

\cmidrule(lr){2-3}\cmidrule(lr){4-5}

BSC  & Payment proof           & 0 (0.0\%)        & 0.0\%   & \mypie{bsccolor}{0.0}    & 0  \\
BSC  & Task linkage            & 0 (0.0\%)        & 0.0\%   & \mypie{bsccolor}{0.0}    & 0  \\
BSC  & No interaction evidence & 29,444 (100.0\%) & 96.3\%  & \mypie{bsccolor}{96.3}   & 28,367  \\

\cmidrule(lr){2-3}\cmidrule(lr){4-5}

Base & Payment proof           & 770 (0.6\%)      & 92.6\%  & \mypie{basecolor}{92.6}  & 713  \\
Base & Task linkage            & 43 (0.04\%)      & 100.0\% & \mypie{basecolor}{100.0} & 43  \\
Base & No interaction evidence & 121,970 (99.3\%) & 92.6\%  & \mypie{basecolor}{92.6}  & 112,921 \\
\bottomrule
\end{tabular}
\end{table}

The results are striking: Sybil feedback is not concentrated in the low-evidence category but pervades all three evidence classes.
On BASE, $92.6\%$ of feedback records carrying a self-declared payment proof are flagged as Sybil, a rate nearly identical to the $92.6\%$ observed among records with no interaction evidence at all.
On ETH, every record with a payment proof or task linkage is Sybil-flagged.
BSC, where all feedback falls into the no-evidence category, shows a $96.3\%$ Sybil rate.

These results expose a fundamental weakness of self-declared provenance: the \texttt{proofOfPayment} field is an \textbf{unverified} off-chain claim that any reviewer can populate with arbitrary data. Consequently, claiming a payment proof provides no meaningful indication of feedback genuineness. Effective verification would instead require the protocol to verify that a declared proof corresponds to an on-chain payment to the rated agent's registered wallet before the feedback was submitted.

\section{Supplementary Details for Agent Functional Profile Analysis}
\label{app:agent_function}

\subsection{Analysis of Agent Functional Profile by Agent Description}
\label{app:agent_function_1}
This section provides methodological details for our analysis of agent functional profile in \S\ref{sec:identity_function}.

\vspace{0.1cm}
\noindent\textbf{Background.}
As discussed in \S\ref{sec:identity_function}, characterizing what agents do downstream ideally requires observing the tasks they perform for users. Such data are largely unavailable: execution occurs off-chain, while on-chain identifiers (e.g., job IDs) reveal little about task content.

We therefore use the free-text \texttt{description} in ERC-8004 registration files as a \textbf{proxy} for intended tasks, asking \emph{what registered agents claim to do}, rather than verifying what they actually do. We group similar descriptions, manually label the groups, and organize them into a two-level functional taxonomy. Results are reported both per agent and per distinct description to distinguish registration volume from descriptive diversity, since a few templates are reused by tens of thousands of agents.

\subsubsection{Data}
We use the ERC-8004 identity registrations on Ethereum, BSC, and Base.
For each agent, we collect their \texttt{name} and the \texttt{description} metadata using the off-chain registration file. There are $173{,}473$ registered agents: $32{,}343$ on ETH, $90{,}145$ on BSC, and $50{,}985$ on BASE.

\vspace{0.1cm}
\noindent\textbf{Quality rules.}
Many agents do not have a description we can analyze. We apply a fixed set of rules, in order, and keep the rest as \emph{usable}. \textit{(i) Registration file unavailable:} the file could not be retrieved or parsed; \textit{(ii) Empty description:} file available but the \texttt{description} filed is empty; \textit{(iii) Gibberish:} text with few letters, unusual vowel ratios, long runs of consonants, or one word repeated (e.g., long digit strings); \textit{(iv) Test or placeholder text:} short texts (at most 12 words) containing words such as \emph{test}, \emph{demo}, \emph{placeholder}, or \emph{hello world}. \textit{(v) Too short:} fewer than 5 words and fewer than 25 characters, unless the text contains a clear function word (e.g., ``An AI agent on Ethereum'').

Overall, $98{,}702$ agents ($56.9\%$) have a usable description. The share differs a lot by chain: $25.3\%$ on ETH, $76.6\%$ on BSC, and $42.1\%$ on BASE.
On ETH and BASE, the main loss is registration files that cannot be retrieved.
All analyses below use the $98{,}702$ agents with usable descriptions.

\vspace{0.1cm}
\noindent\textbf{Text cleaning.}
Before analysis, we remove URLs, wallet addresses, emoji, and markdown. We replace the agent's own name with ``the agent,'' so that two agents with the same text but different names count as the same description. We also remove two phrases that carry no information about function: mentions of ``ERC-8004'' and the stock phrase ``as on-chain attestations''.

\vspace{0.1cm}
\noindent\textbf{Languages.}
Using \texttt{langdetect} (accept a label only when its probability is at least 0.80),
$89.8\%$ of distinct descriptions are English and a further $4.5\%$ are very likely English.
Only $1.3\%$ are clearly in another language, mostly Chinese.
We keep all of them and use a multilingual embedding model.

\subsubsection{Method}
Figure~\ref{fig:pipeline} summarizes the analysis pipeline.

\begin{figure}[htb]
\centering
\begin{tikzpicture}[
  box/.style={draw, rounded corners=2pt, align=center,
              font=\footnotesize, minimum height=9mm,
              text width=2.55cm, inner sep=3pt, fill=white},
  arr/.style={-{Stealth[length=5pt]}, thick}]
\node[box] (a) at (0,0)    {Registered agents\\173{,}473};
\node[box] (b) at (3.2,0)  {Quality rules\\98{,}702 usable};
\node[box] (c) at (6.4,0)  {Text merging\\13{,}983 groups};
\node[box] (d) at (9.6,0)  {Embedding\\UMAP, HDBSCAN};
\node[box] (e) at (9.6,-1.3) {Labeling clusters\\ and templates};
\node[box] (f) at (6.4,-1.3) {Long-tail classifier\\($p \ge 0.7$)};
\node[box] (g) at (3.2,-1.3) {Validation\\Manual audit};
\node[box, fill=black!4] (h) at (0,-1.3) {Functional\\taxonomy};
\draw[arr] (a) -- (b); \draw[arr] (b) -- (c); \draw[arr] (c) -- (d); \draw[arr] (d) -- (e);
\draw[arr] (e) -- (f); \draw[arr] (f) -- (g); \draw[arr] (g) -- (h);
\end{tikzpicture}
\caption{Pipeline overview. Numbers are counts after each step. ``Text merging'' merges exact and near-duplicate descriptions; labeling covers clusters, top templates, and keyword candidates.}
\label{fig:pipeline}
\end{figure}

\noindent\textbf{From agents to description groups.}
We first merge duplicated descriptions. Many agents share exactly the same (cleaned) text.
The $98{,}702$ usable descriptions contain only $14{,}530$ distinct texts, a reduction of $85.3\%$.
Reuse is strongest on BSC, where each distinct text is used by $41.4$ agents on average, compared with $2.2$ on ETH and $2.3$ on BASE.
The largest template (``AI-driven multi-chain trading agent with on-chain reputation'') alone covers $49{,}187$ agents, almost all on BSC.

We further merge highly similar descriptions. Templates are often reused with small edits (e.g., a swapped word).
We embed each distinct text and merge texts whose cosine similarity is at least $0.95$.
This merges $547$ texts and leaves $13{,}983$ \emph{description groups}.
A description group is our unit of clustering.
When we report results ``by distinct description'', we count description groups.

\vspace{0.1cm}
\noindent\textbf{Embeddings.} We then convert description groups into embeddings. We use \texttt{BAAI/bge-m3}, a multilingual sentence-embedding model ($1{,}024$ dimensions, up to $8{,}192$ tokens). The median description has $31$ tokens and the 95th percentile has 133, so no text is truncated.

\vspace{0.1cm}
\noindent\textbf{Clustering.}
We reduce the embeddings to $5$ dimensions with UMAP ($15$ neighbors, cosine distance, minimum distance 0)
and cluster them with HDBSCAN (minimum cluster size $35$, minimum samples $5$, excess-of-mass selection).
Each description group counts once, regardless of how many agents use it, so large templates cannot dominate the clusters.
HDBSCAN finds $58$ clusters and leaves $27.0\%$ unassigned (``noise'').
Cluster sizes range from $35$ to $1{,}914$ description groups (median $93$).

We then assign a noise point to a cluster only when the evidence is strong: at least $60\%$ of its 15 nearest clustered neighbors belong to the same cluster, and their mean cosine similarity is at least $0.75$. This assigns $259$ more groups and leaves $25.2\%$ of description groups unclustered.
For each cluster, we extract keywords with class-based TF-IDF (via \texttt{BERTopic}), after removing words that are common to all agent descriptions (e.g., \emph{agent}, \emph{AI}, \emph{on-chain}, and chain names).

\vspace{0.1cm}
\noindent\textbf{Labeling and taxonomy.}
Labels describe what an agent \emph{claims to do}, following five rules: \textit{(i)} label the stated function, not the text's tone or style; \textit{(ii)} label the service an agent offers, treating payment support (e.g., accepting x402 payments) as a billing feature rather than its function; \textit{(iii)} when several functions are listed, select the one introduced first or emphasized most; \textit{(iv)} assign profiles or personas without a concrete service to \emph{Persona \& Social}, and agent descriptions without a clear function to \emph{Other/Unclear}; and \textit{(v)} assign text unrelated to an agent to \emph{Non-Agent Content}.

For each cluster, we reviewed its keywords, the five descriptions closest to its center, and three random members, and assigned a category and a subcategory. The label then applies to every description group in the cluster.
We also deal with large templates. Because a few templates cover most agents, we reviewed the $50$ largest description groups. Together they cover $77.3\%$ of agents with a usable description. For $19$ of them, the cluster label did not fit the text (or the template was unclustered), and we gave the template its own label. The other $31$ kept their cluster label.

Grouping the labels gives $12$ categories and $51$ subcategories.
Nine categories describe a concrete function. The remaining three (\emph{Persona \& Social}, \emph{Other/Unclear}, \emph{Non-Agent Content}) do not state a concrete service; we keep them as they are a large and informative part of the Registry.

\vspace{0.1cm}
\noindent\textbf{Label propagation for the long tail.}
After the steps above, $3{,}458$ description groups ($5{,}599$ agents) remain unlabeled.
To label the clearer ones, we train a classifier on the description groups that already have a label
($10{,}525$ groups, $12$ categories): a logistic regression with balanced class weights on the description embeddings.
Following rule~(ii), we first remove sentences and phrases that only describe billing (e.g., ``Pay-per-call via x402, USDC on Base'') and re-embed the affected texts.

In 5-fold cross-validation the classifier reaches an accuracy of $0.841$ (macro-F1 $0.791$).
We keep only predictions with probability at least $0.7$, where the classifier labels $53.1\%$ of cases with an accuracy of $0.980$.
The classifier labels $457$ description groups ($800$ agents).
The final long tail is $3{,}001$ description groups ($4{,}799$ agents, $4.9\%$ of agents with a usable description).

\subsubsection{Robustness}
\begin{wraptable}{r}{0.54\textwidth}
\vspace{-\baselineskip}
\centering
\captionsetup{font=small, skip=4pt}
\caption{Stability of clusters and categories. ``Core'' means description groups that are clustered in both runs.}
\label{tab:stability}
\footnotesize
\setlength{\tabcolsep}{3pt}
\begin{tabular*}{\linewidth}{@{\extracolsep{\fill}}ccccc@{}}
\toprule
\multirow{2}{*}{Comparison} & \multicolumn{2}{c}{Clusters (core)} & \multicolumn{2}{c}{Categories (core)} \\
\cmidrule(lr){2-3}\cmidrule(lr){4-5}
 & ARI & NMI & Agreement & ARI \\

\cmidrule(lr){1-1}\cmidrule(lr){2-5}

Seed 1 & 0.986 & 0.978 & 98.9\% & 0.985 \\
Seed 2 & 0.951 & 0.953 & 94.3\% & 0.935 \\
Seed 3 & 0.974 & 0.969 & 97.9\% & 0.967 \\
Seed 4 & 0.966 & 0.964 & 95.8\% & 0.937 \\

\cmidrule(lr){2-3}\cmidrule(lr){4-5}

k-means ($k=58$) & 0.304$^\dagger$ & 0.708$^\dagger$ & 80.5\% & 0.722 \\
\bottomrule
\end{tabular*}
\end{wraptable}

We assess methodological robustness in several aspects.

\vspace{0.1cm}
\noindent\textbf{Random seeds.}
We re-ran UMAP and HDBSCAN with four other random seeds.
On description groups that are clustered in both runs, the cluster assignments agree closely
(adjusted Rand index, ARI: mean $0.969$, range $0.951$--$0.986$; NMI: mean $0.966$).
Including unclustered points, ARI is $0.705$ on average; most of this difference is from points moving in or out of noise.

\vspace{0.1cm}
\noindent\textbf{Category level.}
The taxonomy is what we report, so we also check stability at the category level (see Table~\ref{tab:stability}).
For each alternative clustering, we give every cluster the majority category of its members and compare with our labels.
On commonly clustered description groups, categories agree for $94.3$--$98.9\%$ of groups across seeds (ARI $0.935$--$0.985$).
Even with a different algorithm (k-means with $k=58$), they agree for $80.5\%$ of groups (ARI $0.722$).

\vspace{0.1cm}
\noindent\textbf{Manual audit of label propagation for long-tailed description groups.}
To test the robustness of label propagation for long-tailed description groups, we drew a random sample of $150$ classifier-labeled description groups (drawn at a lower threshold of $0.6$, before fixing the final threshold).
Two annotators labeled each one independently using the guidelines above.
Agreement was substantial (Cohen's $\kappa = 0.66$), and the author resolved all disagreements.
At the $0.6$ threshold, $78.7\%$ of labels were correct ($95\%$ CI $71.4$-$84.5\%$).
Accuracy was clearly lower for predictions with probability up to $0.7$ ($71.2$\%, $n=59$) than above it ($83.5\%$, $n=91$),
and every \emph{Payments \& Commerce} prediction was wrong ($0$ of $7$).
We therefore use a threshold of $0.7$ and do not propagate \emph{Payments \& Commerce}.
The $88$ audited items that fall in this final setting are a random sample of the final classifier labels;
$86.4\%$ of them are correct ($95\%$ CI $77.7$--$92.0\%$). 
Because the classifier only propagates labels to $0.8\%$ of agents, these errors have little effect on the results above.

\begin{figure}[htb]
    \centering
    \begin{minipage}[t]{0.325\linewidth}
    \centering
    \captionsetup{font=scriptsize,justification=raggedright,singlelinecheck=false}
    \includegraphics[width=\linewidth]{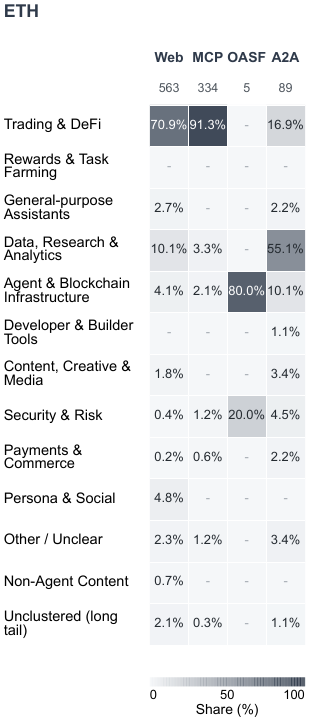}
    \caption{ETH service topics.}
    \label{fig:service_topics_ETH}
    \Description{An ETH service heatmap across Web, MCP, OASF, and A2A, using twelve functional categories and an unclustered long-tail row.}
    \end{minipage}\hfill%
    \begin{minipage}[t]{0.325\linewidth}
    \centering
    \captionsetup{font=scriptsize,justification=raggedright,singlelinecheck=false}
    \includegraphics[width=\linewidth]{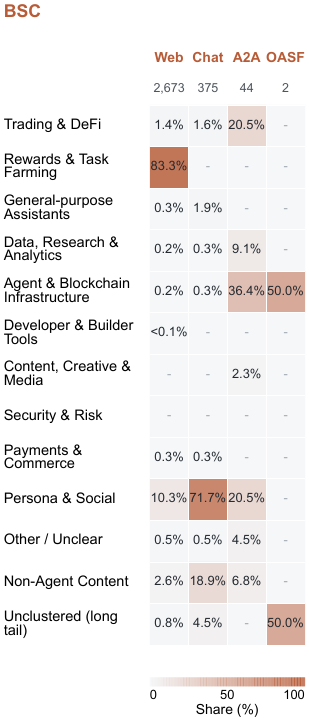}
    \caption{BSC service topics.}
    \label{fig:service_topics_BSC}
    \Description{A BSC service heatmap across Web, Chat, A2A, and OASF, using twelve functional categories and an unclustered long-tail row.}
    \end{minipage}\hfill%
    \begin{minipage}[t]{0.325\linewidth}
    \centering
    \captionsetup{font=scriptsize,justification=raggedright,singlelinecheck=false}
    \includegraphics[width=\linewidth]{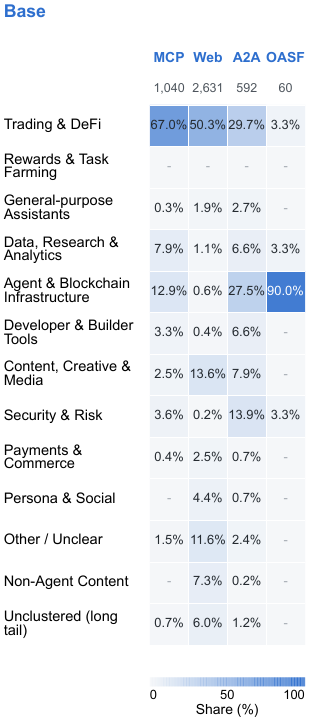}
    \caption{Base service topics.}
    \label{fig:service_topics_BASE}
    \Description{A Base service heatmap across MCP, Web, A2A, and OASF, using twelve functional categories and an unclustered long-tail row.}
    \end{minipage}
\end{figure}

\subsection{Analysis of Agent Functional Profile by Declared Service}
\label{app:agent_function_2}

The registration file's \texttt{description} field is not the only source of information about an agent's functions. Declared service endpoints can also provide more specific descriptions of the capabilities they expose.
While \S\ref{sec: service_and_quality} characterizes declared service types, a declaration alone does not explain the service's business function. We therefore collect service-level text from registration metadata, A2A agent cards and skills, MCP tool catalogs, and service webpages, supplementing platform context with official documentation where needed. We restrict the analysis to the original registration cohort and each chain's four most-declared service types, using service-specific text.

Following the functional taxonomy and labeling principles in Appendix~\ref{app:agent_function_1}, we analyze the collected service-level descriptions through topic grouping, label propagation, and source-based review. We assess the labels against the original service descriptions and supporting evidence.

Figures~\ref{fig:service_topics_ETH}--\ref{fig:service_topics_BASE} present the resulting breakdown across ETH, BSC, and Base using the same category order. Percentages are computed over all description-bearing agent--endpoint records within each service column, rather than all registered agents.
On Ethereum, trading-related functions dominate the retrieved Web and MCP records, accounting for $70.9\%$ and $91.3\%$ respectively, while A2A records skew toward data, research, and analytics ($55.1\%$). On BSC, $83.3\%$ of Web records describe rewards and task farming, showing a different emphasis from the trading-heavy general agent descriptions.
Base again shows a broader functional mix: its Web records are led by trading (50.3\%), followed by content-related services ($13.6\%$), while its OASF records are dominated by agent and blockchain infrastructure-related services ($90\%$). 
Together, these results connect the service-type counts to the functions advertised through those services.

\section{Ethical Considerations}
\label{app:ethics}

Our study relies exclusively on publicly available data: on-chain events from the ERC-8004 Identity and Reputation registries, publicly resolvable registration and feedback files referenced by those events, and public x402-related payments. We do not recruit human subjects, submit transactions or feedback, create agent identities, or intervene in the ecosystem. All off-chain resources are accessed through ordinary read-only requests, and we do not attempt to circumvent access controls.

Although these artifacts are public, blockchain addresses, service endpoints, and off-chain payloads may still be linkable. We therefore avoid intentional deanonymization, do not map addresses to real-world identities, and report aggregate statistics whenever possible. Specific addresses are mentioned only when necessary to document a material security pattern.

To study the security of reputation markets, we analyze public data to identify weaknesses such as feedback manipulation and Sybil-like behavior. We do not attempt to exploit these vulnerabilities or provide attack automation, tooling, or procedural details that could facilitate abuse.

We have also responsibly disclosed our key findings to the ERC-8004 proposer (Ethereum Team). We received their feedback and incorporated the resulting revisions into this paper.

\section{AI Usage Disclosure}
\label{app:AI_usage}
During manuscript preparation and revision, we used GPT-5.5, GPT-6.1 Sol, and Claude Opus 5.5 to polish grammar and wording. All substantive content and research contributions are the authors' original work.

\end{document}